\documentclass[oldversion]{aa}
\usepackage{txfonts}
\usepackage{natbib}
\usepackage{subfigure}
\bibpunct{(}{)}{;}{a}{}{,} 
\usepackage{graphicx}
\usepackage{amssymb}
\usepackage{threeparttable}
\usepackage{rotating}
\usepackage{epsfig}

\usepackage{epstopdf}

\newcommand{\mdot}{\mathrm{M}_{\odot}~\mathrm{yr}^{-1}}
\newcommand{\lum}{\mathrm{erg~s}^{-1}}

\newcommand{\flux}{\mathrm{erg~cm}^{-2}~\mathrm{s}^{-1}}
\newcommand{\cnts}{\mathrm{counts~s}^{-1}}
\newcommand{\nh}{\mathrm{cm}^{-2}}

\newcommand{\sgra}{Sgr~A$^{*}$}
\newcommand{\grsbron}{GRS~1741--2853}

\newcommand{\xmmbron}{XMM~J174457--2850.3}
\newcommand{\nxmmbron}{\#6}
\newcommand{\ascabron}{AX~J1745.6--2901}

\newcommand{\brontwee}{CXOGC~J174535.5--290124}

\newcommand{\ksbron}{KS~1741--293}

\newcommand{\grobron}{GRO J1744--28}

\newcommand{\saxbron}{SAX~J1747.0--2853}

\newcommand{\bronnegen}{CXOGC~J174541.0--290014}
\newcommand{\nbronnegen}{\#8}
\newcommand{\newsource}{XMMU J174654.1--291542}

\newcommand{\andereascabron}{AX~J1742.6--2901}
\newcommand{\nandereascabron}{\#10}

\newcommand{\brondrie}{CXOGC~J174540.0--290005}
\newcommand{\bronvier}{Swift~J174553.7--290347}
\newcommand{\bronviercxo}{CXOGC~J174553.8--290346}
\newcommand{\bronvijf}{Swift~J174622.1--290634}
\newcommand{\bronvijfcxo}{CXOGC~J174622.2--290634}
\newcommand{\bronacht}{CXOGC~J174538.0--290022}
\newcommand{\adcbron}{CXOGC~J174540.0--290031}

\newcommand{\munotransientnew}{CXOGC~J174451.7--285308}

\newcommand{\newswift}{Swift J174535.5--285921}
\newcommand{\newswiftcxo}{CXOGC J174535.6--285928}
\newcommand{\exo}{EXO 0748-676}

\newcommand{\swift}{\textit{Swift}}
\newcommand{\chan}{\textit{Chandra}}
\newcommand{\xmm}{\textit{XMM-Newton}}

\newcommand{\inte}{\textit{INTEGRAL}}
\newcommand{\asca}{\textit{ASCA}}
\newcommand{\rxte}{\textit{RXTE}}
\newcommand{\bepposax}{\textit{BeppoSAX}}
\newcommand{\granat}{\textit{Granat}}
\newcommand{\rosat}{\textit{ROSAT}}
\newcommand{\einstein}{\textit{Einstein}}
\newcommand{\maxi}{\textit{MAXI}}
\newcommand{\suzaku}{\textit{Suzaku}}

\begin{document}

\title{A four-year \xmm/\chan\ monitoring campaign of the Galactic Centre: analysing the X-ray transients}

\titlerunning{\xmm/\chan\ monitoring of the GC}

\author{
N. Degenaar\inst{1,2}\thanks{e-mail: degenaar@umich.edu} \fnmsep \thanks{Hubble fellow}
\and R.~Wijnands\inst{1}
\and E.~M.~Cackett\inst{3}
\and J.~Homan\inst{4}
\and J.~J.~M.~in 't Zand\inst{5}
\and E.~Kuulkers\inst{6}
\and T.~J.~Maccarone\inst{7}
\and M.~van der Klis\inst{1}
}

\authorrunning{N. Degenaar et al.}

\institute{
University of Amsterdam, Postbus 94249, 1098 SJ, Amsterdam, the Netherlands
\and University of Michigan, Department of Astronomy, 500 Church St, Ann Arbor, MI 48109, USA
\and Institute of Astronomy, University of Cambridge, Madingley Road, Cambridge CB3 0HA, UK
\and Center for Space Research, Massachusetts Institute of Technology, 77 Massachusetts Avenue, Cambridge, MA 02139, USA 
\and SRON National Institute for Space Research, Sorbonnelaan 2, 3584 CA, Utrecht, The Netherlands
\and ISOC, ESA/ESAC, Urb. Villafranca del Castillo, P.O. Box 50727, 28080 Madrid, Spain
\and School of Physics and Astronomy, University of Southampton, Southampton SO17 1BJ, UK
}

\date{Received 24 April 2012 / Accepted 8 June 2012}

\abstract {
We report on the results of a four-year long X-ray monitoring campaign of the central 1.2 square degrees of our Galaxy, performed with \chan\ and \xmm\ between 2005 and 2008. Our study focuses on the properties of transient X-ray sources that reach 2--10 keV luminosities of $L_X\gtrsim10^{34}~\lum$ for an assumed distance of 8~kpc. There are 17 known X-ray transients within the field of view of our campaign, eight of which were detected in outburst during our observations: the transient neutron star low-mass X-ray binaries \grsbron, \ascabron, \saxbron, \ksbron\ (all four are also known X-ray bursters), and \grobron\ (a 2.1 Hz X-ray pulsar), and the unclassified X-ray transients \xmmbron, \brontwee\ and \bronnegen. We present their X-ray spectra and flux evolution during our campaign, and discuss our results in light of their historic activity. Our main results include the detection of two thermonuclear X-ray bursts from \saxbron\ that were separated by an unusually short time interval of 3.8~min. Investigation of the lightcurves of \ascabron\ revealed one thermonuclear X-ray burst and a $\sim$1600-s long X-ray eclipse. We found that both \xmmbron\ and \grobron\ displayed weak X-ray activity above their quiescent levels at $L_X\sim10^{33-34}~\lum$, which is indicative of low-level accretion. We compare this kind of activity with the behaviour of low-luminosity X-ray transients that display 2--10 keV peak luminosities of $L_X\sim10^{34}~\lum$ and have never been seen to become brighter. In addition to the eight known X-ray transients, we discovered a previously unknown X-ray source that we designate \newsource. This object emits most of its photons above 2 keV and appears to be persistent at a luminosity of $L_X\sim10^{34}~\lum$, although it exhibits strong spectral variability on a time scale of months. Based on its X-ray properties and the possible association with an infrared source, we tentatively classify this object as a cataclysmic variable. No new transients were found during our campaign, reinforcing the conclusion of previous authors that most X-ray transients recurring on a time scale of less than a decade have now been identified near the Galactic centre. 
}

\keywords{
X-rays: binaries - 
Stars: neutron - 
Accretion, accretion discs - 
Galaxy: centre - 
X-rays: individuals: 
\ascabron, \brontwee, \grsbron, \xmmbron, \bronacht, \ksbron, \grobron, \saxbron, \andereascabron, \newsource.
}

\maketitle 


\section{Introduction}\label{sec:intro}
The region around \sgra, the dynamical centre of our Galaxy, has been observed at various spatial scales and in different energy bands by many past and present X-ray missions. At early times dedicated monitoring campaigns using \einstein\ \citep[][]{watson1981}, \granat\ \citep[][]{churazov1994,pavlinsky1994}, \rosat\ \citep[][]{sidoli01}, \asca\ \citep[][]{sakano02} and \bepposax\ \citep[][]{sidoli99,zand04} have led to the discovery of several X-ray point sources located within the central degrees of our Galaxy. 

More recently, an intensive monitoring campaign carried out with \chan\ between 1999 and 2006 has resolved thousands of distinct X-ray sources in a field of $2^{\circ}\times0.8^{\circ}$ around the Galactic centre \citep[GC;][]{wang2002,baganoff2003,muno03,muno04_apj613,muno2006,muno2009}. Furthermore, starting in 2006 the inner $\sim25'\times25'$ around \sgra\ has been monitored on an almost daily basis with \swift\ \citep[][]{kennea_monit,degenaar09_gc,degenaar2010_gc}, whereas a region subtending many degrees has been regularly scanned by \rxte\ since 1999 \citep[][]{swank2001} and by \inte\ since 2005 \citep[][]{kuulkers07}.

The plethora of X-ray sources found in the direction of the innermost parts of our Galaxy encompasses a variety of objects \citep[e.g.][]{muno04_apj613}. The population of X-ray sources with luminosities of $L_X\lesssim10^{33}~\lum$ (2--10 keV) is thought to be dominated by accreting white dwarfs and active stars \citep[e.g.][]{verbunt1984,verbunt1997,muno2006,rev2009}.\footnote{All fluxes and luminosities quoted in this work refer to the 2--10 keV energy band unless stated otherwise.} The {\it brightest} Galactic X-ray point sources, however, have peak luminosities of $L_X\sim10^{36-39}~\lum$ and can be identified with accreting neutron stars or black holes. Based on the type of companion star, these are classified as either high-mass X-ray binaries (HMXBs; donor mass $\gtrsim10~\mathrm{M_{\odot}}$) or low-mass X-ray binaries (LMXBs; donor mass $\lesssim1~\mathrm{M_{\odot}}$). 

Both LMXBs and HMXBs can be {\it transient}: such systems spend the majority of their time (years/decades) in a quiescent state during which the X-ray luminosity is generally $L_X\lesssim10^{33}~\lum$, while their intensity typically reaches up to $L_X\sim10^{36-39}~\lum$ during short (weeks/months) outburst episodes. This large X-ray variability is ascribed to changes in the mass-accretion rate onto the compact primary. In LMXBs this is thought to be caused by instabilities in the accretion disk or the Roche-lobe overflowing companion star. In HMXBs the compact primary accretes matter that is expelled by the companion star via a wind or a circumstellar disk, and transient behaviour may be the result of variations in the mass-loss rate of the companion or the binary geometry.

\subsection{Low-luminosity X-ray transients}
Repeated observations of the region around \sgra\ with \chan, \xmm\ and \swift\ have revealed a population of transient X-ray sources that have 2--10 keV peak luminosities of $L_X\sim 10^{34-36}~\lum$ \citep[e.g.][]{muno05_apj622,sakano05,porquet05,degenaar09_gc}. Their outburst amplitudes and spectral properties suggest that these objects are X-ray binaries in which the compact object accretes at a very low rate from its companion star, thereby causing a relatively low X-ray luminosity \citep[e.g.][]{pfahl2002,belczynski2004,muno05_apj622,wijnands06}. Earlier X-ray missions already provided a glimpse of such low-luminosity X-ray transients \citep[][]{sunyaev1990,zand1991,maeda1996}, but the current generation of instruments exploiting sensitive and high spatial resolution X-ray imaging have considerably improved our understanding of the number and behaviour of such objects. 

Observations of X-ray binaries accreting at low X-ray luminosities can address several questions related to stellar and binary evolution, as well as accretion flows at low rates. For instance, constraining the number and nature of low-luminosity X-ray transients allows us to gain more insight into the statistics of different source classes, and can serve as an important calibration point for population synthesis models \citep[e.g.][]{pfahl2002,belczynski2004}. Furthermore, the mass-accretion rate averaged over thousands of years, $\langle \dot{M} \rangle _{\mathrm{long}}$, plays an important role in the evolution of LMXBs. First studies have shown that values of $\langle \dot{M} \rangle _{\mathrm{long}} <10^{-13}~\mdot$ put tight constraints on the possible evolutionary paths, and might require an unusual type of binary such as neutron stars accreting from a hydrogen-depleted or planetary companion \citep[][]{king_wijn06}. Monitoring observations are an important tool for estimating the time-averaged mass-accretion rates of low-luminosity transients \citep[][]{degenaar09_gc,degenaar2010_gc}. Repeated non-detections allow us to place more stringent upper limits on their long-term averaged accretion rates and can therefore be as interesting as actual detections. 

Furthermore, thermonuclear X-ray bursts observed from slowly accreting neutron stars have provided important new insight into the physics of nuclear burning on the surface of these compact objects \citep[e.g.][]{cornelisse02,zand05,cooper07,peng2007,degenaar2010_burst}. Moreover, studying the low-luminosity transients gives insight into the accretion process at low mass-accretion rates. Some bright transient X-ray binaries that exhibit accretion outbursts with intensities of $L_X\gtrsim 10^{36}~\lum$ have also been observed to undergo episodes of low-level accretion with X-ray luminosities of $L_X\sim 10^{33-35}~\lum$ \citep[e.g.][]{wijnands2001_1808,wijnands2002_saxj1747,linares2008,sidoli08,degenaar09_gc,degenaar2011_1701,fridriksson2011,romano2011}. This provides an interesting comparison with the outbursts of the low-luminosity X-ray transients.

\subsection{This work}
Dedicated surveys aiming to search for and to identify low-luminosity X-ray transients have the potential to unveil rare types of accreting compact objects, and can provide valuable input for theoretical models. Here, we report on the results from a joint \chan\ and \xmm\ monitoring campaign that covered a region of 1.2 square degrees around \sgra\ and was carried out between 2005--2008. \citet{wijnands06} reported on the initial results of our monitoring observations, discussing the first series of data obtained in 2005 June--July. 

The main goal of this programme is to investigate the X-ray properties (below 10 keV) of transient objects that have low 2--10 keV peak luminosities of $L_X\sim 10^{34-36}~\lum$. Owing to the high concentration of X-ray point sources in the inner square degree around \sgra, as well as sensitivity limitations, such systems are often inaccessible to the current monitoring instruments. Our \chan/\xmm\ campaign provides the means to refine our understanding of currently known low-luminosity systems (e.g. characterise their outburst behaviour, improve estimates of their duty cycles and time-averaged mass-accretion rates), to search for new X-ray transients and to capture low-level accretion activity in bright X-ray binaries.

We present our work as follows. We describe the setup of the programme in Section 2, and proceed by detailing the reduction and analysis procedures in Section 3. The results of our temporal and spectral analysis of ten different X-ray sources are presented on a case-by-case basis in Appendix~\ref{sec:appendix}, while in Section 4 we summarise the main results and highlight a few individual sources. We end in Section 5 where we give an overview of all transient X-ray sources located in the region covered by our campaign. We discuss the implications of our findings for understanding the nature of the GC X-ray transients and low-level accretion activity.

\begin{table}
\begin{threeparttable}
\begin{center}
\caption[]{{Log of the monitoring observations.}}
\begin{tabular}{l l l l l}
\hline
\hline
Field & Obs ID & Date & $t_{\mathrm{exp}}$ (ks) & Observatory \\
\hline 
GC-1 & 6188 & 2005-06-05 & 5.1 & \chan\  \\
GC-2 & 6190 & 2005-06-05 & 5.2 & \chan\  \\	 
GC-3 & 6192 & 2005-06-05 & 5.1 & \chan\  \\
GC-4 & 6194 & 2005-06-05 & 5.1 & \chan\  \\   
GC-5 & 6196 & 2005-06-05 & 5.1 & \chan\  \\    
GC-6 & 6198 & 2005-06-05 & 5.1 & \chan\  \\
GC-7 & 6200 & 2005-06-05 & 5.1 & \chan\  \\
\hline 
GC-1 & 6189 & 2005-10-18 & 4.3 & \chan\ \\
GC-2 & 6191 & 2005-10-20 & 4.3 & \chan\   \\
GC-3 & 6193 & 2005-10-20 & 4.3 & \chan\   \\
GC-4 & 6195 & 2005-10-20 & 4.4 & 	\chan\   \\
GC-5 & 6197 & 2005-10-20 & 4.3 & \chan\  \\
GC-6 & 6199 & 2005-10-20 & 4.4 & \chan\   \\
GC-7 & 6201 & 2005-10-21 & 4.3 & \chan\   \\
\hline  
GC-1 & 0302882501 & 2006-02-27 & 9.1 & \xmm\   \\
GC-2 & 0302882601 & 2006-02-27 & 6.5 & \xmm\   \\
GC-3 & 0302882701 & 2006-02-27 & 6.8 & \xmm\   \\
GC-4 & 0302882801 & 2006-02-27 & 7.5 & \xmm\   \\
GC-5 & 0302882901 & 2006-02-27 & 7.5 & \xmm\   \\
GC-6 & 0302883001 & 2006-02-27 & 7.5 & \xmm\   \\
GC-7 & 0302883101 & 2006-02-27 & 11.3 & \xmm\   \\
GC-1 & 0302883201 & 2006-03-29	& 6.4 & \xmm\ \\
\hline 
GC-1 & 0302883901 & 2006-09-08 & 6.5 & \xmm\   \\
GC-2 & 0302884001 & 2006-09-08 & 6.5 & \xmm\   \\
GC-3 & 0302884101 & 2006-09-08 & 6.5 & \xmm\   \\
GC-4 & 0302884201 & 2006-09-08 & 6.5 & \xmm\   \\
GC-5 & 0302884301 & 2006-09-09 & 6.5 & \xmm\   \\
GC-6 & 0302884401 & 2006-09-09 & 5.5 & \xmm\   \\
GC-7 & 0302884501 & 2006-09-09 & 8.3 & \xmm\   \\
\hline   	
GC-1 & 8531 & 2007-07-24 & 5.1 & \chan\   \\	  
GC-2 & 8532 & 2007-07-24 & 5.1 & \chan\   \\	  
GC-4 & 8533 & 2007-07-24 & 5.1 & \chan\   \\	 
GC-5 & 8534 & 2007-07-24 & 5.1 & \chan\   \\	  
GC-6 & 8535 & 2007-07-24 & 5.1 & \chan\   \\	  
GC-7 & 8536 & 2007-07-24 & 5.1 & \chan\   \\	 
\hline 
GC-1 & 0504940101 & 2007-09-06 & 6.5	 & \xmm\  	\\	
GC-2 &0504940201 & 2007-09-06 & 12.5 & \xmm\  	\\	
GC-4 &0504940401 & 2007-09-06 & 6.5	 & \xmm\  	\\	
GC-5 &0504940501 & 2007-09-06 & 6.5	 & \xmm\   \\		
GC-6 &0504940601 & 2007-09-06 & 6.5	 & \xmm\  	\\	
GC-7 &0504940701 & 2007-09-06 & 6.5	 & \xmm\   \\
\hline 
GC-2 &0511000101	& 2008-03-03 & 8.4	& \xmm\  	\\
GC-2 &0511000301	& 2008-03-03 & 6.5	& \xmm\  	\\
GC-3 &0511000501	& 2008-03-04 & 6.5	& \xmm\  	\\
GC-4 &0511000701	& 2008-03-04 & 6.5	& \xmm\  	\\
GC-5 &0511000901	& 2008-03-04 & 6.5	& \xmm\  	\\
GC-6 &0511001101	& 2008-03-04 & 6.5	& \xmm\  	\\
GC-7 &0511001301	& 2008-03-04 & 6.5	& \xmm\  	\\
\hline 
GC-1 & 9030 & 2008-05-10 & 5.1 & \chan\   \\
GC-2 & 9073 & 2008-05-10 & 5.1 & \chan\   \\	
GC-3 & 9031 & 2008-05-10 & 5.1 & \chan\   \\	
GC-4 & 9032 & 2008-05-10 & 5.1 & \chan\   \\	
GC-5 & 9033 & 2008-05-10 & 5.1 & \chan\   \\
GC-6 & 9074 & 2008-05-11 & 5.1 & \chan\   \\	
GC-7 & 9034 & 2008-05-10 & 5.1 & \chan\   \\
\hline 
GC-1 & 9035 & 2008-07-15 & 5.1 & \chan\   \\
GC-2 & 9036 & 2008-07-15 & 5.1 & \chan\   \\	
GC-3 & 9037 & 2008-07-16 & 5.1 & \chan\   \\	
GC-4 & 9038 & 2008-07-16 & 5.1 & \chan\   \\	
GC-5 & 9039 & 2008-07-16 & 5.1 & \chan\   \\
GC-6 & 9040 & 2008-07-16 & 5.1 & \chan\   \\	
GC-7 & 9041 & 2008-07-16 & 5.1 & \chan\   \\
\hline 
GC-1 &0511000201 & 2008-09-23 & 6.5	& \xmm\  	\\
GC-2 &0511000401 & 2008-09-23 & 4.3	& \xmm\  	\\
GC-3 &0511000601 & 2008-09-23 & 6.5	& \xmm\  	\\
GC-4 &0511000801 & 2008-09-27 & 6.5	& \xmm\  	\\
GC-5 &0511001001 & 2008-09-27 & 6.5	& \xmm\  	\\
GC-6 &0511001201 & 2008-09-27 & 6.5	& \xmm\  	\\
GC-7 &0511001401 & 2008-09-27 & 6.5	& \xmm\  	\\
\hline
\end{tabular}
\label{tab:obs_monit}
\begin{tablenotes}
\item[]
\end{tablenotes}
\end{center}
\end{threeparttable}
\end{table}

\begin{table}
\begin{threeparttable}
\begin{center}
\caption[]{{Log of \chan/ACIS-I follow-up pointings.}}
\begin{tabular}{l l l l}
\hline
\hline
Field & Obs ID & Date & $t_{\mathrm{exp}}$ (ks)\\
\hline 
\grsbron\ & 6311 & 2005-07-01 & 4.0 \\
\grsbron & 6602 & 2007-03-12 & 5.1    \\
\grsbron & 6603 & 2007-04-06 & 4.9    \\
\grsbron & 6604 & 2007-04-18 & 5.1  \\   
\grsbron & 6605 & 2007-04-30 & 5.1   \\      
\grsbron & 6606 & 2007-05-16 & 5.0   \\ 
\hline
\end{tabular}
\label{tab:obs_followup}
\begin{tablenotes}
\item[]
\end{tablenotes}
\end{center}
\end{threeparttable}
\end{table}


\section{Description of the monitoring programme}\label{sec:obs}

\subsection{Observations}
Our choice to monitor the central square degree of our Galaxy was motivated by the fact that this region is populated by nearly 20 known X-ray transients, several of which undergo sub-luminous accretion episodes \citep[][]{muno05_apj633,wijnands06,kennea06_atel920,degenaar09_gc,degenaar2011_newtransient}. The relatively wide field of view (FOV; $\sim30'\times30'$) and large collecting area ($\sim$1\,100$~\mathrm{cm}^{2}$ at 1 keV) of \xmm\ make it an excellent facility for surveying sky regions down to relatively faint flux levels. 

We used the data obtained with the European Photon Imaging Camera (EPIC), which consists of one PN \citep[][]{struder2001_pn} and two MOS \citep[][]{turner2001_mos} detectors that are sensitive in the 0.1--15 keV range and have spectral imaging capabilities. The PN is an array of 12 CCDs ($64\times200$ pixels each), while the MOS units are composed of an array of seven CCDs, each consisting of $600\times600$ pixels. A micrometeorite strike damaged one of the CCDs of the MOS1, which is operated with only six detectors since then \citep[][]{abbey06}.

The \xmm\ observations are complemented by \chan\ pointings that provide high spatial (sub-arcsec) resolution and a very low X-ray background within an energy band of 0.1--10 keV. We chose the High Resolution Camera \citep[HRC;][]{kenter2000_hrc} as our prime \chan\ instrument, because it provides the largest FOV ($\sim30' \times 30'$), comparable in size to \xmm. The HRC-I is a square micro-channel plate detector (made up of $32\,768\times32\,768$ pixels) that has an effective area of $225~\mathrm{cm}^{2}$ at 1 keV and is designed for imaging observations. Because the energy resolution of the HRC is poor, we obtained a few additional pointings with the Advanced CCD Imaging Spectrometer \citep[ACIS;][]{garmire2003_acis} to follow up active transients, aiming to obtain spectral information. The ACIS-I consists of a four-chip imaging array (each having $1\,024\times1\,024$ pixels), providing an effective area of $340~\mathrm{cm}^{2}$ at 1 keV and a FOV of $\sim16'\times16'$.

Our \chan/\xmm\ campaign covers 1.2 square degrees around \sgra, sub-divided into seven different pointing directions \citep[named GC-1, GC-2, GC-3, GC-4, GC-5, GC-6 and GC-7;][]{wijnands06}.\footnote{We note that the naming of the different pointing directions changed during our campaign. The fields that were initially denoted as GC-7, GC-8, GC-9 and GC-10 in the 2005--2006 observations \citep[see][]{wijnands06} were in 2007--2008 renamed GC-4, GC-5, GC-6 and GC-7, respectively. We adopt the latter indications throughout this paper.} Adjacent pointings partially overlapped by a few arcminutes (see Fig.~\ref{fig:chan}). The programme comprises 34 \chan/HRC-I and 35 \xmm/EPIC pointings, carried out in ten different epochs between 2005 June and 2008 September. An overview of the monitoring observations is given in Table~\ref{tab:obs_monit}. Follow-up \chan/ACIS-I  observations were performed in 2005 July (one pointing) and 2007 March--May (a series of five pointings) as listed in Table~\ref{tab:obs_followup}.

\subsection{Sensitivity and X-ray images}
The exposure time of individual observations was typically 5--10~ks. Depending on the spectral properties, a 5 ks \chan/HRC-I pointing can detect sources (near aimpoint) down to a 2--10 keV luminosity of $L_X\sim(3-6)\times10^{33}~(D/\mathrm{8~kpc})^2~\lum$ for photon indices of $\Gamma=1.0-3.0$ and hydrogen column densities of $N_H=(5-10)\times10^{22}~\nh$. The \xmm\ observations are a factor of a few more sensitive. With our programme we reached X-ray luminosities that are a factor of $\sim$$100-1000$ deeper than the sensitivity of wide-field monitoring instruments (e.g. \rxte/ASM, \rxte/PCA, \swift/BAT, \maxi, \inte), which are typically limited to $L_X\gtrsim10^{35}~(D/\mathrm{8~kpc})^2~\lum$ in their instrument passbands. In addition, \xmm\ and \chan\ provide (sub-) arcsecond spatial resolution, while these other instruments typically yield positional uncertainties of tens of arcseconds to arcminutes. 

\begin{figure*}
 \begin{center}
	\includegraphics[width=9.5cm]{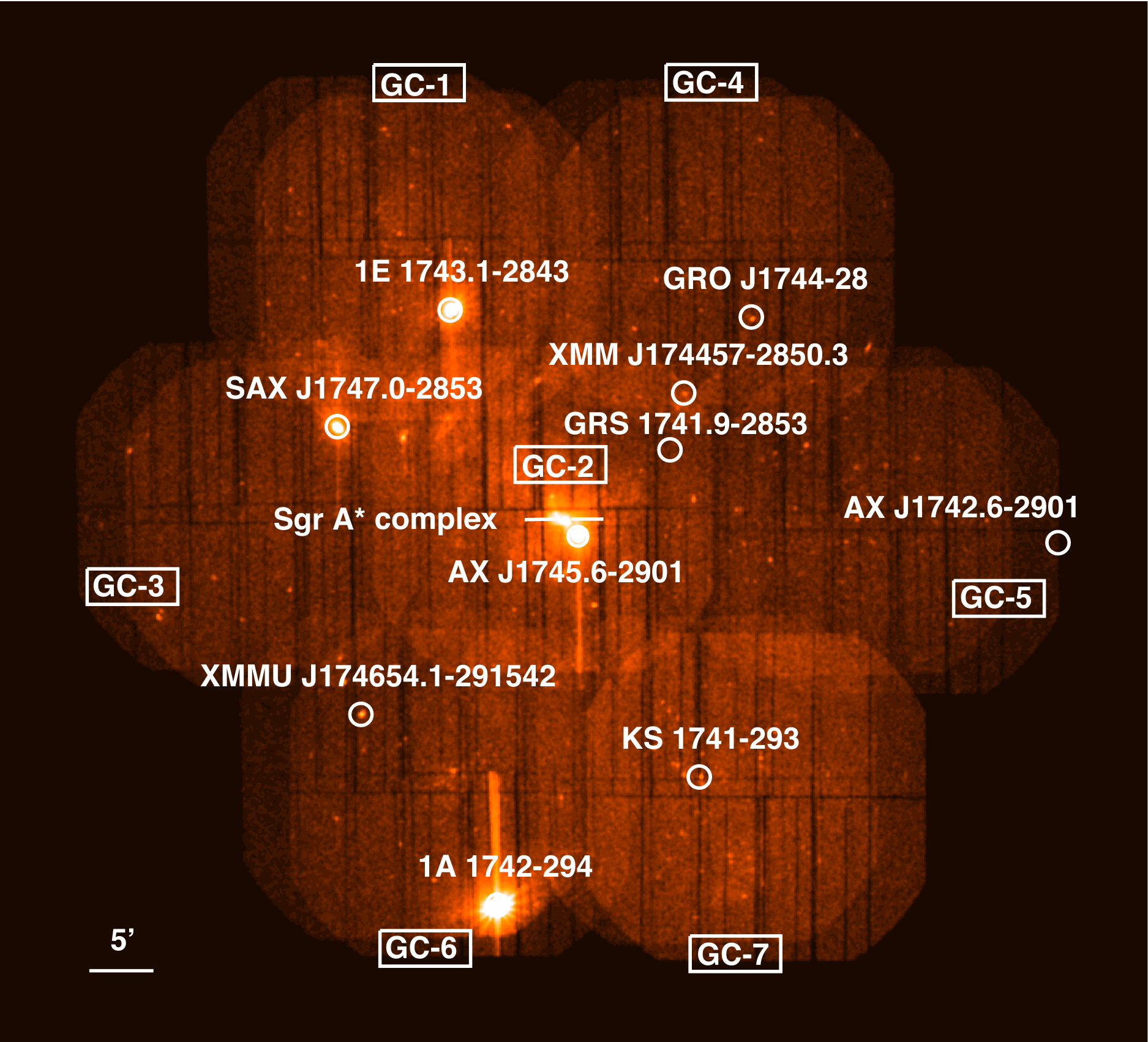}\vspace{+0.3cm}
          \includegraphics[width=9.5cm]{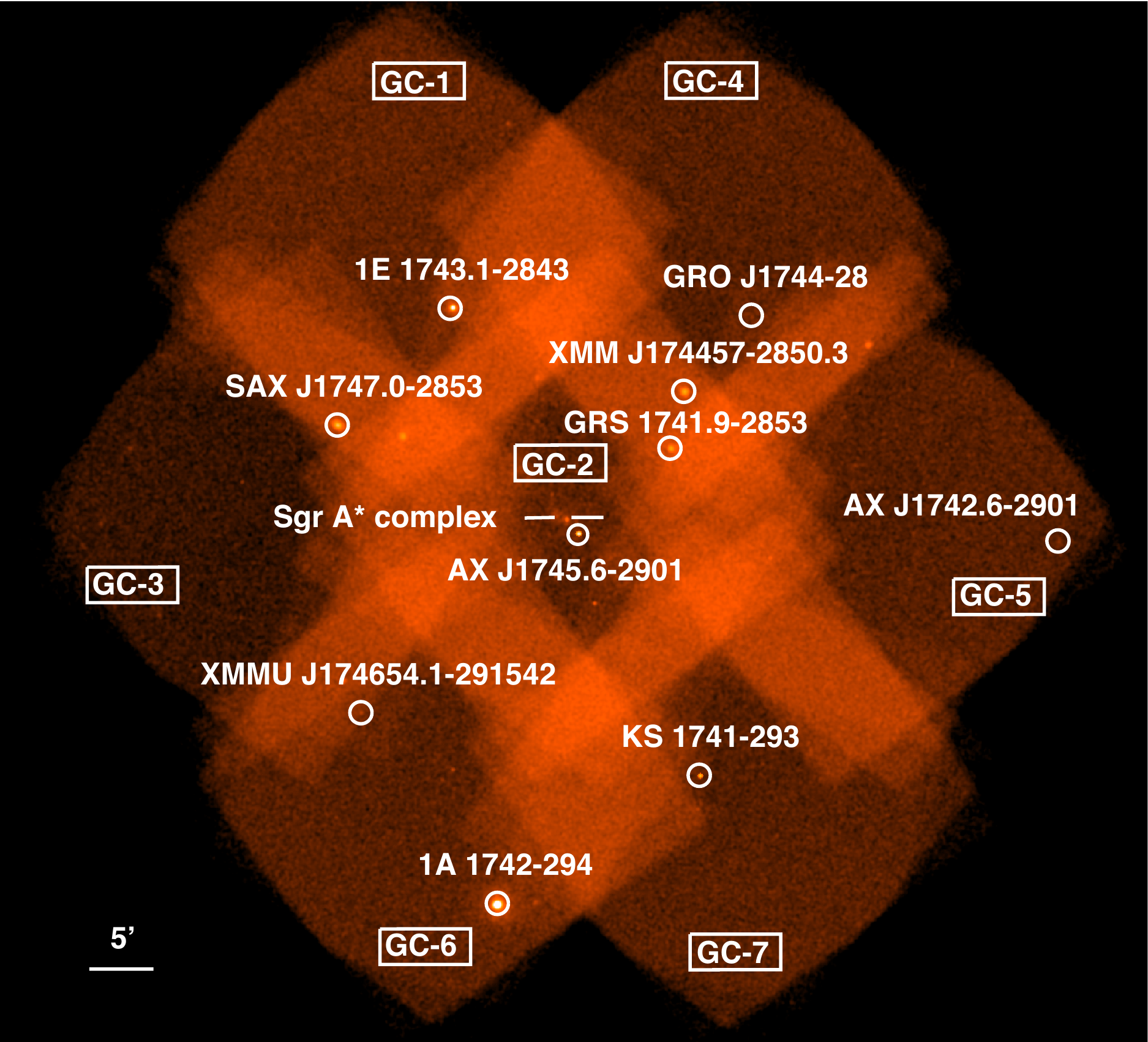}\vspace{+0.3cm}
	\includegraphics[width=9.5cm]{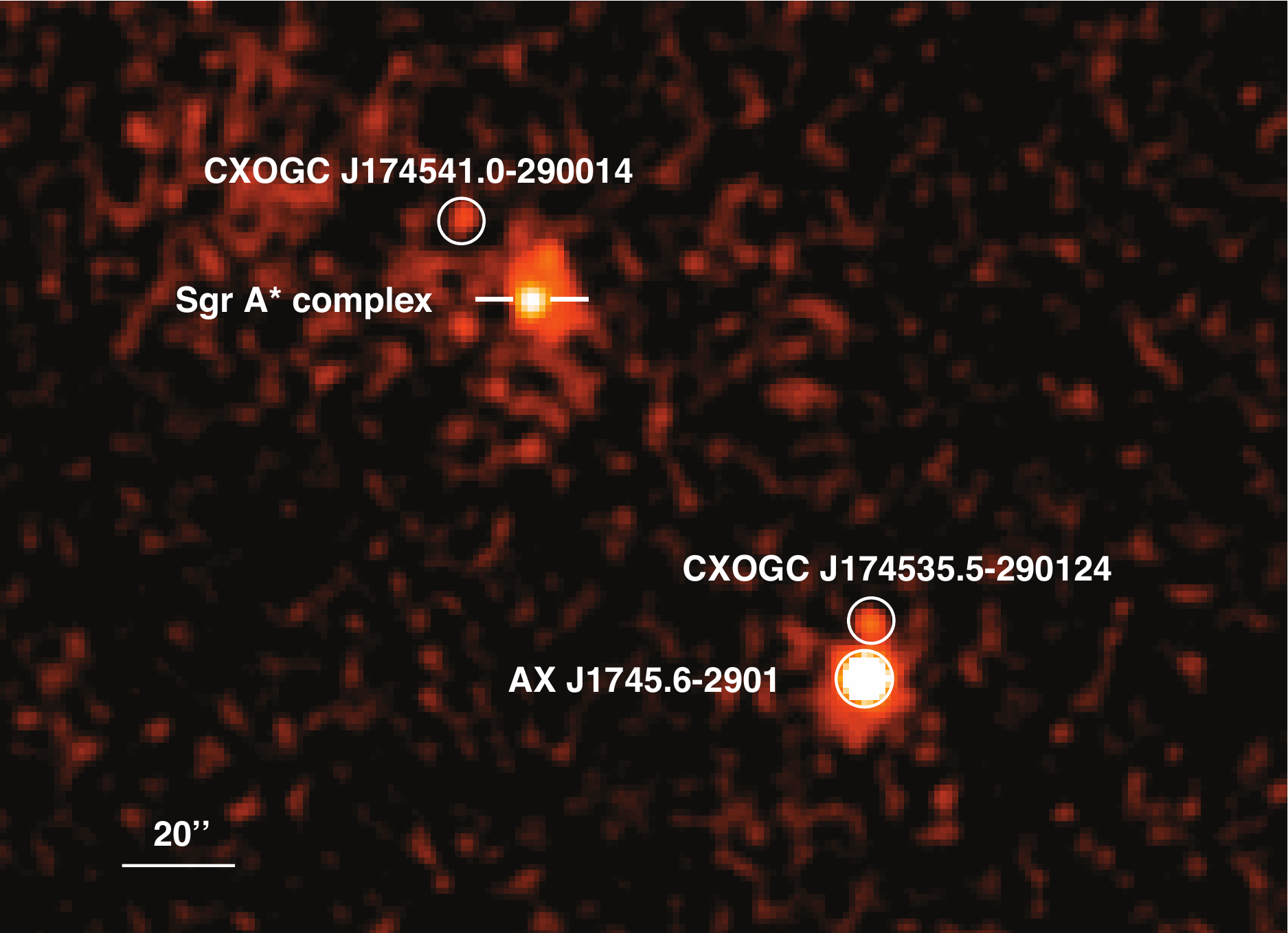}
    \end{center}
    \caption[]{Composite X-ray images of our monitoring campaign (2005--2008). The field names of the different pointing directions are given in boxes. Active transients and persistent X-ray binaries are indicated by circles. Unlabelled X-ray sources can be identified with stars or star clusters. Top: \xmm/PN mosaic. Middle: \chan/HRC-I mosaic. Bottom: \chan/HRC-I image magnified to display the inner $\sim$1.5$'$ around \sgra.}
 \label{fig:chan}
\end{figure*}

The entire programme spanned a period of 39 months (3.25 yr), for a cumulative exposure time of 412.7~ks (168.1 ks with \chan, 244.6 ks with \xmm). Subsequent pointings were separated by 2--10 months (see Table~\ref{tab:obs_monit}). The total exposure time reached in the different pointing directions is $\sim 46-68$~ks. Mosaic images of the \chan/HRC-I and \xmm/PN data are shown in Fig.~\ref{fig:chan}. Apart from diffuse X-ray structures (e.g. around \sgra), these images reveal several X-ray point sources (see also Section~\ref{sec:results}). The locations of active transients and two persistent X-ray binaries (1E 1743.1--2843 and 1A 1742--294) are indicated by circles and the cross-hair in the centre of the images shows the position of the \sgra\ complex. Fig.~\ref{fig:chan} also includes a zoomed \chan/HRC image of the inner $\sim$$1.5'$ around \sgra, where three active X-ray transients were detected.

\section{Data analysis}\label{sec:data_ana}
For the present study we were only interested in (candidate) transient X-ray binaries. We therefore focused on transient X-ray sources that have a 2--10 keV peak luminosity $L_X\gtrsim1\times10^{34}~\lum$ for an assumed distance of 8 kpc, since there will be a high fraction of cataclysmic variables among the fainter objects \citep[][]{verbunt1997,muno2003,muno2009}. We searched for transient sources in our \chan\ and \xmm\ data by comparing images of different epochs with one another. 

The objects detected in our observations were correlated with the SIMBAD database to identify the known X-ray sources in our sample based on positional coincidence. Furthermore, we overlaid the positions of sources found in our campaign on an optical image from the Digital Sky Survey (DSS) and an infra-red image from the Two Micron All Sky Survey (2MASS), to filter out likely foreground objects (e.g. active stars). Spectral information obtained from our \xmm\ and \chan/ACIS-I observations also aids to identify transients that are located near or beyond the GC (i.e., at a distance of $D\gtrsim8$~kpc). These sources will appear relatively hard in X-rays, since the softer photons (below $\sim2$~keV) will be strongly absorbed by the interstellar medium in the direction of the GC (hydrogen column densities of several times $10^{22}$~atoms$~\nh$ are typical in this region). X-ray sources with detectable emission below 2 keV are likely to be foreground X-ray active stars or cataclysmic variables that are located within a few kiloparsecs from the Sun. 

To characterise the X-ray spectra and to calculate source fluxes, we fitted the obtained spectral data between 0.5--10 keV using \textsc{XSpec} \citep[v. 12.6;][]{xspec}. We used a simple powerlaw model (POWERLAW), modified by interstellar absorption (PHABS). For this we used the default abundances and cross-sections available in \textsc{XSpec}. Using the tool \textsc{grppha}, we grouped the spectra of the brightest sources to contain a minimum number of 20 photons per bin, whereas fainter objects were binned into groups of at least 10 or 5 photons. For a small number of counts we also fitted the unbinned spectra without background subtraction using Cash-statistic (CSTAT in \textsc{XSpec}). Since this yielded spectral parameters and fluxes that were consistent with those obtained using the minimum $\chi^2$-method, we only report on the results obtained using the latter.

Whenever a source was detected during multiple observations, we fitted the spectra simultaneously with the hydrogen column density tied between the individual observations. We converted the deduced unabsorbed 2--10 keV fluxes into luminosities by adopting a distance of 8 kpc, unless better distance estimates were available for sources, e.g. as inferred from type-I X-ray burst analysis. Finally, we created long-term lightcurves for each transient source detected during our campaign. 

The detailed data reduction and analysis procedures for both satellites are discussed in the next sections. Our analysis includes the \chan\ observations performed in 2005 June (HRC) and July (ACIS), which were previously reported by \citet{wijnands06}.


\subsection{\chan}\label{subsec:data_chan}
\subsubsection{Data reduction and source detection}\label{}
The \chan\ data were reduced and analysed using the \textsc{ciao} tools (v. 4.2). The ACIS-I observations were carried out in the faint data mode with the nominal frame time of 3.2~s. As an initial step, we reprocessed the HRC and ACIS level-1 data files following the standard data preparation procedures.\footnote{http://cxc.harvard.edu/ciao/guides.}  Each individual pointing was inspected for periods of unusually high background. No significant background flares were found during our \chan\ observations, therefore we used all data for the subsequent analysis. 

We searched the data for X-ray sources by employing the WAVDETECT tool with the default "Mexican Hat" wavelet \citep[][]{freeman2002}. To take into account sensitivity variations across the chips, we generated an exposure map for each observation, evaluated at an energy of 4 keV (the approximate energy at which we expect to detect the largest number of photons for X-ray binaries). 
For each HRC-I observation, we generated images with a binning of 4, 16 and 32 pixels and ran the detection algorithm on each of the separate images with the default input parameters. This approach allowed us to cover a range of source sizes, accommodating the variation of the point spread function (PSF) as a function of off-axis angle. We adopted a recommended significance threshold that is approximately the inverse of the total number of pixels in the image ($1\times 10^{-8}$, $1\times 10^{-7}$ and $1\times 10^{-6}$ for HRC images binned by a factor of 4, 16 and 32 respectively), which should correspond to about one expected spurious source per image \citep[][]{freeman2002}. We compiled a master source list for each observation by combining the objects detected at each image resolution. 

The ACIS-I images were binned by a factor of 2 and we employed WAVDETECT using a detection significance threshold of $1.5\times 10^{-5}$. We ran the detection routine separately for the 0.5--2 and 2--10 keV bands, to be able to distinguish between soft and hard X-ray sources. 

\subsubsection{Count rates, lightcurves and spectra}\label{}
We extracted net count rates and lightcurves for each source employing the tool DMEXTRACT (0.5--10 keV). We employed extraction regions centred on the positions found by the WAVDETECT routine and containing $\sim$95\% of the source counts. Depending on the brightness of the sources and their offset from the aimpoint of the observations, this corresponded to extraction radii of $\sim$$2-25''$. Background events were collected from a source-free region that had a radius of three times that of the source region. We visually inspected lightcurves with bin times of 5, 10 and 100 s to search for features such as thermonuclear X-ray bursts. 

For the ACIS-I data, we extracted source and background spectra using PSEXTRACT. Redistribution matrices (rmf) and ancillary response files (arf) were subsequently generated using the tasks MKACISRMF and MKARF, respectively. 
Since the HRC provides poor energy resolution, we converted the HRC-I count rates to 2--10 keV unabsorbed fluxes employing \textsc{pimms} (v. 4.1) and using either the spectral information deduced from our \chan/ACIS and \xmm\ observations, or values reported in literature (see Table~\ref{tab:spec}). If a transient source was not detected, we obtained a $2\sigma$ Bayesian statistical upper limit on the source count rate using the \textsc{ciao} tool ASPRATES.

Both \ascabron\ and \grsbron\ caused pile-up in the ACIS data. When left uncorrected, pile-up typically causes the broad-band count rate to be underestimated and the spectrum to become harder. In an attempt to circumvent these effects we used an iterative approach in which we extracted source photons from annular extraction regions with increasingly large parts of the core PSF excluded. Once the spectral parameters remained unchanged after increasing the annular radius, we assumed that the piled-up inner regions (which distort the spectral shape) were sufficiently excluded. The necessary aperture correction to the arf file was administered using ARFCORR.


\subsection{\xmm}\label{subsec:data_xmm}
\subsubsection{Data reduction and source detection}\label{}
In all \xmm\ observations the EPIC cameras were operated in full window mode. Data reduction and analysis was carried using the Science Analysis Software (\textsc{sas}; v. 10.0.0) and following standard analysis threads.\footnote{Listed at http://xmm.esac.esa.int/sas/current/documentation/threads/} Starting with the original data files, we reprocessed the MOS and PN data using the tools EMPROC and EPPROC, respectively. In order to asses the background conditions in each of the \xmm\ observations, we extracted the full-field lightcurve for pattern 0 events with energies of $\gtrsim 10$~keV for the MOS, and between 10--12~keV for the PN. This revealed that some of our observations contained background flares. We excluded these episodes by selecting only data with high-energy count rates below $0.5~\cnts$ for the MOS and below $1.0~\cnts$ for the PN. 

Source detection was carried out with the task EDETECT$\_$CHAIN, adopting the default detection likelihood of 10. We searched in two different energy bands of 0.5--2 and 2--10 keV for the PN and the MOS2. We did not include the MOS1 for source detection, since one of the CCD units was damaged by a micrometeoroid strike \citep[][]{abbey06}.\footnote{ We note that although the MOS1 was not included for source detection, we did use it for the extraction of data products: spectra and lightcurves were obtained from both MOS cameras, as well as the PN (see Section~\ref{subsubsec:dataproducts}).} 

\subsubsection{Count rates, lightcurves and spectra}\label{subsubsec:dataproducts}
We extracted count rates for all objects in our source list using the task EREGIONANALYSE (0.5--10 keV). We also employed this tool to determine the optimum extraction region (achieving the highest signal to noise ratio) for source lightcurves and spectra. This yielded source regions with radii of $\sim$$10-100''$ and a typical enclosed energy fraction of $\sim$$85-95$ percent. For the extraction of background events we used regions with a radius three times larger than that of the source, positioned on a source-free portion of the CCD. For the observations in which our transient sources were not detected, we obtained a $2\sigma$ upper limit on the count rate using EREGIONANALYSE. 

We created background-corrected lightcurves at a resolution of 5, 10 and 100~s for the PN and both MOS cameras using the tools EVSELECT and EPICLCCORR. Source and background spectra, as well as the associated rmf and arf files, were generated using the meta task ESPECGET. The spectral data were fitted within \textsc{XSpec} with all model parameters tied between the three EPIC detectors. 

During our observations, both \ascabron\ and \saxbron\ became bright enough to cause pile-up in the EPIC instruments (see Section~\ref{sec:results}). We used the \textsc{sas} task EPATPLOT to evaluate the level of pile-up in the MOS and PN data, using annular regions of increasing size. Once the observed pattern distribution matched the expected one, we chose that annular size to extract source photons.

\begin{table*}
\begin{threeparttable}[htb]
\begin{center}
\caption[]{{Spectral parameters and obtained X-ray fluxes for (candidate) transient X-ray binaries when detected.}}
\begin{tabular}{c l l l l l l l l l}
\hline
\hline
\# & Source name & Date & Instr. & Field &   $N_{\mathrm{H}}$ & $\Gamma$ & $\chi_{\nu}^2$ (d.o.f.) & $F_{\mathrm{X,\mathrm{abs}}}$ & $F_{\mathrm{X,\mathrm{unabs}}}$\\
& {\it class} & &  & &  ($10^{22}~\nh$) & & &  \multicolumn{2}{c}{($10^{-12}~\flux$)} \\
\hline 
1 & \grsbron\ & &  & &    $11.4\pm1.1$ & &  $0.84$ (176) && \\
& {\it bursting neutron star LMXB} & 2005-06-05 & HRC-I & GC-2 &  & 2.0 fix & & $49.5\pm1.5$ & $98.6\pm3.1$  \\
& & 2005-07-01 & ACIS-I & &   & $2.0\pm0.3$ & & $12.0\pm0.6$ & $24.5\pm2.0$ \\
& & 2007-03-12 & ACIS-I &   & & $1.5\pm0.2$ & & $20.5\pm0.8$ & $37.3\pm2.4$ \\
& & 2007-04-06 & ACIS-I &  &  & $1.6\pm0.4$ & & $4.55\pm0.40$ & $8.38\pm0.49$ \\
2 &\ascabron & &  & &   $21.8\pm0.3$ & & 1.25 (4185) && \\
& {\it bursting neutron star LMXB} & 2006-02-27 & PN & GC-2 &   & $1.9\pm0.1$ & & $21.3\pm0.3$ & $60.2\pm2.2$ \\
& & 2007-03-12 & ACIS-I &  &    & $1.6\pm0.1$ & & $139\pm2$ & $353\pm8$ \\
& & 2007-04-06 & ACIS-I &  &    & $1.8\pm0.1$ & & $95.8\pm2.1$ & $258\pm7$ \\
& & 2007-04-18 & ACIS-I &  &   & $1.6\pm0.1$ & & $149\pm3$ & $375\pm8$ \\
& & 2007-04-30 & ACIS-I &  &    & $1.8\pm0.1$ & & $115\pm2$ & $311\pm8$ \\
& & 2007-05-16 & ACIS-I &  &    & $1.6\pm0.1$ & & $61.1\pm1.7$ & $154\pm4$ \\
& & 2007-07-24 & HRC-I & GC-2 &    & 2.0 fix & & $144\pm3$ & $400\pm8$ \\
& & 2007-09-06 & PN & GC-2 &   & $2.6\pm0.1$ & & $113\pm1$  & $426\pm2$ \\
& & 2008-03-03 & PN & GC-2 &    & $2.5\pm0.1$  & & $147\pm0.4$ & $532\pm2$ \\
& & 2008-05-10 & HRC-I & GC-2  &  & 2.0 fix & & $61.7\pm1.8$ & $172\pm5$ \\
& & 2008-07-15 & HRC-I & GC-2  &  & 2.0 fix & & $19.2\pm1.1$ & $53.4\pm3.2$ \\
3 & \saxbron\ & &  & &  $9.5\pm0.2$ & & $1.13$ (1413) && \\
& {\it bursting neutron star LMXB} & 2005-10-20 & HRC-I & GC-3 &    & 2.6 fix & & $58.6\pm1.5$ & $128\pm3$ \\
& & 2006-02-27 & PN & GC-3 &   & $2.0\pm0.1$ & & $208\pm1$ & $391\pm5$ \\
& & 2006-09-08 & PN & GC-3 &    & $2.6\pm0.1$ & & $50.2\pm0.5$ & $112\pm1$ \\
4 & \ksbron\ & &  & &   $16.6\pm1.8$ & & $1.06$ (113) && \\
& {\it bursting neutron star LMXB} & 2007-09-06 & PN & GC-7&   & $1.8\pm0.3$ & & $7.75\pm0.25$ & $17.8\pm1.9$ \\
& & 2008-05-10 & HRC-I & GC-7  &  & 1.8 fix & & $8.27\pm0.83$ & $18.5\pm1.8$ \\
& & 2008-07-16 & HRC-I & GC-7  &  & 1.8 fix & & $63.9\pm0.2$ & $143.4\pm3.9$ \\
5 & \grobron\ & &  & &  $9.4\pm3.0$ & & $1.13$ (67) & &\\
& {\it pulsating neutron star LMXB} & 2005-10-20 & HRC-I & GC-4 &  & 2.8 fix & & $0.64\pm0.15$ & $1.47\pm0.36$ \\
& & 2006-02-27 & PN & GC-4  &  & $2.8\pm1.0$ & & $0.36\pm0.05$ & $0.85\pm0.45$ \\
& & 2006-09-08 & MOS & GC-4   &  & 2.8 fix & & $0.08\pm0.03$ & $0.19\pm0.05$ \\
& & 2007-09-06 & MOS & GC-4   &  & $3.2\pm1.1$ & &$0.25\pm0.03$ & $0.66\pm0.20$ \\
& & 2008-03-04 & PN & GC-4 &    & $2.9\pm1.0$ & & $0.26\pm0.04$ & $0.63\pm0.25$ \\
& & 2008-09-27 & PN & GC-4 &   & $2.5\pm0.7$ & & $1.16\pm0.13$ & $2.49\pm0.79$ \\
6 & \xmmbron & &  & &    7.5 fix &  & $0.51$ (14) && \\
& {\it unclassified} & 2005-06-05 & HRC-I & GC-4  &  & 1.5 fix & & $67.8\pm1.8$ & $103\pm3$ \\
& & 2005-07-01 & ACIS-I  &  & & 1.5 fix & & $0.26\pm0.07$ & $0.39\pm0.11$ \\
& & 2006-02-27 & PN & GC-4  &  & 1.5 fix & & $0.35\pm0.07$ & $0.53\pm0.09$ \\
& & 2006-09-08 & PN & GC-4  &  & 1.5 fix & & $0.35\pm0.05$ & $0.53\pm0.08$ \\
& & 2007-03-12 & ACIS-I &   &  & 1.5 fix & & $0.11\pm0.04$ & $0.17\pm0.05$ \\
& & 2007-04-06 & ACIS-I &   &  & $1.3\pm1.0$ & & $0.77\pm0.36$ & $1.15\pm0.32$ \\
& & 2007-04-18 & ACIS-I &   &  & 1.5 fix & & $0.22\pm0.04$ & $0.34\pm0.05$ \\
& & 2007-04-30 & ACIS-I &   &  & $1.7\pm1.1$ & & $0.39\pm0.13$ & $0.62\pm0.07$ \\
& & 2007-05-16 & ACIS-I &   &  & 1.5 fix & & $0.39\pm0.02$ & $0.56\pm0.06$ \\
& & 2007-09-06 & PN & GC-4 & & 1.5 fix & & $0.40\pm0.07$ & $0.61\pm0.10$ \\
7 & \brontwee\ & &  & &    $30.4\pm8.6$ & & $0.90$ (57) && \\		
& {\it unclassified} & 2005-10-20 & HRC-I & GC-2 &    & 2.0 fix & & $0.33\pm0.30$ & $1.34\pm1.10$ \\
& & 2006-09-08 & PN & GC-2 &    & $1.7\pm0.8$ & & $0.93\pm0.81$ & $3.12\pm2.12$ \\
& & 2007-03-12 & ACIS-I &  &    & 2.0 fix & & $0.60\pm0.11$ & $2.06\pm0.38$ \\
& & 2007-04-06 & ACIS-I &  &    & 2.0 fix & & $0.65\pm0.12$ & $2.25\pm0.37$ \\
& & 2007-04-18 & ACIS-I &  &    & 2.0 fix & & $0.60\pm0.11$ & $2.06\pm0.38$ \\
& & 2007-04-30 & ACIS-I &  &    & 2.0 fix & & $0.43\pm0.06$ & $1.50\pm0.17$ \\
& & 2008-05-10 & HRC-I & GC-2 &  & 2.0 fix & &  $0.33\pm0.30$ & $1.34\pm1.10$ \\
& & 2008-07-15 & HRC-I & GC-2  &  & 2.0 fix & &  $0.35\pm0.28$ & $1.36\pm1.08$ \\
& & 2008-09-23 & PN & GC-2  &  & $2.5\pm1.2$ & &  $0.53\pm0.84$ & $2.84\pm1.53$ \\
8 & \bronnegen & 2005-10-20 & HRC-I & GC-2 & 18.8 fix & 2.0 fix & & $0.55\pm0.28$ & $1.40\pm0.70$ \\
& {\it unclassified} &  & & &  &  & &  &  \\
9 & \newsource & &  & & $5.2\pm1.2$ & & 1.07 (159) && \\
& {\it likely weak persistent} & 2006-02-27 & PN & GC-6 &    & $0.3\pm0.3$ & & $2.48\pm0.10$ & $3.02\pm0.17$ \\
& & 2006-09-08 & PN & GC-6 &    & $1.8\pm0.5$ & & $0.72\pm0.21$ & $1.06\pm0.11$ \\
& & 2007-09-06 & PN & GC-6 &	  & $0.7\pm0.4$ & & $1.02\pm0.25$ & $1.29\pm0.11$ \\
& & 2008-03-04 & PN & GC-6 &    & $1.3\pm0.4$ & & $1.04\pm0.15$ & $1.43\pm0.12$ \\
& & 2008-09-27 & PN & GC-6 &    & $1.7\pm0.5$ & & $0.65\pm0.14$ & $0.93\pm0.07$ \\
10 & \andereascabron & 2008-07-16 & HRC-I & GC-5 & 1.6 fix & 2.9 fix & & $0.22\pm 0.08$ & $0.27\pm 0.10$ \\
& {\it likely weak persistent} &  & & &  &  & &  &  \\ 
\hline
\end{tabular}
\label{tab:spec}
\begin{tablenotes}
\item[]Note. -- Quoted errors represent 90\% confidence levels. $F_{\mathrm{X,\mathrm{abs}}}$ and $F_{\mathrm{X,\mathrm{unabs}}}$ denote the absorbed and unabsorbed 2--10 keV model fluxes. The hydrogen column density of \xmmbron\ (\nxmmbron) was fixed to the value reported by \citet{degenaar09_gc}. For \bronnegen\ (\nbronnegen) we adopted the spectral parameters given by \citet{muno04_apj613} and for \andereascabron\ (\nandereascabron) those reported by \citet{degenaar2011_asca}, since these objects were only detected in our \chan/HRC observations.
\end{tablenotes}
\end{center}
\end{threeparttable}
\end{table*}


\section{Results summary and highlights}\label{sec:results}
Each of our \chan\ and \xmm\ observations reveals tens of distinct X-ray sources. Amongst the detected objects are the Arches cluster and \sgra, which are both complexes of X-ray point sources combined with diffuse emission structures. Several of the X-ray point sources found in our observations could be identified with known stars or had clear DSS/2MASS counterparts, which renders them likely foreground sources. 

We detected two persistent X-ray binaries during our campaign: 1E 1743.1--2843 and 1A 1742--294. The former is an LMXB black hole candidate \citep[e.g.][]{porquet03}, whereas 1A 1742--294 is a neutron star LMXB that displays type-I X-ray bursts \citep[e.g.][]{pavlinsky1994}. Our monitoring observations caught a total of six type-I X-ray bursts from this source.\footnote{The type-I X-ray bursts from 1A 1742--294 are detected in the \chan/HRC observations 6200 and 9040, and in the \xmm\ observations 0302884401, 0504940601 and 0511001101. The latter contains two bursts that are separated by $\sim$15~min. Five of the X-ray bursts had a duration of $\sim$50~s, but the one detected in observation 0302884401 was considerably longer and lasted $\sim$250~s.} Since the focus of our present work lies on transient X-ray sources, we do not discuss these two persistent X-ray binaries in more detail.

During our programme we detected activity from eight previously known X-ray transients, several of which exhibited multiple outbursts during our campaign. The results of our spectral and temporal analysis are presented in detail for each of the individual sources in Appendix~\ref{sec:appendix}, where we also touch upon their long-term X-ray behaviour. In addition to the eight previously known transients, we found two X-ray sources that were detected only in a subset of our observations. Although this possibly indicates a transient nature, we argue that both are likely weak persistent X-ray sources with intensities close to the detection limit of our observations (see Section~\ref{subsec:persistent}). 

The results of our spectral analysis of these ten X-ray sources are summarised in Table~\ref{tab:spec}. This table gives an overview of all observations in which a particular source was detected. For clarity we have assigned source numbers that correspond to the individual subsections of Appendix~\ref{sec:appendix}. All fluxes listed in Table~\ref{tab:spec} and reported elsewhere in this work are given for the 2--10 keV energy band. Unless stated otherwise, luminosities were calculated from the unabsorbed flux by assuming a source distance of 8 kpc. All quoted errors refer to 90\% confidence levels. Some of the detected transients were within the FOV of two different pointing directions. In these cases we report only the information extracted from the observations in which the source is nearest to the aim point. 
In the following sections we highlight some of the results of our campaign.

 \begin{figure*}
 \begin{center}
           \includegraphics[width=8cm]{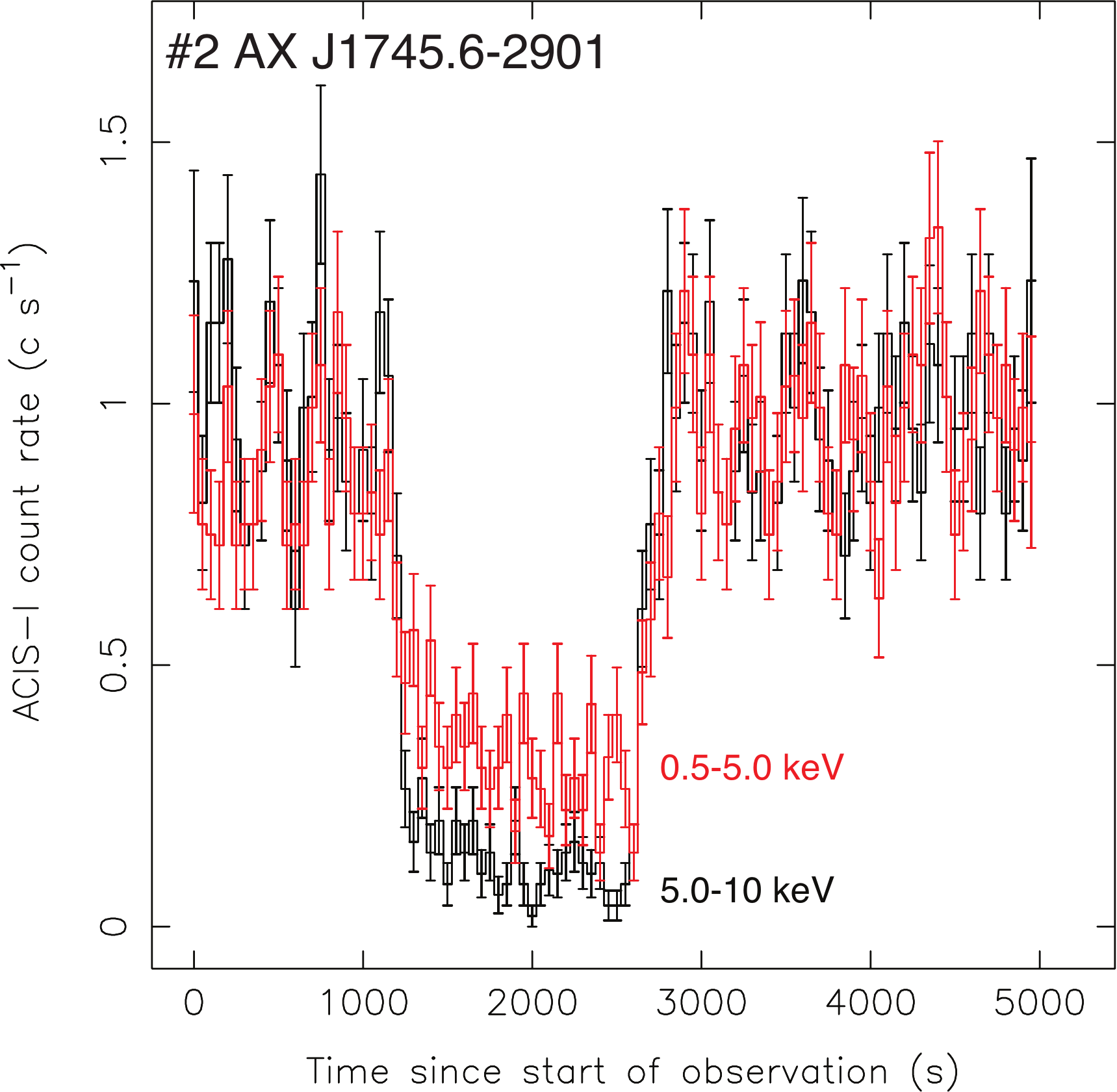}\hspace{1.0cm}
          \includegraphics[width=8cm]{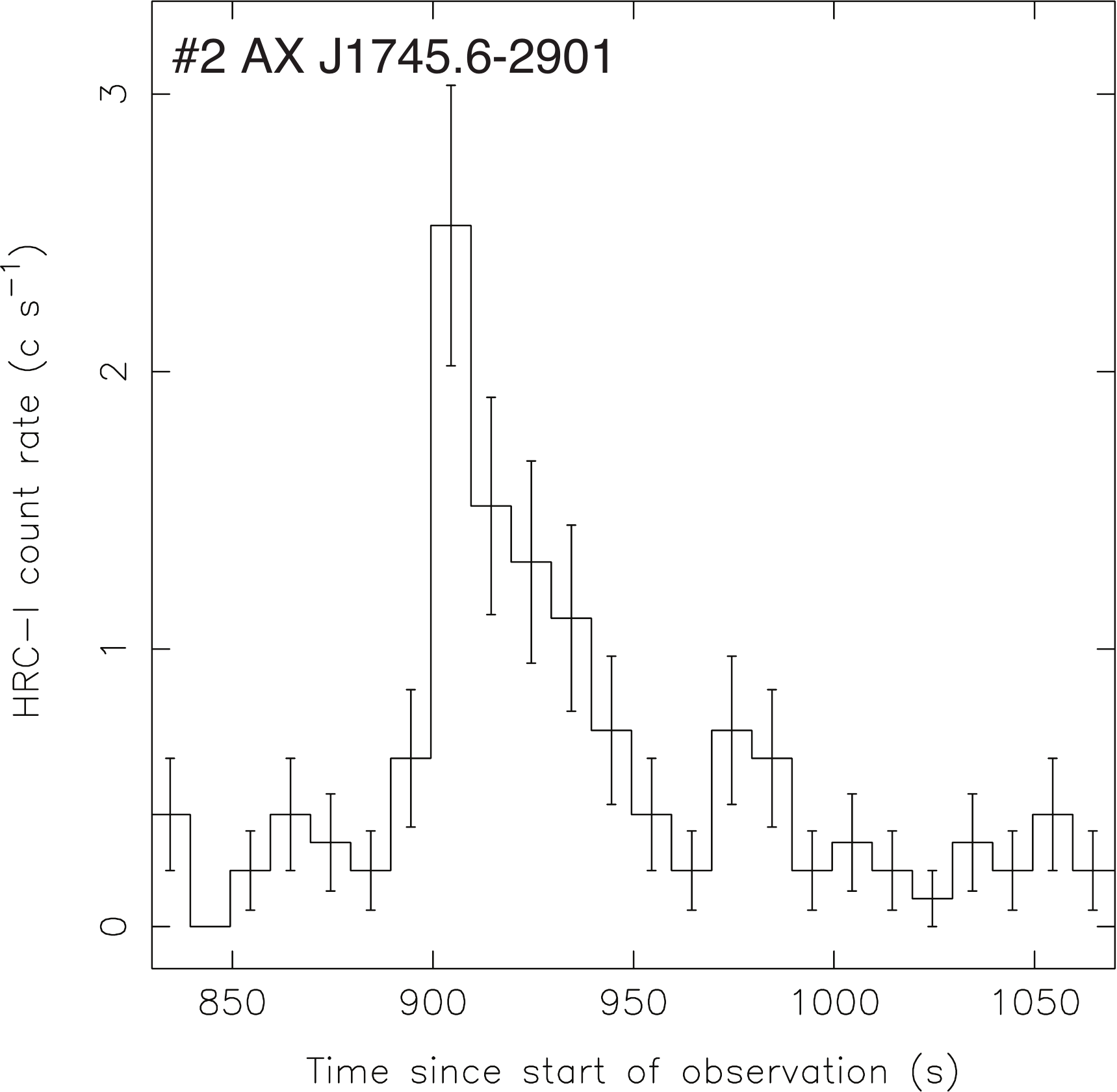}
    \end{center}
    \caption[]{Lightcurve features of the neutron star LMXB \ascabron (0.5--10 keV). Left: Binned 50-s lightcurve of the \chan/ACIS observation of 2007 April 6 (obs ID 6603) showing a $\sim$1600-s X-ray eclipse in two energy bands: 0.5--5.0 keV (red) and 5.0--10 keV (black). Right: Binned 10-s lightcurve of the \chan/HRC observation of 2008 May 10 (obs ID 9073) showing a $\sim$50-s flare that is likely a type-I X-ray burst. }
 \label{fig:ascalc}
\end{figure*} 


\subsection{Lightcurve features of \ascabron}\label{subsec:ascalc}

The transient neutron star LMXB \ascabron\ was active during several of our observations and exhibited two distinct accretion outbursts during our campaign (2006 and 2007--2008; see Appendix~\ref{subsec:ascabron}). Investigation of the X-ray lightcurves revealed two prominent features that we discuss below. 

\subsubsection{X-ray eclipse}
During the \chan/ACIS-I observation performed on 2007 April 6, we detected a strong reduction in the X-ray flux of \ascabron\ (left panel of Fig.~\ref{fig:ascalc}). This event has every characteristic of an X-ray eclipse, which is caused by temporal obscuration of the central X-ray emitting region by the companion star. X-ray eclipses allow for a direct determination of the orbital period and can be used to study any possible evolution of the binary orbit \citep[e.g.][]{wolff2008c}. Only a handful of neutron star LMXBs are known to display X-ray eclipses. 

We barycentred the lightcurve using the tool BARYCEN to determine the time at which the eclipse occurred. The eclipse started on 2007 April 6 around 16:03 UTC and had a total duration of $\sim$1600~s. The ingress of the eclipse lasted for $\sim$400~s, while the egress time was $\sim$200~s. The eclipse was not total, with a residual count rate of $\sim$20\% of the out-of-eclipse emission (0.5--10 keV). This profile matches the description of the eclipses that were seen by \asca\ and the mid-eclipse time is consistent within the uncertainties of the ephemeris derived by \citet[][]{maeda1996}. This leaves no doubt that this X-ray transient seen active during our campaign is \ascabron. Other authors also reported on X-ray eclipses with similar characteristics seen during \xmm\ (2007 March--April) and \chan\ (2008 May) observations \citep[][]{porquet07,heinke08}. A detailed study of the \asca-detected eclipses suggested that the binary is seen at an inclination angle of $\sim$$70^{\circ}$ and harbours a G-dwarf companion star in an $\sim$8.4-h orbit \citep[][]{maeda1996}. 

We extracted lightcurves in two different energy bands to investigate whether the eclipse properties are energy-dependent. We compare the 0.5--5.0 and 5.0--10 keV energy bands, because this yields similar intensities for the non-eclipsed emission. Figure~\ref{fig:ascalc} (left) compares the eclipse profiles. The duration, ingress and egress time is similar for both energy ranges, but the eclipse is deeper in the harder band. At the base of the eclipse the intensity in the 0.5--5.0 keV band is $\sim$30\% of the out-of-eclipse emission, whereas this is only $\sim$10\% for the 5.0--10 keV band. This may indicate that the softer X-ray photons come from a region that is more extended than the emission site of the harder photons. Similar eclipse behaviour has been observed for other LMXBs and explained in terms of a dust-scattering halo \citep[e.g.][]{homan2003,ferrigno2011}.

\subsubsection{X-ray burst}
The \chan/HRC data obtained on 2008 May 10 revealed an X-ray flare from \ascabron\ that started around 2008 May 11 at 00:04 UT and had a duration of $\sim$50~s (right plot in Fig.~\ref{fig:ascalc}). The fast rise and exponential decay shape, combined with the fact that this source is a known X-ray burster, strongly suggest that this event was a type-I X-ray burst. These are bright flashes of thermal X-ray emission that are caused by unstable thermonuclear burning of the accreted matter on the surface of the neutron star. Unfortunately, the HRC data does not allow for a spectral confirmation. 

Using $N_H=21.8\times10^{22}~\nh$ (see Table~\ref{tab:spec}), and assuming a blackbody temperature typically seen for type-I X-ray bursts ($kT_{\mathrm{BB}}=2-3$~keV), the observed HRC peak count rate translates into a 0.01--100 keV luminosity of $L_X\sim5\times10^{37}~\lum$ for an assumed distance of $D=8$~kpc. The duration and peak intensity of the flare match other thermonuclear bursts detected from \ascabron\ \citep[][]{maeda1996,degenaar09_gc}. The duration suggests that helium is ignited in a hydrogen-rich environment \citep[cf.][]{galloway06}. The persistent accretion luminosity at the time of the X-ray burst was $L_X\sim1\times10^{36}~(D/8~\mathrm{kpc})^2~\lum$ (see Table~\ref{tab:spec}), corresponding to $\sim 0.5\%$ of the Eddington rate of a neutron star.


\subsection{Transient nature and X-ray bursts of \saxbron}\label{subsec:saxlc}

\subsubsection{Quiescent luminosity}
The neutron star LMXB \saxbron\ has been detected on numerous occasions and by different satellites ever since its discovery in 1998 (Appendix~\ref{subsec:saxbron}). \citet{wijnands2002_saxj1747} observed the source with \chan\ in between two bright outburst and detected it at a 0.5--10 keV luminosity of $L_X\sim 2\times10^{35}~(D/6.7~\mathrm{kpc})^2~\lum$. Since this is 2--3 orders of magnitude higher than the typical quiescent X-ray luminosity of neutron star LMXBs, the transient nature of \saxbron\ was cast in doubt \citep[][]{wijnands2002_saxj1747}. However, an apparent quenching of type-I X-ray bursts suggested that the accretion was suppressed at least during some intervals \citep[][]{werner2004}.

\saxbron\ was active during several of our monitoring observations, but there were also epochs during which the source went undetected (Appendix~\ref{subsec:saxbron}). We infer upper limits on the 2--10 keV luminosity of $L_X\lesssim(4-40)\times 10^{33}~(D/6.7~\mathrm{kpc})^2~\lum$ for individual observations. This favours a classification as transient X-ray source. We further improve these constraints by using archival \chan/ACIS-I observations performed on 2006 October 22 (obs ID 7157, $\sim15$~ks exposure) and 2007 February 14 (obs ID 7048, $\sim39$~ks). The combined ACIS-I image shows an excess of photons at the location of \saxbron\ with a significance of $2.2\sigma$. For an absorbed powerlaw model with $N_H=9.5\times10^{22}~\nh$ and $\Gamma=2.6$ (see Table~\ref{tab:spec}), the net source count rate of $\sim6\times10^{-5}~\cnts$ translates into a 2--10 keV luminosity of $L_X\sim2\times10^{31}~(D/6.7~\mathrm{kpc})^2~\lum$. This clearly establishes the transient nature of \saxbron. During the \chan\ observations presented by \citet{wijnands2002_saxj1747} the source was thus likely detected in a low-level accretion state.

\subsubsection{Short recurrence time X-ray bursts}
\begin{figure}
 \begin{center}
          \includegraphics[width=8.0cm]{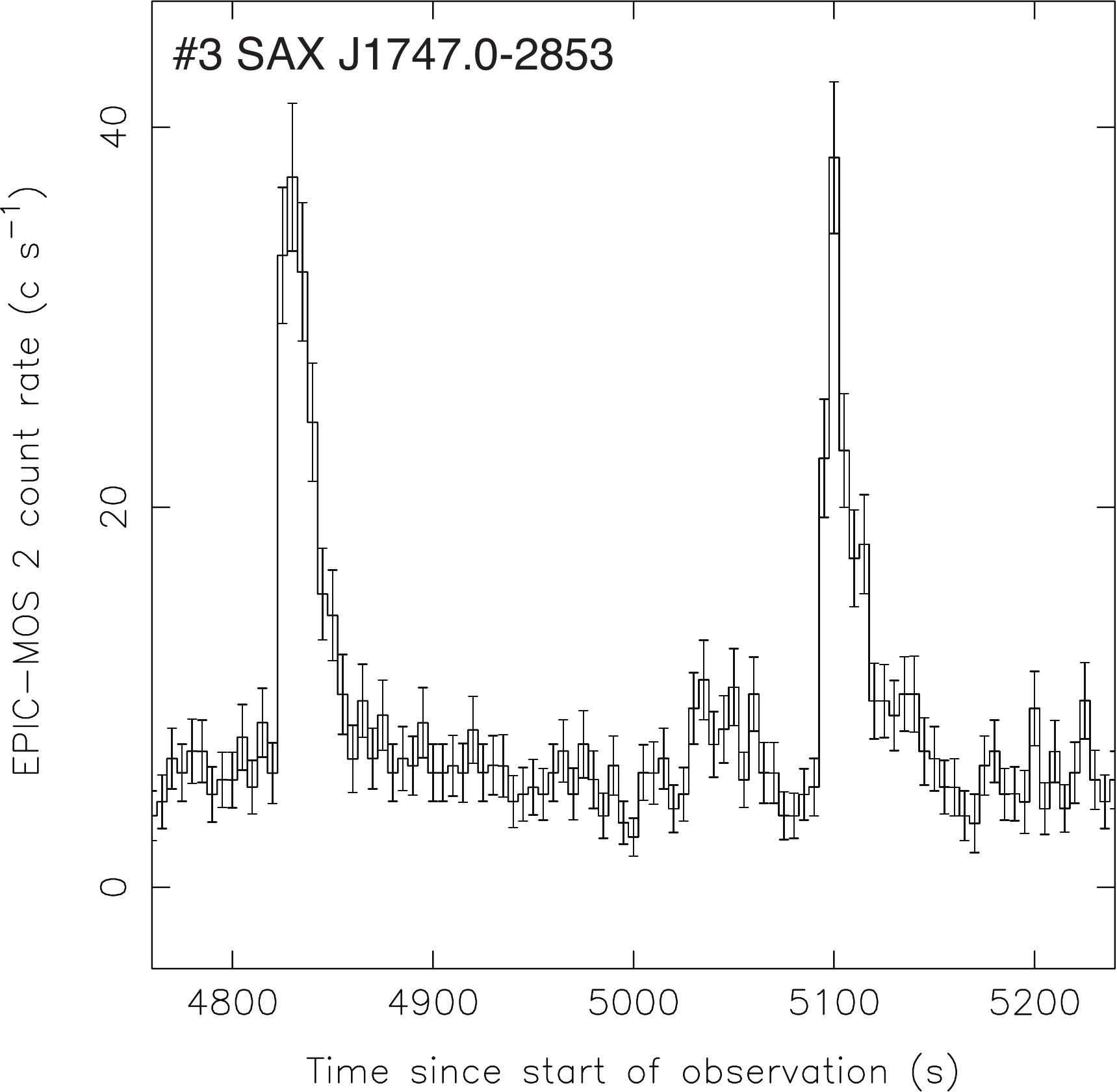}     
    \end{center}
    \caption[]{\xmm/EPIC-MOS2 0.5--10 keV lightcurve of \saxbron\ obtained on 2006 February 27 (obs ID 0302882701) using time bins of 5 s. The image displays an interval of $\sim$400~s during which two type-I X-ray bursts of similar duration ($\sim$40~s) and peak intensity occurred. }
 \label{fig:saxj1747bursts}
\end{figure}

\saxbron\ is a known X-ray burster and we detected two such events during our campaign. The lightcurve extracted from the 2006 February \xmm\ observation revealed a pair of type-I X-ray bursts occurring around 07:57 and 08:01 UT (Fig.~\ref{fig:saxj1747bursts}). The time elapsed between the end of the first and the start of the second burst is $230$~s (3.8~min). Both show the typical fast rise, exponential decay shape and are of similar duration ($\sim$40~s) and peak intensity (Fig.~\ref{fig:saxj1747bursts}). We analysed the spectra of the X-ray bursts using only the MOS cameras, because the PN switched off during the bursts (possibly due to the high count rate). To avoid pile-up we extracted events from an annulus with inner radius of $20''$ and an outer radius of $40''$. 

Limited by statistics, we cannot perform time-resolved spectroscopy; therefore we extracted the full-burst spectra (spanning $\sim$40~s of data). A spectrum obtained from an interval of 100~s preceding the first burst was used as a background to account for the underlying persistent emission. We fitted both burst spectra to an absorbed blackbody model with the hydrogen column density fixed to the value obtained from fitting the persistent emission ($N_{H}=9.5\times10^{22}~\nh$; Table~\ref{tab:spec}). For the first X-ray burst this yields a blackbody temperature of $kT_{\mathrm{bb}}=1.8\pm0.4$~keV and an emitting radius of $R_{\mathrm{bb}}=5.2\pm0.9$~km. For the second burst we obtain comparable values of $kT_{\mathrm{bb}}=1.4\pm0.3$~keV and $R_{\mathrm{bb}}=7.4\pm1.0$~km. 

Extrapolation of the blackbody fits to the energy range of 0.01--100 keV yields an estimate of the bolometric luminosity of $L_{\mathrm{bol}}\sim3.6\times10^{37}$ and $3.1\times10^{37}~(D/6.7~\mathrm{kpc})^2~\lum$ for the first and second burst, respectively. Using the count rate to flux conversion factor inferred from fitting the average burst spectra, we estimate that the bolometric peak luminosities of both bursts reached $L_{\mathrm{bol}}\sim1.0\times10^{38}~(D/6.7~\mathrm{kpc})^2~\lum$. This is close to the Eddington limit for neutron stars. Fitting the spectral data of the persistent emission shows that the source was accreting at $\sim$1\% of the Eddington rate when the burst doublet occurred. The duration of the bursts ($\sim$40~s) is typical of helium ignited in a hydrogen-rich environment \citep[cf.][]{galloway06}.

The short time-interval of 3.8~min between the two consecutive type-I X-ray bursts is amongst the shortest measured for burst doublets.\footnote{We note that the bursts observed from \saxbron\ contained within the \rxte\ catalogue of \citet{galloway06} have recurrence times of hours. However, \citet{keek2010} study an extended data sample, including bursts detected with \bepposax, and report on the detection of three burst pairs with short recurrence times from \saxbron\ (compared to 57 single bursts). The waiting times for these burst pairs were $\sim$11, 13 and 42 min, and these occurred when the persistent luminosity of the source was $\sim$1, 1 and 5\% of the Eddington rate, respectively (L. Keek, private communication).} Other bursts with similarly short recurrence times were reported from the neutron star LMXBs IGR J17480--2446 \citep[3.3 min;][]{motta2011}, 4U 1705--44 \citep[3.8~min;][]{keek2010}, 4U 1636--536 \citep[5.4~min;][]{linares2009}, 4U 1608--52 \citep[4.3--6.4~min;][]{galloway06} and \exo\ \citep[6.5~min;][]{boirin2007,galloway06}. Type-I X-ray bursts repeating within minutes present a challenge to our understanding of burst physics. 

Current theoretical models predict that over 90\% of the accreted hydrogen/helium is burned during a type-I X-ray burst, which implies that it would take at least a few hours to accumulate enough matter to power a new burst \citep[][]{woosley2004}. This is at odds with the detection of X-ray bursts that have short recurrence times and thus suggests that some of the initial fuel is preserved after ignition of the first burst \citep[][]{galloway06,keek2010}. One possible explanation for the occurrence of burst doublets could be that the matter is confined to certain parts of the neutron star surface, e.g. the magnetic poles, and that the bursts of a pair ignite at different locations. Alternatively, the bursts might ignite in separate layers that lie on top of each other. In this scenario the first burst causes unburned fuel to be mixed down to larger depth, which can cause the ignition of a new burst \citep[e.g.][]{keek2009}.

A systematic study of short recurrence time bursts shows that, on average, the second burst of a pair is shorter, less bright, cooler, and less energetic than the first \citep[][]{boirin2007,keek2010}. This is suggestive of a reduced hydrogen content after the first burst has ignited and would favour the explanation that the bursts are resulting from different envelope layers rather than different areas of the neutron star \citep[][]{boirin2007,keek2010}. In contrast, the two type-I X-ray bursts observed from \saxbron\ were of similar duration ($\sim40$~s) and similar intensity, which points towards a similar fuel content.

 \begin{figure*}
 \begin{center}
\includegraphics[width=9.7cm]{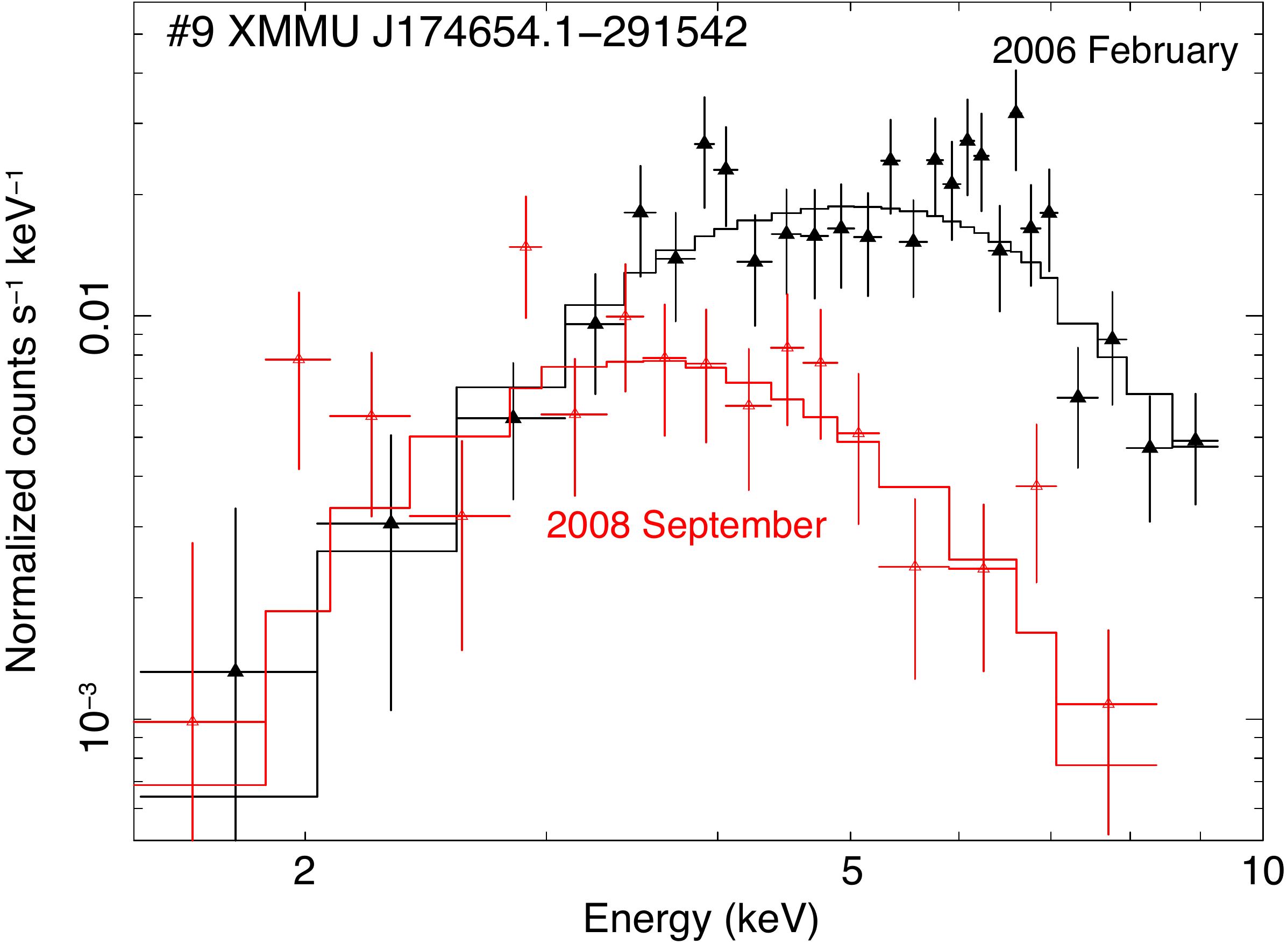} \hspace{0.3cm}
 \includegraphics[width=7.9cm]{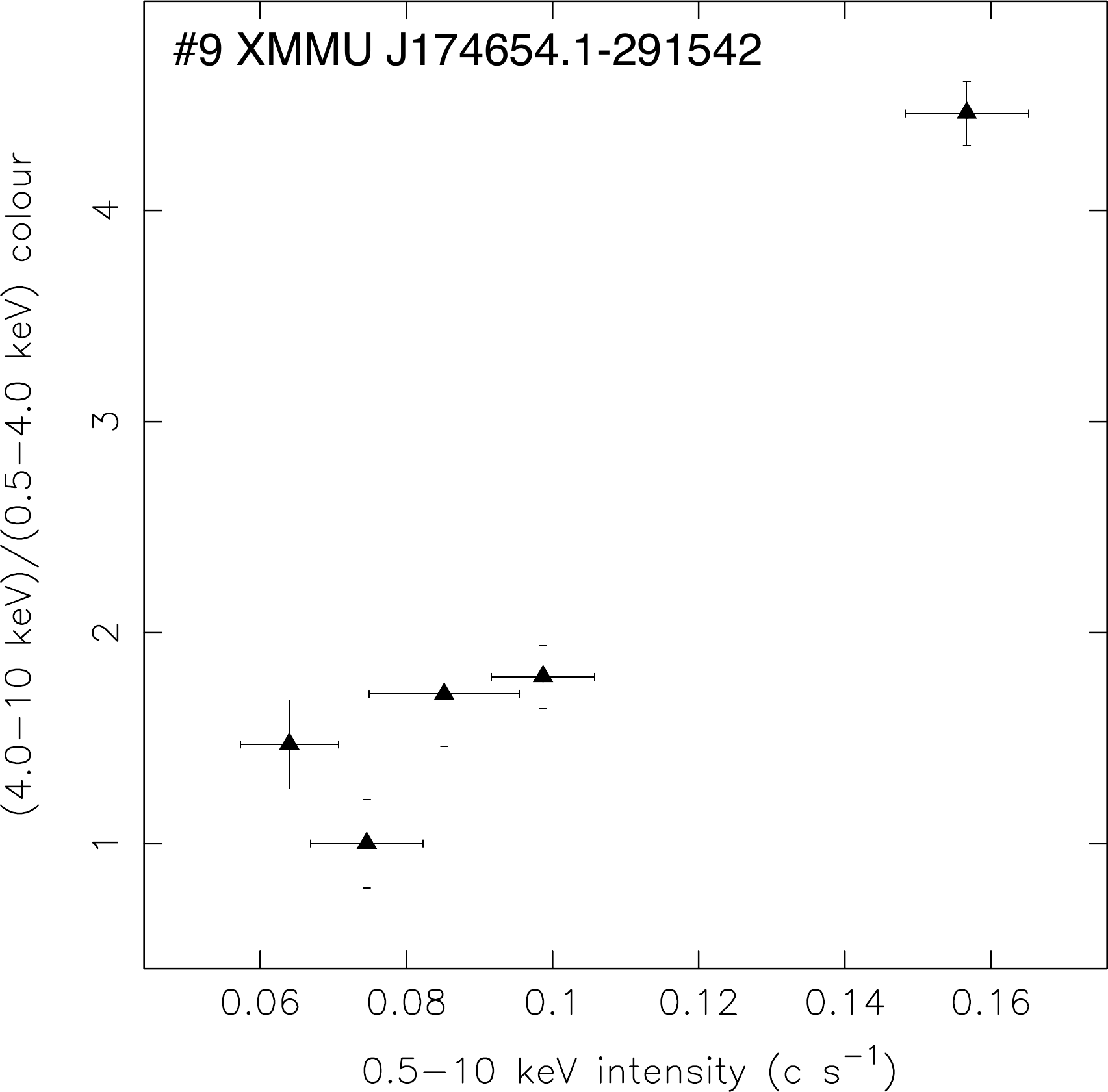}        
    \end{center}
    \caption[]{Results for the new X-ray source \newsource. Left: A comparison of \xmm/PN spectra obtained on two different epochs. Right: Colour versus intensity plot using all \xmm/PN data of the field GC-6.}
 \label{fig:newsource}
\end{figure*}


\subsection{Low-level accretion activity}\label{subsec:weak}
Two of the X-ray transients covered by our monitoring observations were detected at low X-ray luminosities of $L_X\sim10^{33-34}~\lum$. This is a factor of $\sim$$10-100$ above their quiescent levels, yet well below their maximum X-ray intensities of $L_X\gtrsim10^{36}~\lum$. 
Below we briefly discuss our findings, which indicate that these sources likely exhibit low-level accretion activity outside their regular (i.e., bright) accretion outbursts. We discuss this in more detail in Section~\ref{subsec:lowlum_activity}.

\subsubsection{\xmmbron} 
\xmmbron\ is a faint, unclassified transient source that displays a quiescent luminosity of $L_X\sim10^{32}~\lum$ and exhibits X-ray outbursts with a peak luminosity of $L_X\sim10^{36}~\lum$ (Appendix~\ref{subsec:xmmbron}). We detected one such outburst during our campaign (in 2005). Several of our observations, however, detect the source at an intensity of $L_X\sim (1-9) \times 10^{33}~\lum$ (see Table~\ref{tab:spec} and Fig.~\ref{fig:xmmbron}). This is a factor 10--100 higher than its quiescent emission level, yet considerably weaker than the full outbursts it displays. 

As detailed in Appendix~\ref{subsec:xmmbron}, our \chan/\xmm\ monitoring observations support previous findings that the bright episodes of this source last only shortly and that the source spends most of its time at luminosities that are a factor $\sim$10--100 above its quiescent level. This peculiar behaviour has led to the speculation that this unclassified X-ray source is possibly a wind-accreting X-ray binary \citep[][]{degenaar2010_gc}, although some neutron star LMXBs also appear to display low-level accretion activity (Section~\ref{subsec:lowlum_activity}). One such example is \grobron\ (see below).

\subsubsection{\grobron}
The neutron star LMXB \grobron, also known as ``the bursting pulsar'' (Appendix~\ref{subsec:grobron}), undergoes outbursts that reach up to $L_X\sim10^{37-38}~\lum$ \citep[][]{woods1999} and has a quiescent luminosity of $L_X\sim(1-3)\times10^{33}~\lum$ \citep[][]{wijnands2002_gro1744,daigne2002}. \grobron\ is detected during all five \xmm\ observations of GC-4. During the first four data sets, the source intensity was found to be $L_X\sim(1.5-6.5)\times10^{33}~\lum$. Within the errors this is consistent with the source being in quiescence. During the final observation (2008 September), however, the source intensity was enhanced to $L_X\sim1.9\times10^{34}~\lum$ (Appendix~\ref{subsec:grobron}). Indications of enhanced activity above the quiescent level have been reported from this source on other occasions as well \citep[][]{muno07_atel1013}.

 \begin{figure*}
 \begin{center}
	\includegraphics[width=8.0cm]{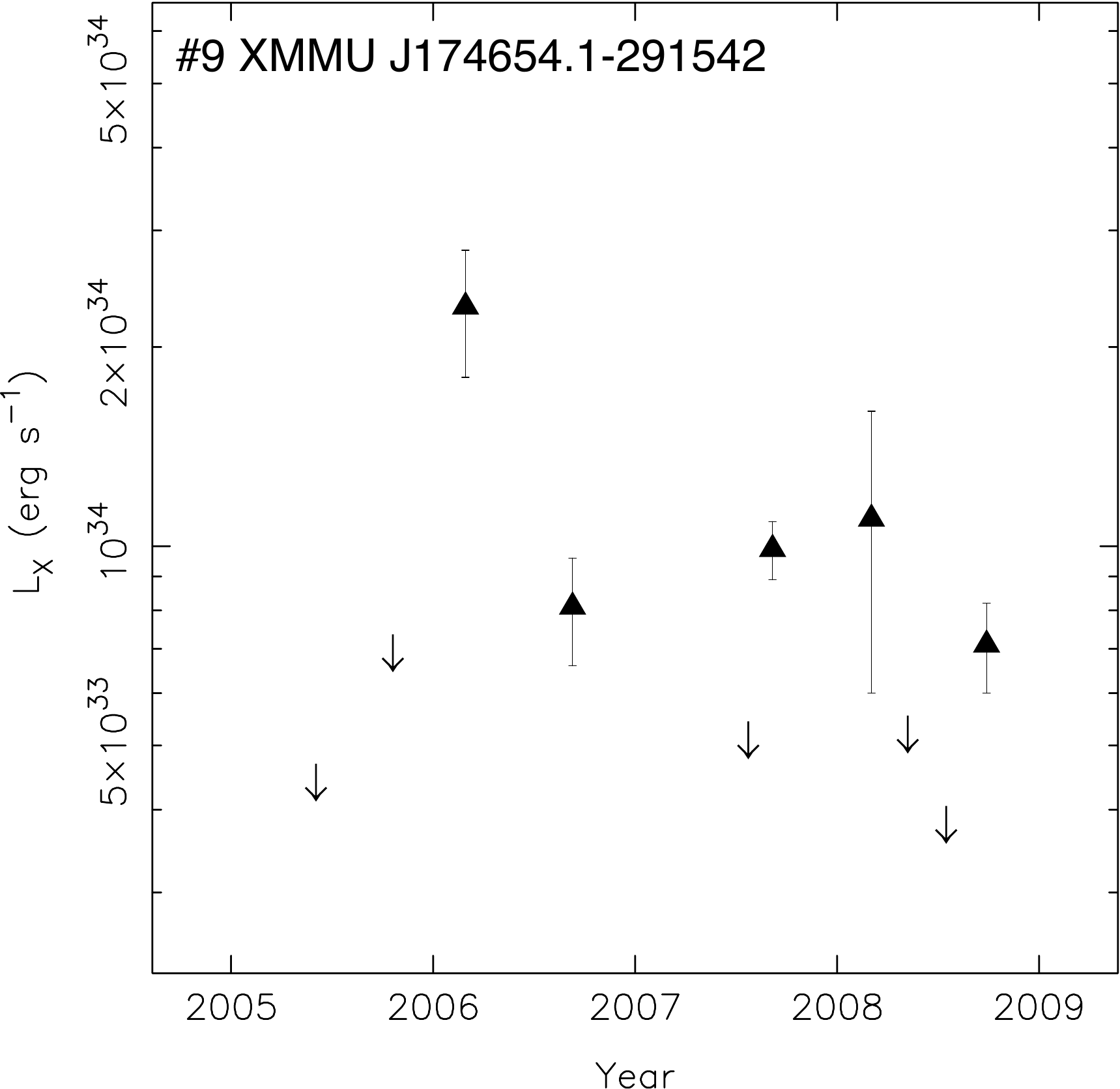}\hspace{0.5cm}
		\includegraphics[width=8.0cm]{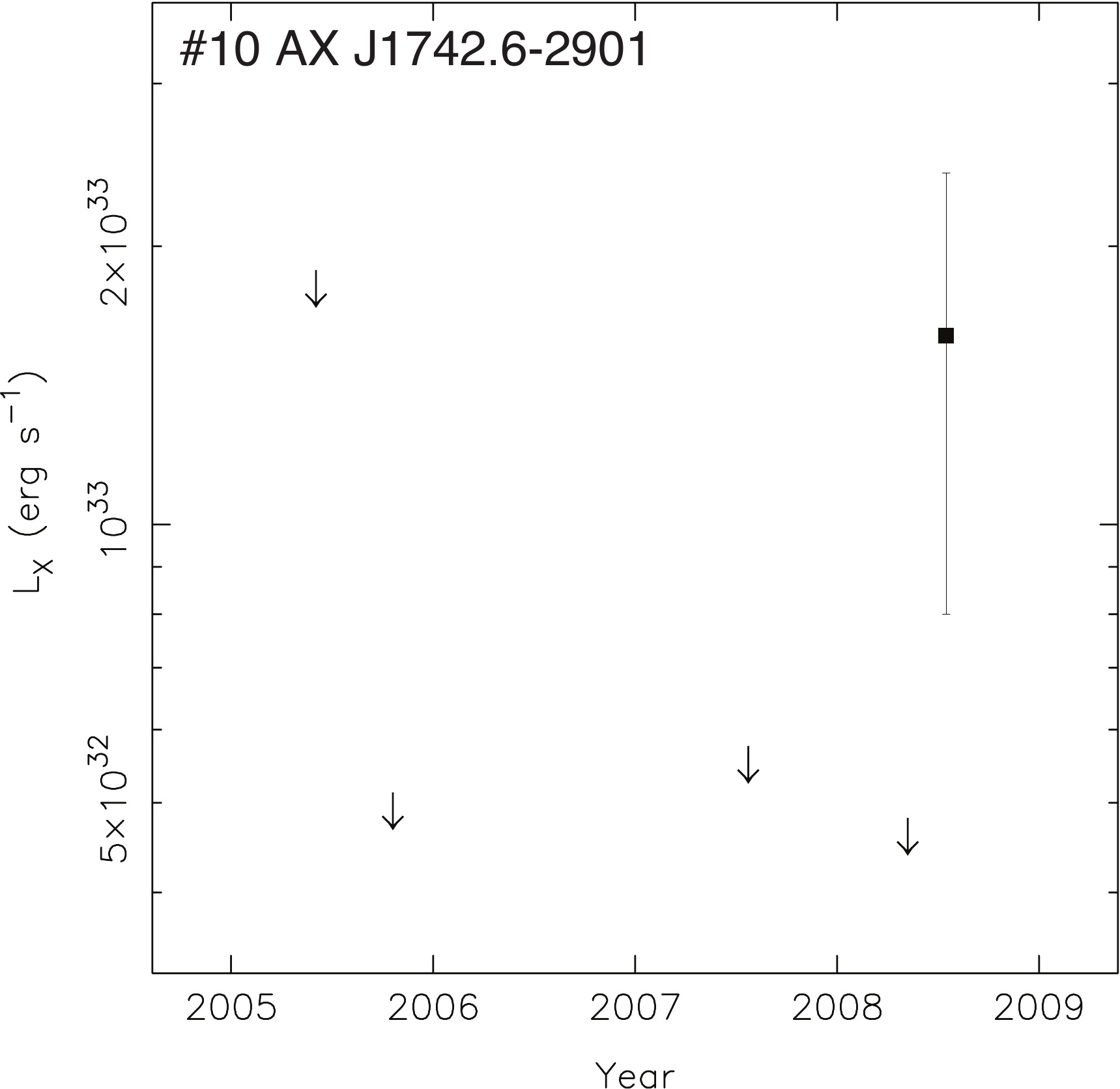}
    \end{center}
    \caption[]{Evolution of the X-ray luminosity of the two likely weak persistent X-ray sources \newsource\ (left) and \andereascabron\ (right) during our campaign. Triangles indicate \xmm\ observations, whereas the square and upper limit symbols represent \chan/HRC data.}
 \label{fig:persistent}
\end{figure*} 


\subsection{Two weak persistent objects}\label{subsec:persistent}

\subsubsection{XMMU~174654.1--291542: a new X-ray source}\label{subsec:newsource}
In all \xmm\ observations covering the field GC-6, we detect an X-ray point source at $\alpha$=$17^{h}46^{m}54.15^{s}$, $\delta$=$-29^{\circ}15'42.6''$ (J2000) with an uncertainty of $4''$. This object is only detected above 2 keV and has no counterpart in the SIMBAD astronomical database or in DSS/2MASS images.\footnote{We note that the source region was not covered by the \chan\ survey of \citet{muno2009}.} We designate this new X-ray source \newsource. Although the source is not detected in the individual HRC-I exposures, it is clearly visible by eye when all data are merged. This allows us to refine its position to $\alpha$=$17^{h}46^{m}54.47^{s}$, $\delta$=$-29^{\circ}15'44.0''$ (J2000) with an uncertainty of $1.5''$. 

Fitting all \xmm\ spectra simultaneously to an absorbed powerlaw yields $N_{\mathrm{H}}=(5.2\pm1.2)\times10^{22}~\nh$ and photon indices varying between $\Gamma=0.3-1.8$ (see Table~\ref{tab:spec}). This suggests that there is considerable spectral variation between the different data sets. This is illustrated by Fig.~\ref{fig:newsource} (left), which compares the PN spectra obtained in 2006 September and 2008 May. If the different spectra are fitted individually, we obtain a considerable spread in hydrogen column densities of $N_{\mathrm{H}}=(1.6-9.0)\times10^{22}~\nh$ and powerlaw indices of $\Gamma=0.1-1.8$, although the errors on both parameters are large. 

We also fitted the spectra simultaneously with the powerlaw index tied between the different observations, whereas the hydrogen column density was left to vary. This resulted in $\Gamma=1.0\pm0.3$ and $N_{\mathrm{H}}=(2.5-10.2)\times10^{22}~\nh$, with typical errors of $\sim 2.0 \times10^{22}~\nh$ (for $\chi_{\nu}=1.04$ for 159 d.o.f.). Regardless of the chosen approach, we obtain 2--10 keV luminosities lying in a range of $L_X\sim(0.7-3.0)\times10^{34}~\lum$ (assuming $D=8$~kpc).

Additional evidence in support of spectral variations is provided by Fig.~\ref{fig:newsource} (right), where we plot the ratio of PN source counts in the 4--10 keV and 0.5--4 keV energy bands versus the count rate over the full 0.5--10 keV range, using all five \xmm\ observations. The biggest outlier in this graph concerns the 2006 September data set, during which the count rate in the full energy band was a factor $\sim$2 higher than for the other observations. The count rates in the soft 0.5--4 keV band are similar for all five observations, so this difference can be completely attributed to the intensity in the harder 4--10 keV band.

\newsource\ is not detected in the individual HRC-I observations, which constrains the source luminosity during those epochs to $L_X\lesssim(4-7)\times10^{33}~\lum$. In the merged HRC image, the source is weakly detected at a count rate of $(3\pm1)\times10^{-3}~\cnts$. For $N_{\mathrm{H}}=5.2\times10^{22}~\nh$ and $\Gamma=1.0$, this implies a luminosity of $L_X=6.3\times10^{33}~\lum$. Given the small difference between the \xmm\ detections and the \chan\ upper limits, we cannot asses whether this object is truly transient or rather a persistent source that displays intensity variations by a factor of a few (see left panel of Fig.~\ref{fig:persistent}). 

We searched through archival data to shed more light on the long-term variability of this newly identified X-ray source. We found two \chan/ACIS pointings that cover the source region, carried out on 2006 November 2 and 2008 May 10 (obs IDs 7163 and 9559 with exposure times of $\sim$14.3 and $\sim$ 14.8~ks, respectively). During both observations \newsource\ is one of the brightest objects in the field, yielding $\sim$350--400 net photons per observation. We extracted source and background spectra for these two data sets. 

There are no prominent spectral differences between the two archival \chan\ observations. We obtain similar photon indices of $\Gamma=0.5\pm0.5$ and hydrogen column densities of $N_H= (2.5\pm 1.8) \times 10^{22}$ and $(2.8\pm 1.7) \times 10^{22}~\nh$ for the 2006 and 2008 data, respectively. The inferred luminosity is $L_X\sim1\times10^{34}~\lum$ for both observations. The archival \chan\ observations thus detect \newsource\ at similar intensity levels as seen during our monitoring programme. This likely points towards a persistent nature. However, all detections occurred within a time frame of $\sim$2.5~yr and we cannot exclude that the source is a weak transient that underwent a long outburst. 

The strong spectral variation that we observe for \newsource\ is reminiscent of the behaviour seen for systems that harbour a compact primary accreting from the wind of a companion star. Wind accretion occurs e.g. in neutron star or black hole HMXBs, symbiotic X-ray binaries (a subclass of LMXBs that contain a neutron star and an M-type giant) or symbiotics in which the accreting object is a white dwarf \citep[e.g.][]{luna2007,masetti2007,heinke09,romano2011}. In such systems the changing spectral properties are attributed to intrinsic variations in the absorption column density due to the wind of the companion star. 

Our spectral analysis of \newsource\ indicates that there might be a large spread in hydrogen column densities between the different observations. The obtained spectral index of $\Gamma =1.0\pm0.3$ is also quite typical of wind-accreting binaries, whereas Roche-lobe overflowing LMXBs usually have softer X-ray spectra. Its X-ray spectral properties thus suggest that \newsource\ may harbour a compact object accreting from the wind of its companion star. 

We found a possible infrared counterpart in the {\it Spitzer}/IRAC catalogue that is consistent with our position of \newsource\ \citep[][]{ramirez2008}. This object is relatively bright with an apparent magnitude of 10.7~mag at 3.6 micron. A quick comparison with WISE data \citep[][]{wise} of the recurrent nova T Coronae Borealis (T CrB), which has an absolute 3.6-micron magnitude of approximately $-4$~mag, suggests that the {\it Spitzer} source might be a white dwarf binary located at $D\sim6$~kpc. If this object is associated with \newsource, our X-ray source might thus indeed be a symbiotic. At a distance of 6 kpc, the X-ray fluxes inferred from our campaign translate into luminosities of $L_X\sim(4-10)\times10^{33}~\lum$.


\subsubsection{The unclassified X-ray source \andereascabron}\label{subsec:andereascabron}
During the final series of \chan/HRC observations, performed in 2008 July, we detected an X-ray point source located at $\alpha=17^{h}42^{m}42.58^{s}$, $\delta=-29^{\circ}02'04.8''$ (J2000). The source is weak (a total number of $\sim$35 net counts) and detected at a large offset angle ($\sim$$18'$) from the aimpoint of the observation, which strongly decreases the accuracy of the source localisation. Using equation 5 of \citet{hong2005}, we tentatively estimate a positional uncertainty of $\sim$$9''$. 

Within our estimated error, the \chan\ coordinates are consistent with the \swift\ localisation of \andereascabron. This unclassified X-ray source (not to be confused with the neutron star LMXB \ascabron\ discussed in Section~\ref{subsec:ascalc}) was discovered during \asca\ observations of the GC in 1998 September \citep[][]{sakano02}. The \asca\ source was tentatively associated with the \rosat\ object 2RXP J174241.8--290215, which was detected at a count rate of $(1.9\pm0.3)\times10^{-2}~\cnts$ during a 2-ks PSPC observation in 1992 March.\footnote{The Second ROSAT Source Catalog of Pointed Observations (2RXP) is available at http://www.mpe.mpg.de/xray/wave/rosat/rra.}  

\andereascabron\ was detected only once during our campaign, with a HRC-I count rate of $(6.8\pm2.5)\times10^{-2}~\cnts$. The source is located just outside the FOV of our \xmm\ pointings (see Fig.~\ref{fig:chan}) and was not covered by the \chan\ survey of \citet{muno2009}. However, the source region was observed with \swift/XRT as part of a \swift\ follow-up programme of unclassified \asca\ sources \citep[][]{degenaar2011_asca}. During the XRT observations, performed on 2008 March 7 (i.e., four months prior to our \chan/HRC detection), a single X-ray source was detected within the FOV. The coordinates inferred from the \swift\ data coincide with our \chan/HRC position. Analysis of the XRT spectral data yielded $N_{\mathrm{H}}=(1.6\pm0.8) \times10^{22}~\nh$ and $\Gamma=2.9\pm1.1$, resulting in a luminosity of $L_X\sim2.8\times10^{33}~\lum$ \citep[][]{degenaar2011_asca}.

We used the reported \swift/XRT spectral parameters to convert our observed \chan/HRC count rate into 2--10 keV fluxes (see Table~\ref{tab:spec}). Assuming a distance of 8 kpc, the unabsorbed flux translates into a luminosity of $L_X\sim2.1\times10^{33}~\lum$. For these spectral parameters the \rosat/PSPC and \asca/GIS count rates correspond to $L_X\sim9.2\times10^{33}$ and $5.1\times10^{33}~\lum$, respectively. Given that \andereascabron\ has been detected at comparable intensity levels with \rosat\ (1992), \asca\ (1998), \swift\ and \chan\ (both 2008), and that the source is close to the detection limit of our HRC observations (see Fig.~\ref{fig:persistent}), we consider it likely that this is not a transient object but rather a weak persistent X-ray source that displays a factor of a few variability and peaks near $L_X\sim10^{34}~\lum$. We cannot classify the source further.


\subsection{Possible transient reported by \citet{wijnands06}}\label{subsec:softsource}
\citet{wijnands06} reported on the possible discovery of a new sub-luminous X-ray transient, which was detected during our HRC-I observation of 2005 June 5 at a count rate of $\sim8\times10^{-3}~\cnts$. This object is also detected in the \xmm\ observations carried out in 2008 September, displaying a count rate of $\sim (1.9\pm 0.1)\times10^{-2}~\cnts$, whereas the other \xmm\ observations yield $2\sigma$ intensity upper limits of $\lesssim 7\times10^{-3}~\cnts$. This suggests a possible transient nature. However, the \xmm\ observations reveal that the source is detected only below 2 keV. It is therefore not a candidate transient X-ray binary, but more likely a flaring star.


\section{Discussion}\label{sec:discussion}
We have presented the results of a four-year monitoring campaign of the GC carried out with the \chan\ and \xmm\ observatories. We have covered a field of 1.2 square degrees around \sgra, which was targeted on ten different epochs between 2005 June and 2008 September. Our study focused on the behaviour of transient X-ray sources that reach 2--10 keV peak luminosities $\gtrsim1\times10^{34}~\lum$ when in outburst. We detected activity of eight previously known X-ray transients during our campaign. On average, six of these were seen active each year (see Table~\ref{tab:spec}). We studied their X-ray spectra and long-term lightcurves. All are highly absorbed ($N_{\mathrm{H}}\gtrsim1\times10^{22}~\nh$), indicating source distances near or beyond the GC.

We detected type-I X-ray bursts from the neutron star LMXBs \ascabron\ and \saxbron\ (see Sections~\ref{subsec:ascalc} and~\ref{subsec:saxlc}). For the former we also observed a $\sim$1600-s long eclipse in one of the X-ray lightcurves, which had similar properties as the eclipses that had previously been seen by \asca\ \citep[][]{maeda1996}. For \saxbron\ we found a pair of a type-I X-ray bursts that have a waiting (recurrence) time of only 3.8~min. Their similar peak intensity and duration suggests that the fuel that powered both bursts was likely of similar composition. This is an important pointer to understand the mechanism that is responsible for the unusually short recurrence time. We determined the quiescent luminosity of \saxbron\ of $L_X\sim2\times10^{31}~(D/6.7~\mathrm{kpc})^2~\lum$, which firmly establishes its classification as X-ray transient \citep[cf.][]{werner2004}. 

We uncovered episodes of low-level accretion in a luminosity range of $L_X\sim10^{33-34}~\lum$ from two X-ray transients that are known to reach considerably higher intensities of $L_X\gtrsim10^{36}~\lum$ in full outburst (\grobron\ and \xmmbron; see Section~\ref{subsec:lowlum_activity}). Two other transient sources  that were detected during our campaign (\brontwee\ and \bronacht) exhibit outbursts with similar intensities of $L_X\sim10^{34}~\lum$, but have never been observed in a brighter state (see Appendices~\ref{subsec:brontwee} and~\ref{subsec:bronnegen}).

In addition to the eight known transient systems, we detected two weak unclassified X-ray sources that, despite being undetected at some epochs during our campaign, are likely persistent and may be variable by a factor of a few (see Section~\ref{subsec:persistent}). The first is a previously unknown X-ray source that we designate \newsource. Inspection of archival data suggests that the source may be persistent at a luminosity of $L_X\sim10^{34}~\lum$ (assuming $D= 8$~kpc). Despite its relatively steady X-ray intensity, we observed significant changes in the source spectrum on a time scale of months. Based on its X-ray spectral properties and the possible association with a {\it Spitzer}/IRAC infrared object, we tentatively classify this new X-ray source as a cataclysmic variable. The unclassified X-ray source \andereascabron\ was detected with \rosat, \asca, \swift, and \chan\ at similar intensity levels of $L_X\sim10^{34}~\lum$ between 1992 and 2008, which suggests that the source is likely persistent.


\subsection{The population of Galactic centre X-ray transients}\label{subsec:pop}
In Table~\ref{tab:sources} we list all known X-ray transients that are located in the region covered by our campaign. This table is an update from \citet{wijnands06}, in which we have included three new transients that were discovered by \swift\ in 2006 \citep[\bronvier\ and \bronvijf;][]{degenaar09_gc} and in 2011 \citep[\newswift;][]{degenaar2011_newtransient}. For each source we list the angular distance from \sgra, the minimum and maximum X-ray luminositiy, and any other relevant information that characterises the source. We indicate which sources were active during our campaign and whether the transients displayed multiple outbursts in the past decade.

There are five confirmed and two candidate LMXBs amongst the 17 transients covered by our campaign (Table~\ref{tab:sources}). The confirmed LMXBs contain a neutron star primary, as demonstrated by the detection of X-ray pulsations (\grobron) or type-I X-ray bursts (\grsbron, \ascabron, \ksbron\ and \saxbron). The candidate LMXB \adcbron\ displays eclipses in its X-ray lightcurve that indicate a high inclination and an orbital period of 7.9 h. This source has been suggested to harbour a black hole based on the detection of strong radio emission \citep[][]{muno05_apj633,porquet05_eclipser}. Similarly, the transient X-ray source 1A 1742--289 that erupted in 1975 has been tentatively classified as a black hole LMXB based on its X-ray properties and the apparent association with a strong radio source \citep[][]{davies1976,branduardi1976}.

The remaining ten X-ray transients are unclassified, although their large outburst amplitudes (a factor $>100$) and spectral properties render it likely that these are accreting neutron stars or black holes. A lack of infrared counterparts with $K<15$ mag, in turn suggests that these transients are either LMXBs or HMXBs with companions fainter than a B2\,V star \citep[][]{muno05_apj622,mauerhan2009}. Tentative X-ray pulsations were reported for two of the unclassified transients: \xmmbron\ \citep[$\sim$5~s;][]{sakano05} and XMMU J174554.4--285456 \citep[$\sim$172~s;][]{porquet05}. These spin periods fall within the range of HMXBs. However, both results require confirmation \citep[][]{sakano05,porquet05}. 

It is remarkable that the ten unclassified transients all have maximum 2--10 keV luminosities of $L_X\lesssim10^{36}~\lum$ and are thus fainter than the seven confirmed LMXBs \citep[disregarding the LMXB \adcbron, which has been argued to be intrinsically brighter and possibly strongly obscured;][]{muno05_apj633}. Apart from the low outburst intensity, the X-ray spectra, outburst profiles and recurrence times are not notably different from the brighter transients \citep[see Table~\ref{tab:sources}, see also][]{muno04_apj613,degenaar09_gc}. If they are X-ray binaries, their sub-luminous character requires an explanation. 

Based on statistical arguments, \citet{wijnands06} argued that it is unlikely that all these objects are viewed at high inclination (thereby reducing their observed X-ray emission) and that most of them must be intrinsically faint. It has been proposed that these transients might consist of compact objects accreting from the wind of a main sequence companion \citep[][]{pfahl2002}. This would be a relatively inefficient process that may account for the observed low X-ray luminosities. An alternative explanation is that these are LMXBs with tight orbits and unusual donor stars (e.g. a white dwarf, brown dwarf or planet) that can only accommodate a small accretion disk \citep[][]{king_wijn06,zand09_J1718}. Comparing the low-luminosity transients with weak outbursts observed from the brighter transients suggests that there might be examples of both these types of objects amongst the Galactic centre transients (see Section~\ref{subsec:lowlum_activity}). 

Out of the 17 transients listed in Table~\ref{tab:sources}, we have detected eight in an active state during our campaign. Moreover, extensive monitoring of the GC region during the past decade has shown that 11 of these transients (i.e., 65\%) recurred between 1999 and 2012 (see the column labelled "Rec?"), six of which even experienced three or more distinct outbursts during this epoch (the neutron star LMXBs \grsbron, \ascabron, \ksbron\ and \saxbron, and the unclassified transients \brontwee\ and \xmmbron). 

No new X-ray transients with a 2--10 keV peak luminosity of $L_X\gtrsim1\times10^{34}~\lum$ were found during our campaign. This confirms suggestions from previous authors that the majority of sources that recur on time scales less than a decade, and undergo outbursts of at least a few days, have now been identified in this region \cite[][]{zand04,muno2009,degenaar2010_gc}. Although a new transient was discovered recently in July 2011 \citep[][]{degenaar2011_newtransient,degenaar2011_newtransient2,chakrabarty2011,sevillat2011}, previous discoveries date back to 2006 in spite of extensive monitoring with various X-ray satellites \citep[][]{degenaar09_gc}.


\subsection{Quiescent luminosities}\label{subsec:quiescence}
In addition to the maximum 2-10 keV intensity reached during outburst, Table~\ref{tab:sources} also includes information on the quiescent luminosity of the 17 transients covered by our campaign. The listed values correspond to the minimum X-ray luminosity observed for each source and were taken from the literature. Only for \saxbron, \ksbron\ and \bronvijf\ we did not find reported quiescent properties. We therefore searched the \chan\ and \xmm\ data archives for observations that covered the quiescent states of these three transients. We presented the results for \saxbron\ in Section~\ref{subsec:saxlc}, whereas the other two are discussed below. 

We note that the new transient that was discovered in 2011 (\newswift) is likely associated with \newswiftcxo\ \citep[][]{chakrabarty2011}. This faint X-ray source was discovered in the \chan\ survey of the GC, during which it was detected at a luminosity of $L_X\sim3 \times10^{31}~\lum$ and did not display any variability on long or short time scales \citep[][]{muno2009}. \newswiftcxo\ likely represents the quiescent counterpart of the new 2011 transient and therefore we quote its luminosity as the quiescent level of \newswift.

\noindent
{\bf \ksbron:} To determine the quiescent luminosity of \ksbron, we used archival \chan\ observations performed in 2001, 2006 and 2007 (obs IDs 2267, 2272, 7038, and 8459) that amount to a total exposure time of $\sim$56~ks. The composite spectrum was obtained using the tool COMBINE$\_$SPECTRA and yields a 2--10 keV quiescent luminosity of $L_X\sim5\times10^{32}~\lum$ for an assumed distance of 8~kpc. The quiescent spectral properties of \ksbron\ will be discussed in more detail in a separate work.\\ 

\noindent
{\bf \bronvijf}: We found a total of seven archival \chan/ACIS-I observations that covered the position of \bronvijf. This source exhibited an outburst in 2006 that was first detected with \swift\ in mid-May. In late June, the source went undetected by \swift/XRT, indicating an upper limit on the 2--10 keV quiescent luminosity of a few times $10^{33}~\lum$ \citep[][]{degenaar09_gc}. We found that the source is weakly detected in \chan\ observations carried out on 2006 July 4 (obs ID 6642, $\sim$5~ks) and 2006 August 24 (obs ID 7037, $\sim$40~ ks). 

By fitting the spectra of these two \chan\ observations simultaneously with the \swift\ outburst data, we found a joint value of $N_H=(10.5 \pm 3.7)\times 10^{22}~\nh$ and photon indices of $\Gamma=3.0\pm1.0$ (\swift), $4.0\pm2.8$ and $6.6\pm2.6$ (\chan; 2006 July and August, respectively). It appears that the spectrum was softening, although the source continued to be detected above 2 keV (at lower energies the source photons are strongly absorbed). The corresponding 2--10 keV luminosities are $L_X\sim 1\times10^{34}$, $\sim4\times10^{33}$ and $\sim1\times10^{33}~\lum$, respectively. 

We can obtain additional constraints on the quiescent luminosity of \bronvijf\ by using archival \chan\ data obtained in 2001 and 2004 (obs IDs 2282, 2291, 4683, and 4684; total exposure time of $\sim$120~ks). This allows us to estimate a pre-outburst upper limit on the 2--10 keV luminosity of $L_X\sim(0.7-1)\times10^{32}~\lum$ (for an absorbed powerlaw model with $N_H=10.5\times 10^{22}~\nh$ and $\Gamma=3-6$, assuming $D=8$~kpc). \\

\noindent
There are caveats to consider when interpreting the quiescent luminosities listed in Table~\ref{tab:sources}. Although it is generally assumed that accretion is strongly suppressed during quiescence, it may not have fully come to a halt. Indeed, several transient neutron star and black hole X-ray binaries have been found to display considerable variation in their quiescent X-ray luminosity, which is attributed to residual accretion \citep[for recent examples, see][]{kong2002,hynes2004,cackett2010,cackett2011,fridriksson2011,degenaar2012_1745}. This implies that some of the minimum X-ray luminosities listed in Table~\ref{tab:sources} might correspond to low-level accretion. Literature values often only reflect a snapshot of the quiescent state and might therefore not necessarily represent the absolute minimum X-ray luminosity of these transient sources.

It is worth noting that several transients have quiescent luminosities in the range of $L_X\sim10^{32-33}~\lum$ (see Table~\ref{tab:sources}). Examples include the neutron star LMXBs \grsbron, \ascabron, \ksbron\ and \grobron. Transient neutron star LMXBs often have quiescent X-ray spectra that are composed of a soft, thermal component that dominates the spectrum below $\sim$2~keV, accompanied by a hard emission tail that can be described by a simple powerlaw. Since any soft thermal emission is strongly absorbed by the high hydrogen column densities inferred for our sources, this implies that they must have considerable hard powerlaw emission in quiescence \citep[cf.][]{jonker2003}. We note that detections and upper limits of about $L_X\sim10^{31}~\lum$ (such as obtained for \saxbron) do not provide strong constraints on the temperature of the neutron star, since the soft thermal emission is so heavily absorbed.


\subsection{Low-level accretion activity}\label{subsec:lowlum_activity}

Although several transients covered by our campaign are known to exhibit outbursts with 2--10 keV peak luminosities of $L_X>10^{36}~\lum$ (see Table~\ref{tab:sources}), these sources also display activity well below that level. We already discussed in Section~\ref{subsec:weak} that the unclassified transient \xmmbron\ and the pulsating neutron star LMXB \grobron\ were both detected during our campaign at a luminosity of  $L_X\sim10^{33-34}~\lum$, which is intermediate between their quiescent state and their full X-ray outbursts. For the former, \swift\ monitoring observations of the GC have shown that this low-luminosity state can persist for several weeks \citep[][]{degenaar2010_gc}. Indeed, this X-ray transient is more often detected at a low X-ray luminosity than in full outburst.

The two neutron star LMXBs \grsbron\ and \ksbron\ have both shown short, weak outbursts ($t_{\mathrm{ob}}\lesssim1$~week, $L_X\lesssim10^{34-35}~\lum$) that preceded longer and brighter accretion episodes with a duration of several weeks/months and 2--10 keV luminosities peaking at $L_X\sim10^{36-37}~\lum$ (see Appendices~\ref{subsec:grsbron} and \ref{subsec:ksbron}). Albeit on a different scale, one can argue that the neutron star LMXB \ascabron\ displayed something similar; a major outburst with a 2--10 keV peak luminosity of $L_X\sim6\times10^{36}~\lum$ was detected from this system in 2007--2008, whereas a few months earlier (in 2006) it underwent an accretion episode that had a duration and intensity that were both a factor $\sim$5 lower \citep[see also][]{degenaar2010_gc}. Finally, the neutron star LMXB \saxbron\ was found to display activity at $L_X\sim10^{35}~\lum$ on several occasions, which is also well below its maximum outburst intensity \citep[][]{wijnands2002_saxj1747,campana2009}. Another firm indication of low-level accretion activity is the detection of type-I X-ray bursts from \saxbron\ and \ksbron\ at accretion luminosities of $L_X\sim10^{35}~\lum$ (\citealt{chenevez2009}; \citealt{linares2011}; see also \citealt{kuulkers2008} for similar examples). 

Taken together, it appears that all five neutron star LMXBs covered by our campaign exhibited multiple luminous outbursts with $L_X \gtrsim10^{36}~\lum$ in the past decade, but also frequently display low-level accretion activity at $L_X \sim10^{33-35}~\lum$. For \grsbron, \ksbron\ and \ascabron\ it was found that the weaker outbursts are shorter. The disk instability model, which is thought to provide the framework to explain the transient behaviour of LMXBs, provides a possible explanation \citep[e.g.][]{king98,lasota01}. Sub-luminous outbursts may occur when only a small fraction of the accretion disk becomes ionized and accreted onto the compact object, whereas a larger part of the disk is consumed during the more luminous outbursts \citep[see also][]{degenaar2010_gc}. 

Our campaign covered two unclassified X-ray transients that appear active at levels of $L_X\sim10^{33-34}~\lum$ without becoming brighter. Despite being frequently active, both \brontwee\ (Appendix~\ref{subsec:brontwee}) and \bronnegen\ (Appendix~\ref{subsec:bronnegen}) have never been observed at a  luminosity $L_X\gtrsim5\times10^{34}~\lum$ \citep[][]{muno05_apj622}. Strikingly, as many as ten out of 17 transients covered by our campaign have 2--10 keV peak luminosities that are well below $L_X\sim10^{36}~\lum$ and have never been observed in a brighter state (see Table~\ref{tab:sources}). 

These very-faint X-ray transients are possibly connected to the "burst-only" sources: a group of neutron star LMXBs that became detectable only when exhibiting a type-I X-ray burst, while no persistent emission could be detected down to a limit of $L_X\sim10^{36}~\lum$ \citep[][]{cornelisse02,wijnands06,campana09}. In analogy with the weak outbursts seen for the brighter neutron star LMXBs, we may interpret the low X-ray luminosities in terms of a small amount of matter accreted from the disk. The fact that these sources have never been observed in a brighter state could in turn imply that the entire disk (and hence the binary orbit) is relatively small. It is of note that it takes a long time to accumulate enough fuel to power an X-ray burst when the mass-accretion rate is very low. These are therefore expected to occur much less frequently than for higher accretion rates \citep[e.g.,][]{zand09_J1718,degenaar2010_burst}. 

Long-lasting activity at $L_X\sim10^{33}~\lum$, such as seen for the unclassified transient \xmmbron, is difficult to accomodate within the disk instability model. This behaviour might find a more natural explanation in terms of wind-accretion \citep[see the discussion in][]{degenaar2010_gc}. Similarly, some of the other unclassified transients (e.g. \brontwee\ and \bronacht) appear to spend long times at relatively low intensity levels \citep[][]{muno05_apj622,degenaar2010_gc}. This might be an indication that in these systems the companion star does not fill its Roche lobe, but supplies matter to the compact object via a wind. On the other hand, there are a few examples of LMXBs in which the donor star does overflow its Roche lobe and that appear to persist at low X-ray luminosities of $L_X\sim10^{34-35}~\lum$ \citep[e.g.][]{delsanto07,zand09_J1718,campana09,degenaar2010_burst}. At least one of these harbours a hydrogen-rich companion and must thus have a relatively large orbit \citep[][]{degenaar2010_burst}. The sub-luminous character therefore remains a puzzle.

\subsection{Ultra-faint X-ray transients}\label{subsec:ultrafaint}
While our present work focused on transient X-ray sources with luminosities $L_X\gtrsim1\times10^{34}~\lum$, we detected a number of objects that appear variable by a factor of $\gtrsim5-10$ but remain below $L_X\sim5\times10^{33}~\lum$ (assuming $D=8$~kpc). We found several such sources in our \xmm\ observations. They are hard (most photons emitted above 2 keV) and have no DSS/2MASS counterparts, which effectively rules out that these are foreground stars. 

Two such examples are \munotransientnew\ and CXOGC J174423.4--291741 from the \chan\ catalogue of \citet{muno2009}. Both objects were detected once during our campaign at 2--10 keV luminosities of $L_X\sim3\times10^{33}~\lum$. \munotransientnew\ is indicated by \citet{muno2009} as exhibiting long-term variability by a factor of $\sim$70. CXOGC J174423.4--291741, on the other hand, is listed in this catalogue as a weak persistent source displaying a luminosity of $L_X\sim10^{32}~\lum$ (using the conversion factor from photon to energy flux quoted by these authors). 

As discussed in Section~\ref{subsec:lowlum_activity}, several confirmed and candidate X-ray binaries display activity at similar intensity levels. Furthermore, \citet{heinke09_vfxt} found a transient object with a peak luminosity of $L_X\sim6\times10^{33}~\lum$ in the globular cluster M15, which is likely a neutron star LMXB. Although a significant fraction of these ``ultra-faint X-ray transients" found in our \xmm\ data might be accreting white dwarfs \citep[][]{verbunt1997}, this suggests that there could also be X-ray binaries amongst them.

\begin{table*}
\begin{threeparttable}[htb]
\begin{center}
\caption[]{{List of X-ray transients with 2--10 keV peak luminosities $L_X\gtrsim10^{34}~\lum$ that are located in the region covered by this campaign.}}
\begin{tabular}{l l l l l l l}
\hline
\hline
Source name & Offset from  & $L_X^{\mathrm{max}}$ &  $L_X^{\mathrm{min}}$ & Comments & Rec? & Ref.\\
 & \sgra\ ($'$) & ($\mathrm{erg~s^{-1}}$)  & ($\mathrm{erg~s^{-1}}$)  & & & \\
\hline 
\adcbron & 0.05 & $1 \times 10^{35}$ & $\lesssim 2 \times10^{31}$ & LMXB black hole candidate, radio source & No & 1,2,3\\
& &  &  & $P_{\mathrm{orb}}=7.9$~h, high inclination &  & \\
\bronnegen* & 0.31 & $1 \times 10^{34}$ & $\lesssim 8 \times10^{31}$ & unclassified & Yes & 2 \\
\brondrie & 0.37 & $2 \times 10^{35}$ & $\lesssim 4 \times10^{31}$ & unclassified & Yes & 2,4\\	
\bronacht & 0.44 & $2 \times 10^{35}$ & $\sim 1 \times10^{33}$ & unclassified & Yes & 2,5\\	
1A 1742--289 & 0.92 & $7 \times 10^{38}$ &  $\sim 5 \times10^{31}$ & LMXB black hole candidate, radio source & No & 6,7,19\\
\brontwee* & 1.35 & $3 \times 10^{34}$ & $\lesssim 9 \times10^{30}$ & unclassified & Yes & 2,4,5\\
\newswift & 1.36 & $8 \times 10^{34}$ & $\sim 3\times10^{31}$ & unclassified (\newswiftcxo) & No & 19,20,21,22 \\
\ascabron* & 1.37 & $6 \times 10^{36}$ & $\sim 5\times10^{32}$ & neutron star LMXB (burster), $P_{\mathrm{orb}}=8.4$~h & Yes & 4,5,8 \\
\bronvier & 4.50 & $2 \times 10^{35}$ & $\sim 7 \times10^{31}$ & unclassified (\bronviercxo)& No & 4 \\	
XMMU J174554.4--285456 & 6.38 & $8 \times 10^{34}$ & $\lesssim 2 \times10^{31}$ & unclassified, possible 172-s X-ray pulsar & No & 2,23\\
\grsbron* & 10.00 & $1 \times 10^{37}$ & $\sim 1\times10^{32}$ & neutron star LMXB (burster), $D=7.2$~kpc & Yes & 4,5,9,10 \\
\bronvijf & 11.04 & $7 \times 10^{34}$ &  $\lesssim 7 \times10^{31}$ & unclassified (\bronvijfcxo) & No & 4\\	
XMM J174544--2913.0 & 12.56 & $5 \times 10^{34}$ & $\lesssim 1\times10^{32}$ & unclassified & No & 11 \\
\xmmbron* & 13.78 & $2\times 10^{36}$ &  $\sim 1\times10^{32}$ & unclassified, possible 5-s X-ray pulsar & Yes & 4,5,11 \\
\saxbron* & 19.55 & $4 \times 10^{37}$ & $\sim 2 \times10^{31}$ & neutron star LMXB (burster), $D=6.7$~kpc & Yes & 12,13 \\
\grobron* & 21.71 & $3 \times 10^{38}$ & $\sim 2\times10^{33}$ & neutron star LMXB (0.5-s X-ray pulsar)  & Yes & 14,15,16\\
 &  &  &  & $P_{\mathrm{orb}}=11.8$~d &  & \\
\ksbron* & 22.09 & $5 \times 10^{36}$ &  $\sim 5\times10^{32}$ & neutron star LMXB (burster) & Yes & 17,18\\
\hline
\end{tabular}
\label{tab:sources}
\begin{tablenotes}
\item[]Note.-- Table updated from \citet{wijnands06}. Sources marked by an asterisk were detected in an active state during our campaign. $L_X^{\mathrm{max}}$ and $L_X^{\mathrm{min}}$ represent the peak and quiescent luminosities, respectively. These values were taken from the literature and converted into the 2--10 keV energy band, or determined in this work. A distance of 8 kpc was assumed, except for \grsbron\ and \saxbron, which have estimated distances of 7.2 and 6.7 kpc, respectively.
The column labelled "Rec?" indicates whether a source recurred (i.e., displayed more than one distinct outburst) between 1999 and 2012. 
References: 1=\citet{muno05_apj633}, 2=\citet{muno05_apj622}, 3=\citet{porquet05_eclipser}, 4=\citet{degenaar09_gc}, 5=\citet{degenaar2010_gc}, 6=\citet{davies1976}, 7=\citet{branduardi1976}, 8=\citet{maeda1996}, 9=\citet{muno03_grs}, 10=\citet{trap09}, 11=\citet{sakano05}, 12=\citet{werner2004}, 13=\citet{wijnands2002_saxj1747}, 14=\citet{giles1996}, 15=\citet{wijnands2002_gro1744}, 16=\citet{daigne2002}, 17=\citet{zand1991}, 18=\citet{cesare2007}, 19=\citet{muno2009}, 20=\citet{degenaar2011_newtransient}, 21=\citet{degenaar2011_newtransient2}, 22=\citet{chakrabarty2011}, 23=\citet{porquet05}.
\end{tablenotes}
\end{center}
\end{threeparttable}
\end{table*}

\begin{acknowledgements}
This work was supported by the Netherlands organisation for scientific research (NWO) and the Netherlands Research School for Astronomy (NOVA). ND is supported by NASA through Hubble Postdoctoral Fellowship grant number HST-HF-51287.01-A from the Space Telescope Science Institute, which is operated by the Association of Universities for Research in Astronomy, Incorporated, under NASA contract NAS5-26555. RW acknowledges support from a European Research Council (ERC) starting grant. This research has made use of data obtained from the Chandra Data Archive and the Chandra Source Catalog, and software provided by the \chan\ X-ray Center (CXC) in the application package CIAO. This work was in part based on observations obtained with \xmm, an ESA science mission with instruments and contributions directly funded by ESA Member States and NASA. We also made use of the \swift\ public data archive. 
\end{acknowledgements}


\begin{appendix}

\setcounter{figure}{0}
\renewcommand{\thefigure}{A.\arabic{figure}}

\section{Results on individual sources}\label{sec:appendix}
In this appendix we describe the properties of the eight X-ray transients that were detected during our campaign in more detail. For each source we describe the X-ray spectra and lightcurves of individual observations, as well as the flux evolution during our campaign. In particular, we attempt to constrain the luminosity, duration and recurrence time of the observed outbursts, sometimes using literature reports and archival X-ray observations. We also briefly review the historic outburst behaviour of each source, as well as any activity reported elsewhere during and after our campaign.

As mentioned in Section~\ref{sec:intro}, the inner $\sim$$25' \times 25'$ around \sgra\ (corresponding to field GC-2 in our campaign) has been covered with \swift's XRT starting in 2006 \citep[][]{kennea_monit,degenaar09_gc,degenaar2010_gc}. Those quasi-daily observations provide a partial overlap with our campaign. For sources located in this region we will therefore compare reports on the \swift\ data with results from our \chan/\xmm\ observations. Sources located at larger angular distances ($\gtrsim$$20'$) are covered by the \inte\ Galactic bulge monitoring programme, which provides observations every few days during two $\sim$4-month windows per year \citep[][]{kuulkers07}. 

We summarised the results of our spectral analysis in Table~\ref{tab:spec}. Below, the individual sources are discussed in the same order as they appear in Table~\ref{tab:spec}, so that the numbering of the following sections corresponds to the numbered labels in that table. To directly compare the different sources all spectra are plotted in the same energy range (2--10 keV) and all long-term lightcurves on the same intensity scale ($L_X=10^{31-37}~\lum$). In all plots, triangles represent \xmm\ observations, while bullets and squares are used for \chan/ACIS and HRC data, respectively. The last two objects listed in Table~\ref{tab:spec} are likely weak persistent X-ray sources and were already discussed in detail in Section~\ref{subsec:persistent}.


\subsection{\grsbron}\label{subsec:grsbron}
\noindent{\bf Brief historic overview:} This transient neutron star LMXB was discovered by the \granat\ observatory in 1990 \citep[][]{sunyaev1990} and has been detected in an active state many times since then \citep[for a detailed overview, see][]{trap09}. The source is known to display type-I X-ray bursts \citep[e.g.][]{cocchi99}, from which a distance of 7.2 kpc can be inferred \citep[][]{trap09}. It is frequently active and typically exhibits outbursts that reach a 2--10 keV peak luminosity of $L_X\sim10^{36-37}~(D/7.2~\mathrm{kpc})^2~\lum$, and last for a few weeks \citep[e.g.][]{degenaar2010_gc}. In quiescence, the source is detected at $L_X\sim1\times10^{32}~(D/7.2~\mathrm{kpc})^2~\lum$ \citep[][]{muno03_grs}.  \\

\noindent{\bf Activity during our campaign:} We detected activity from \grsbron\ in two different epochs (in 2005 and 2007). The source was first detected in outburst during the \chan/HRC observations performed on 2005 June 5 and was also seen in the follow-up ACIS pointing of 2005 July 1 \citep[][]{wijnands06}. Over the one month time span separating these two observations, the source intensity decreased by a factor $\sim$4 (see Table~\ref{tab:spec}) from $L_X\sim6\times10^{35}$ to $1.5\times10^{35}~(D/\mathrm{7.2~kpc})^2~\lum$. This may indicate that the activity was ceasing. The rise of this outburst was caught by \inte\ in 2005 in early April \citep[][]{kuulkers07}. If the decrease in flux signalled by the \chan\ data was due to a transition towards quiescence, the duration of the 2005 outburst was therefore $\sim$13~weeks. 

\begin{figure}
 \begin{center}
\includegraphics[width=8.0cm, angle=0]{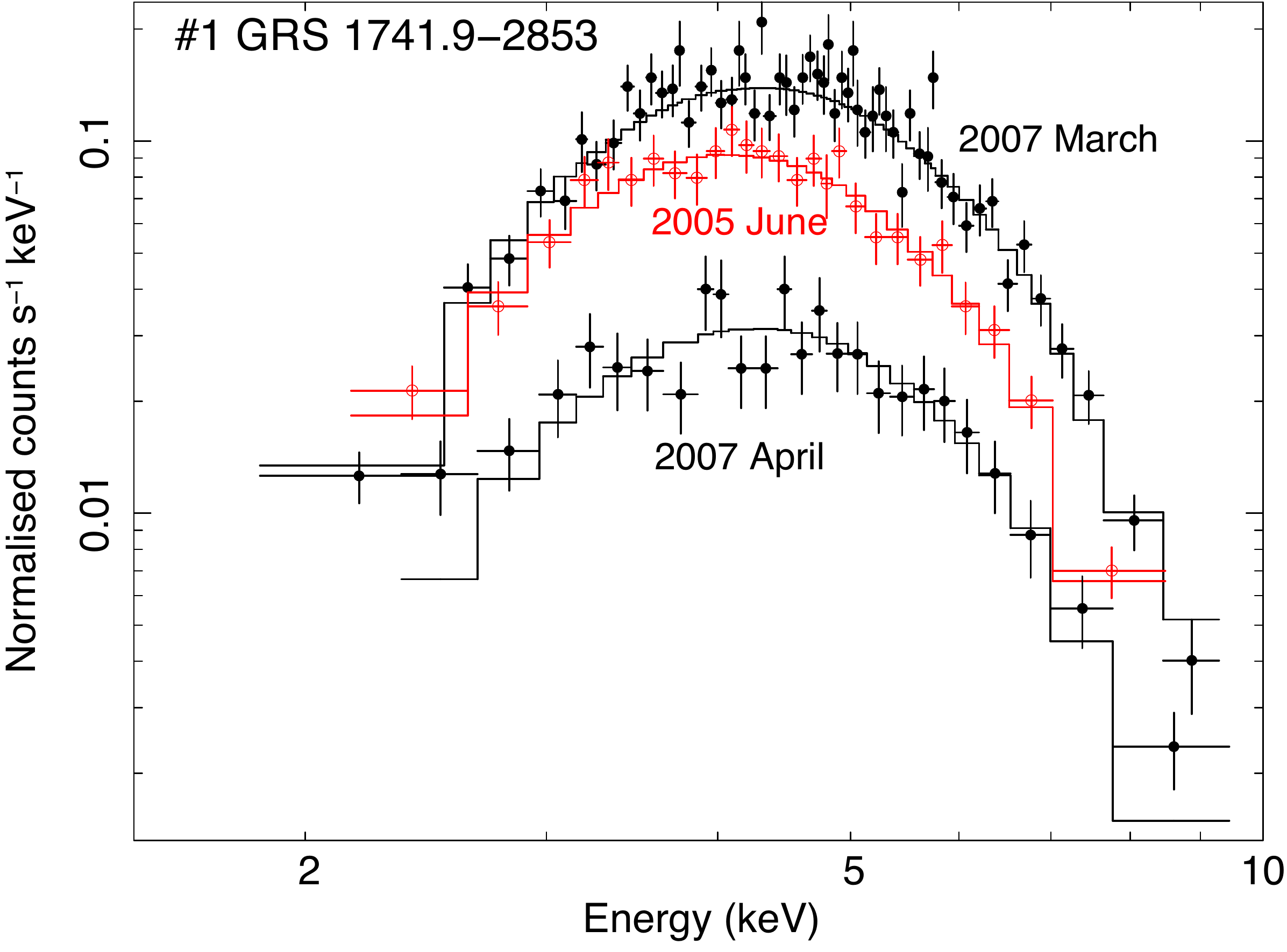}\vspace{0.1cm}
\includegraphics[width=8.0cm]{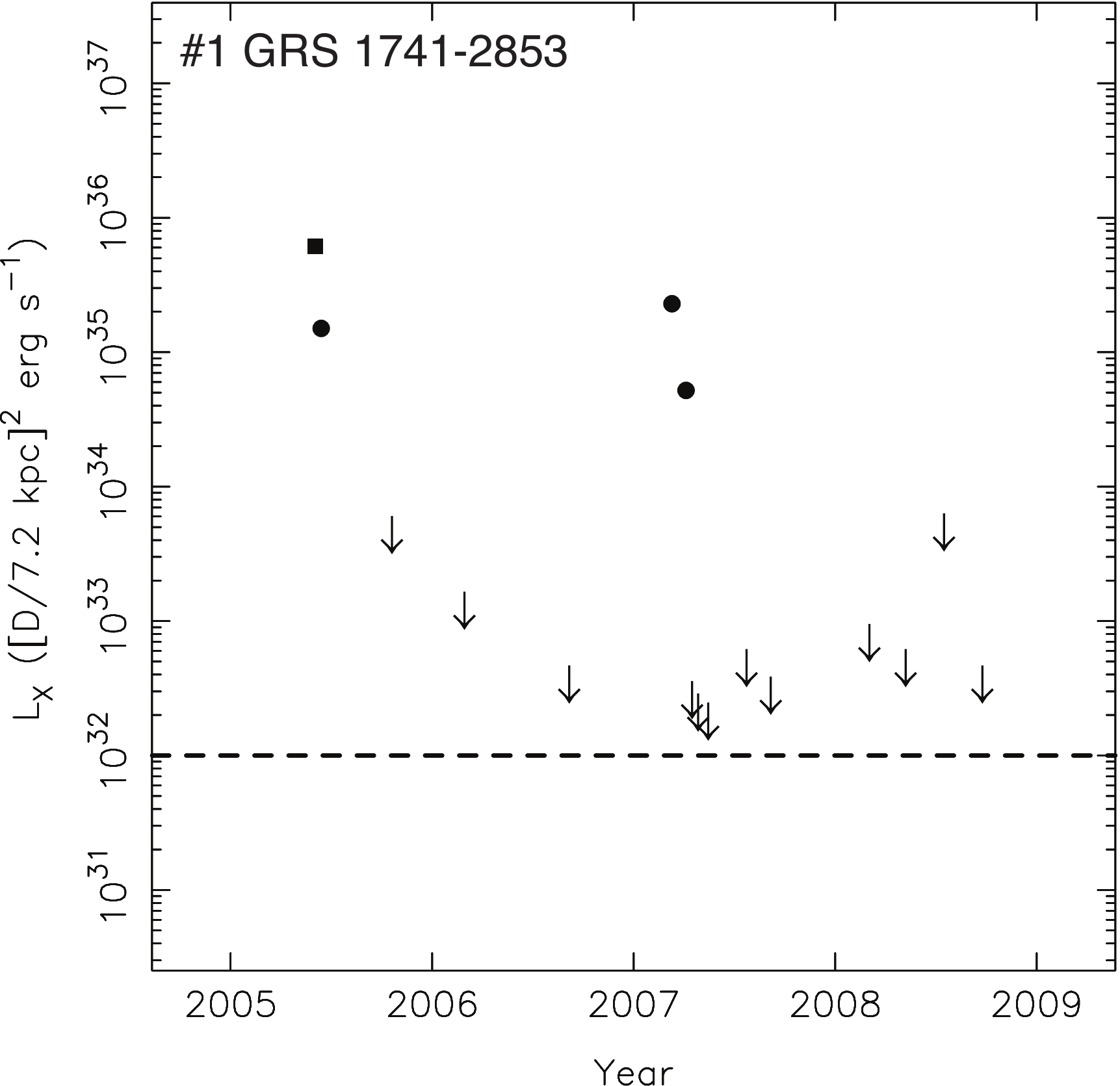}
    \end{center}
    \caption[]{Background-corrected \chan/ACIS spectra (top) and 2--10 keV luminosity evolution (bottom) of the bursting neutron star LMXB \grsbron. In the lightcurve the square indicates \chan/HRC data and the bullets \chan/ACIS measurements. The upper limit symbols represent a $2\sigma$ confidence level and the horizontal dashed line indicates the quiescent luminosity of the source.  }
 \label{fig:grsbron}
\end{figure}

We found the source again in outburst during our \chan\ observations of 2007 March 12, when it was bright enough to cause pile-up of the ACIS instrument. We therefore extracted the source spectrum using a $10''-40''$ annulus, avoiding the inner piled-up part of the  PSF (see Section~\ref{subsec:data_xmm}). The inferred luminosity was $L_X\sim2\times10^{35}~(D/\mathrm{7.2~kpc})^2~\lum$. During the subsequent observation performed on 2007 April 6, \grsbron\ had nearly faded by one order of magnitude (see Table~\ref{tab:spec}), whereas the source was not detected on April 18 with an upper limit of $L_X\lesssim3\times10^{32}~\lum$. This indicates that the source had returned to the quiescent state at that time (see Fig.~\ref{fig:grsbron}). The 2007 outburst of \grsbron\ was covered by different satellites \citep[][]{kuulkers07_atel1005,wijnands07_atel1006,muno07_atel1013,porquet07}. \swift/XRT observations detected the source with a peak luminosity of $L_X\sim1.5\times10^{36}~(D/\mathrm{7.2~kpc})^2~\lum$ and constrain the outburst duration to be $\gtrsim13$~weeks \citep[][]{degenaar2010_gc}. \\

\noindent{\bf X-ray spectra:} A joint fit to the ACIS spectral data obtained during the two outbursts of 2005 and 2007 (Fig.~\ref{fig:grsbron}) yields $N_H=(11.4\pm1.1)\times10^{22}~\nh$ and $\Gamma \sim1.5-2.0$ (see Table~\ref{tab:spec}). These values are comparable to those found for other outbursts of \grsbron\ \citep[][]{muno03_grs,trap09,degenaar09_gc,degenaar2010_gc}. The count rate detected during the 2005 June \chan/HRC pointing and upper limits inferred from observations in which the source was not detected, were converted to 2--10 keV unabsorbed fluxes using $\Gamma=2.0$ and $N_{\mathrm{H}}=11.4\times10^{22}~\nh$. \\

\noindent{\bf Constraints on the weak 2006 outburst:} Apart from outbursts reaching $L_X\gtrsim10^{36}~(D/\mathrm{7.2~kpc})^2~\lum$, \grsbron\ also undergoes low-level accretion activity. \swift/XRT monitoring observations of the GC exposed a weak, short outburst from \grsbron\ between 2006 September 14--20, during which the source did not become brighter than $L_X\sim7\times10^{34}~(D/7.2~\mathrm{kpc})^2~\lum$ \citep[2--10 keV;][]{degenaar09_gc}. This is about three orders of magnitude lower than the maximum outburst luminosity exhibited by this source, yet still a factor $\gtrsim100$ above its quiescent level. 

The source region is covered by one of our \xmm\ observations on 2006 September 8, which is just one week before the sub-luminous outburst detected by \swift/XRT. During these observations, \grsbron\ was not detected and we can infer a $2\sigma$ upper limit on the PN count rate of $\lesssim 3\times10^{-3}~\cnts$. Using \textsc{pimms} with $N_{\mathrm{H}}=11.4\times10^{22}~\nh$ and $\Gamma=2.0$, we can estimate that the 2--10 keV luminosity of \grsbron\ was $L_X\lesssim 3\times10^{32}~(D/\mathrm{7.2~kpc})^2~\lum$ at that time. Therefore, one week prior to the peculiar short 2006 outburst there were no indications of enhanced activity above the quiescent level. \\

\noindent{\bf Activity after our campaign:} \swift\ monitoring observations detected a new outburst from \grsbron\ in 2009 September--November, which had a duration of $\sim$4--5~weeks and reached up to $L_X\sim1\times10^{37}~(D/\mathrm{7.2~kpc})^2~\lum$ \citep[][]{degenaar09_gc}. The source was again active for $\sim$12~weeks in 2010 July--October at an average luminosity of $L_X\sim4\times10^{35}~(D/\mathrm{7.2~kpc})^2~\lum$ \citep[][]{degenaar2010_atel_grs_xmm}. This further underlines the frequent activity of this neutron star LMXB, as is illustrated by the results from our \chan/\xmm\ monitoring campaign. It was noted by \citet{degenaar2010_gc} that despite differences in duration and maximum intensity, the fluency of the bright outbursts of \grsbron\ are very similar (with the exception of the unusual faint, short 2006 outburst).

\subsection{\ascabron}\label{subsec:ascabron}
\noindent{\bf Brief historic overview:} Another transient that was frequently detected during our monitoring campaign is located at an angular distance of $\sim$$1.5'$ from \sgra, at a position consistent with the \chan\ localisation of \ascabron\ \citep[][]{heinke08}. This neutron star LMXB was discovered in 1993 by the \asca\ observatory, exhibits type-I X-ray bursts and its X-ray lightcurve displays eclipses that recur every $\sim$8.4~h, corresponding to the orbital period of the binary \citep[][]{maeda1996,kennea1996}. The source is detected in quiescence at a luminosity of $L_X\sim10^{32}~\lum$ \citep[][]{degenaar09_gc}. Following detections by \asca\ in 1993 and 1994, \ascabron\ was never reported in outburst again until 2006, despite extensive monitoring of the source region. \\

\noindent{\bf Activity during our campaign and outburst constraints:} We detected two distinct outbursts from \ascabron\ (in 2006 and 2007--2008). The source was first active during our \xmm\ observations of 2006 February 27, when it displayed a luminosity of $L_X\sim4.6\times10^{35}~\lum$ (assuming a distance of $D=8$~kpc). In the subsequent observation performed on 2006 September 8, the source was not detected with an upper limit of $L_X\lesssim 3 \times10^{33}~\lum$ (assuming $N_H=21.8~\times10^{22}~\nh$ and $\Gamma=2.0$). This is consistent with results obtained with \swift/XRT, which indicated that the source was active for three months between 2006 February and June, but resided in quiescence thereafter \citep[][]{degenaar09_gc}. 

It is unclear when the 2006 outburst of \ascabron\ started, since the position of the Sun with respect to the GC rendered this region unobservable between 2005 November and 2006 February. Since the source was not detected during our 2005 monitoring observations, the outburst must have started after 2005 October 20 (see Table~\ref{tab:obs_monit}). The time span between the \chan\ observations and the first detection of \ascabron\ on 2006 February 24 \citep[with \swift;][]{kennea06_atel920} is 4 months. This constrains the outburst duration to 3--7 months. 

\begin{figure}
 \begin{center}
\includegraphics[width=8.0cm,angle=0]{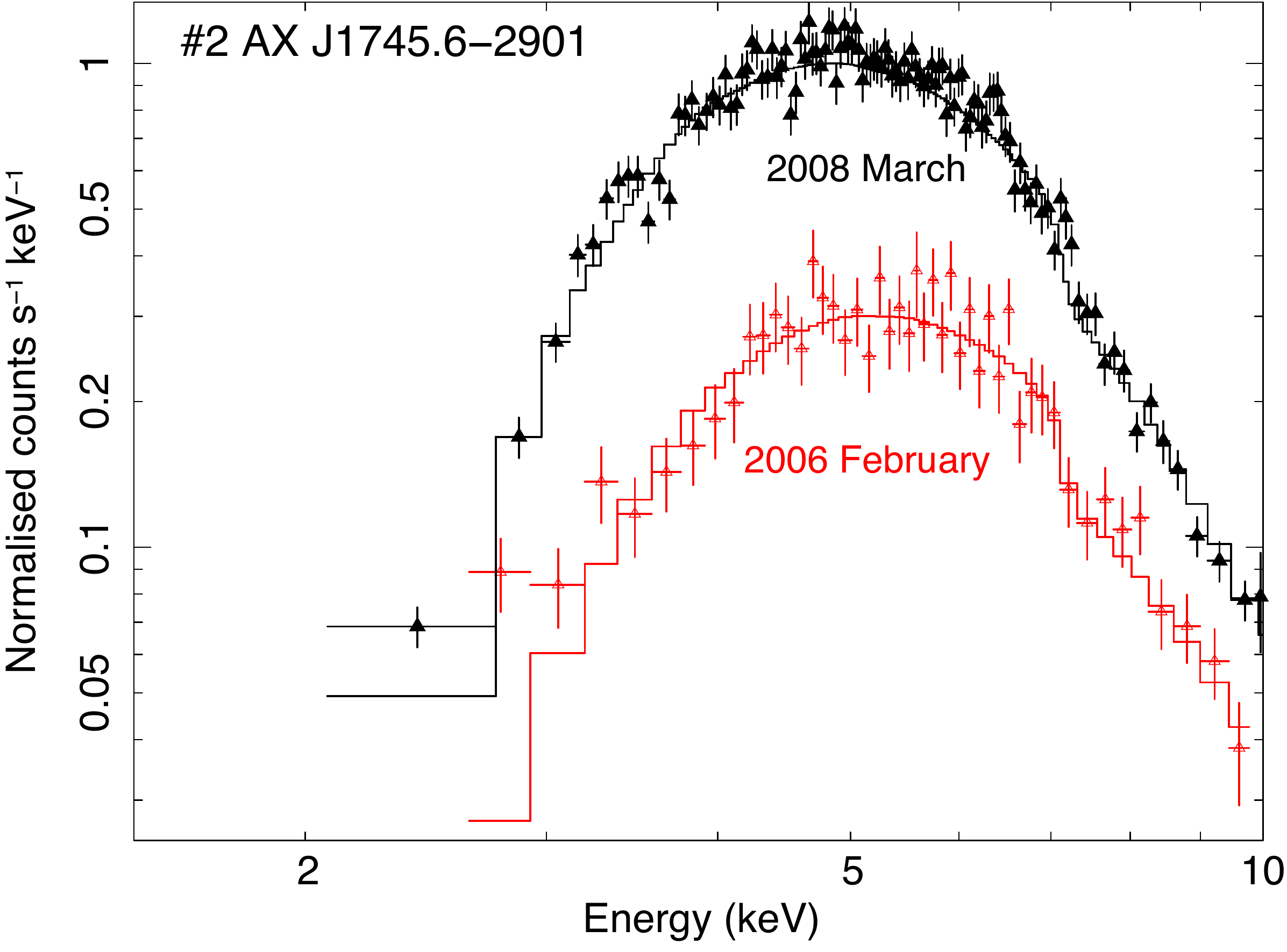}\vspace{0.1cm}
\includegraphics[width=8.0cm]{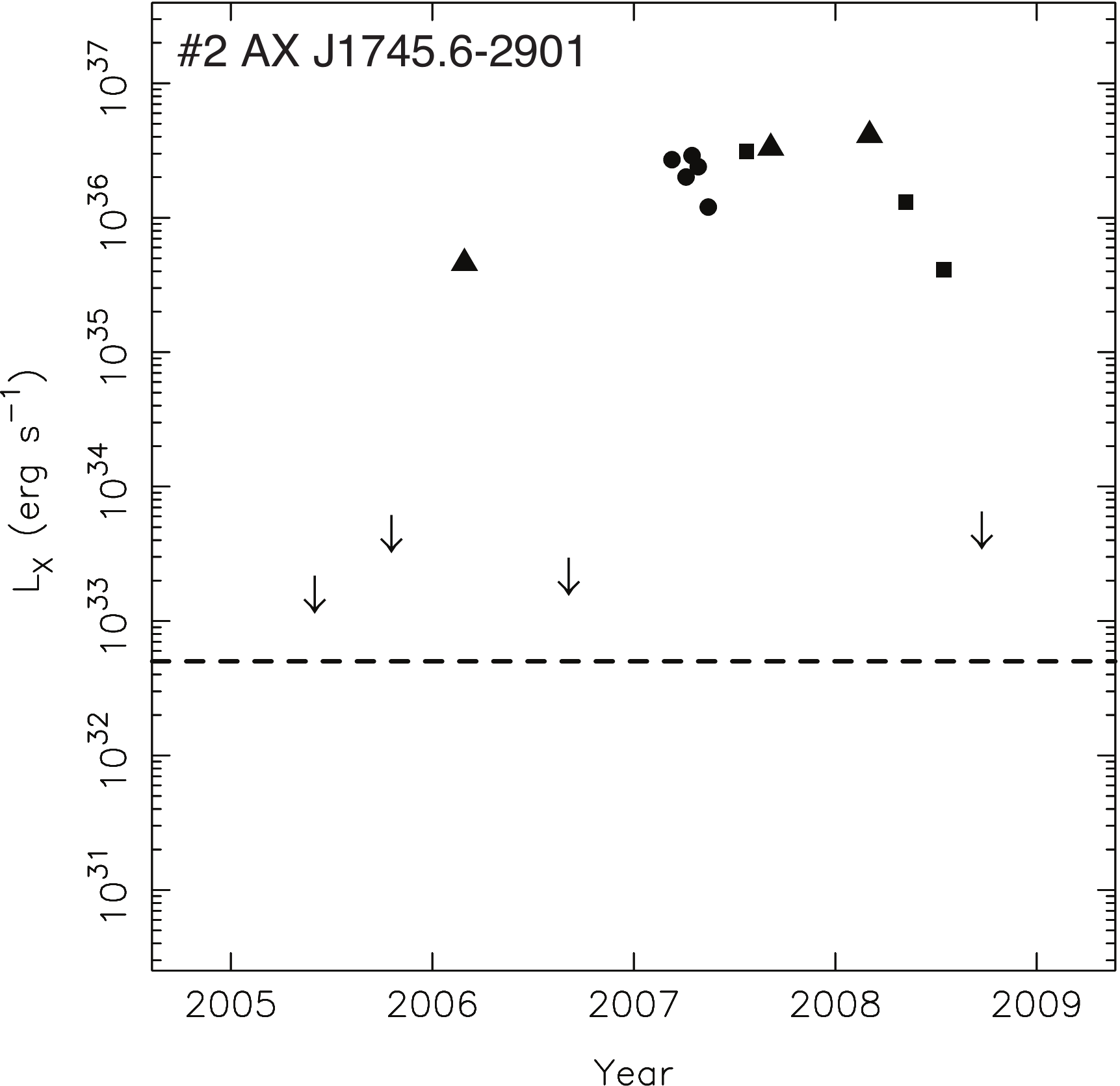}
    \end{center}
    \caption[]{Background-corrected \xmm/PN spectra (top) and 2--10 keV luminosity evolution (bottom) of the bursting neutron star LMXB \ascabron. In the lightcurve squares (HRC) and bullets (ACIS) are used for \chan\ data, whereas the triangles represent \xmm\ observations. The upper limit symbols represent a $2\sigma$ confidence level and the horizontal dashed line indicates the quiescent luminosity of the source.  }
 \label{fig:ascabron}
\end{figure}

Renewed activity of the source was reported in 2007 February, as seen by various instruments \citep[][]{kuulkers07_atel1005,wijnands07_atel1006,porquet07,degenaar09_gc}. \ascabron\ is detected at similar luminosities of $L_X\sim (1-4)\times10^{36}~\lum$ in all our observations carried out between 2007 February and 2008 May (see Table~\ref{tab:spec}). We picked up a likely type-I X-ray burst and an X-ray eclipse during these observations (see Section~\ref{subsec:ascalc}). The source intensity had decreased by nearly a factor 10 in 2008 July, and it went undetected in 2008 September. This indicates that the source had returned to the quiescent state (see Fig.~\ref{fig:ascabron}). 

The \swift/XRT observations of the GC also suggest that \ascabron\ was continuously active since 2007 February, until it returned to quiescence in 2008 in early September \citep[][]{degenaar2010_gc}. The \swift\ observations suggest that this long outburst must have commenced between 2006 November 3 and 2007 March 6 (when the GC was unobservable due to Sun-angle constraints). This constrains the total duration of the outburst to $18-21$ months ($1.5-1.75$~yr). The 2007--2008 outburst was a factor of $\sim$5 longer and a factor of $\sim$5 more luminous than the 2006 outburst \citep[Table~\ref{tab:spec} and Fig.~\ref{fig:ascabron}, see also][]{degenaar2010_gc}. \\


\noindent{\bf X-ray spectra:} Spectral analysis of the \chan/ACIS and \xmm\ data obtained during the 2007--2008 outburst of \ascabron\ is complicated by pile-up. We attempted to circumvent the expected effect on spectral shape and source flux by extracting source event from an annular region with a radius of $10''-40''$. Fig.~\ref{fig:ascabron} displays the \xmm/PN spectra obtained in 2006 February and 2008 March, which represent the two different outbursts of the source. We converted upper limits and HRC-I count rates into unabsorbed 2--10 keV fluxes using $\Gamma=2.0$ and $N_{\mathrm{H}}=21.8\times10^{22}~\nh$.\\

\noindent{\bf Activity after our campaign:} \ascabron\ was again reported to be active in 2010 June \citep[][]{degenaar2010_atel_grs_xmm,degenaar2010_atel_asca}. The outburst persisted until 2010 late-October, but the activity had ceased when the \swift\ monitoring observations resumed in 2011 early-February. The 2010 outburst thus had a duration of $\sim$4--7~months. Both the length and the average 2--10 keV luminosity of this outburst are very similar to the 2006 activity.

\subsection{\saxbron}\label{subsec:saxbron}
\noindent{\bf Brief historic overview:} The neutron star LMXB \saxbron\ was initially discovered with \bepposax\ in 1998, although \granat\ may have detected activity from the source already in 1991 \citep[][]{grebenev2002}. Its neutron star nature has been established by detection of type-I X-ray bursts \citep[][]{zand1998_saxj1747,sidoli99}, which results in a distance estimate of $D=6.7$~kpc \citep[][]{galloway06}. Following its discovery, several accretion outbursts with 2--10 keV luminosities of $L_X\sim10^{36-37}~(D/6.7~\mathrm{kpc})^2~\lum$ have been detected with different X-ray instruments \citep[][]{wijnands2002_saxj1747,natalucci2004,werner2004,markwardt2004, deluit2004}. \\

\noindent{\bf Activity during our campaign and outburst constraints:} We detected \saxbron\ in outburst during three consecutive observations performed on 2005 October 20 \citep[\chan/HRC][]{wijnands2005_atel637,wijnands05_atel638}, 2006 February \citep[][]{wijnands06_atel892} and 2006 September (the latter are both \xmm\ observations). \inte\ detected hard X-ray activity from the source at similar epochs \citep[2005 October and 2006 February;][]{kuulkers2005,chenevez2006}. Our 2006-February observation revealed a pair of type-I X-ray bursts with a recurrence time of only 3.8~min (see Section~\ref{subsec:saxlc}).  

\begin{figure}
 \begin{center}
\includegraphics[width=8.0cm,angle=0]{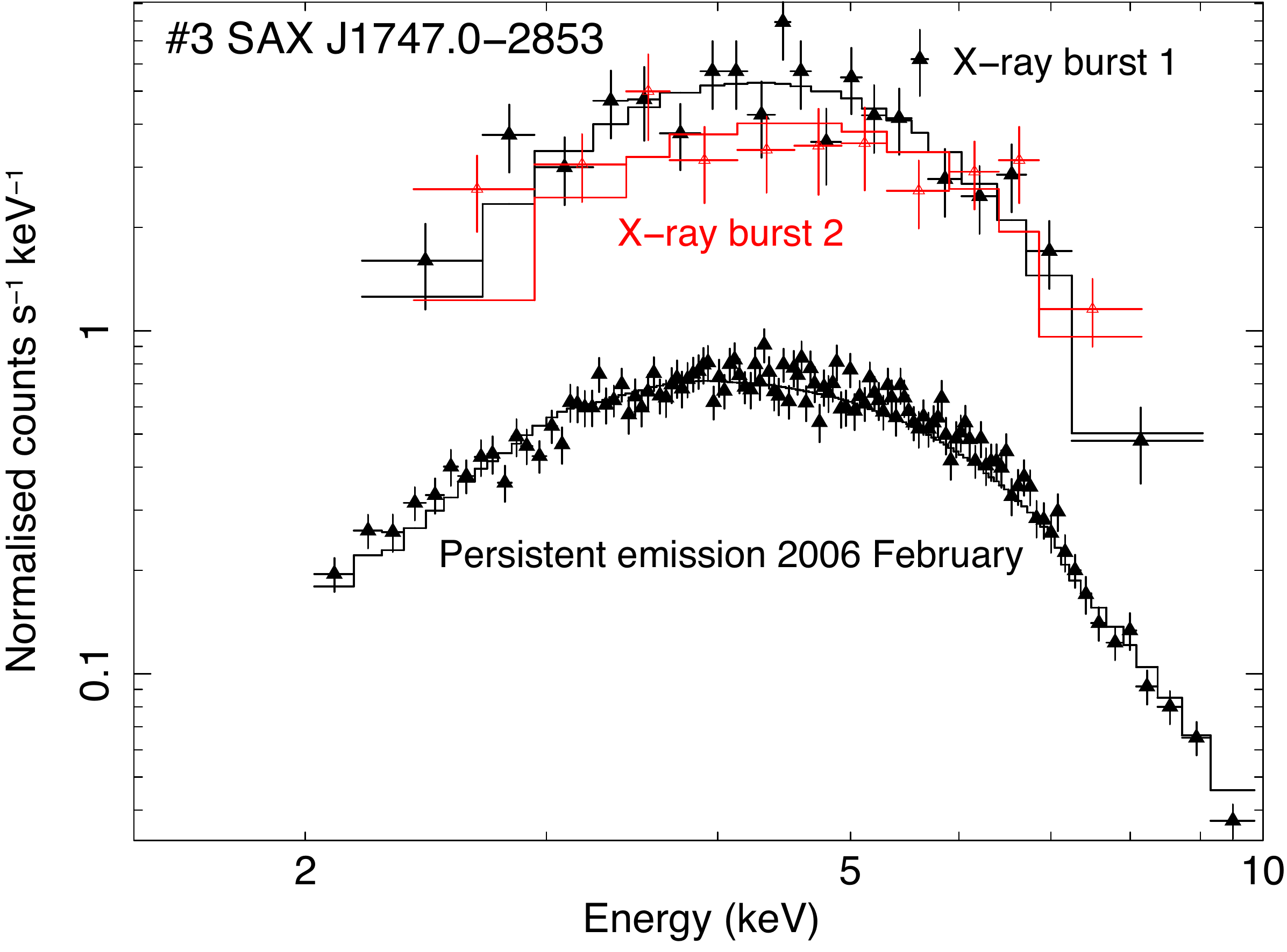}\vspace{0.1cm}
\includegraphics[width=8.0cm]{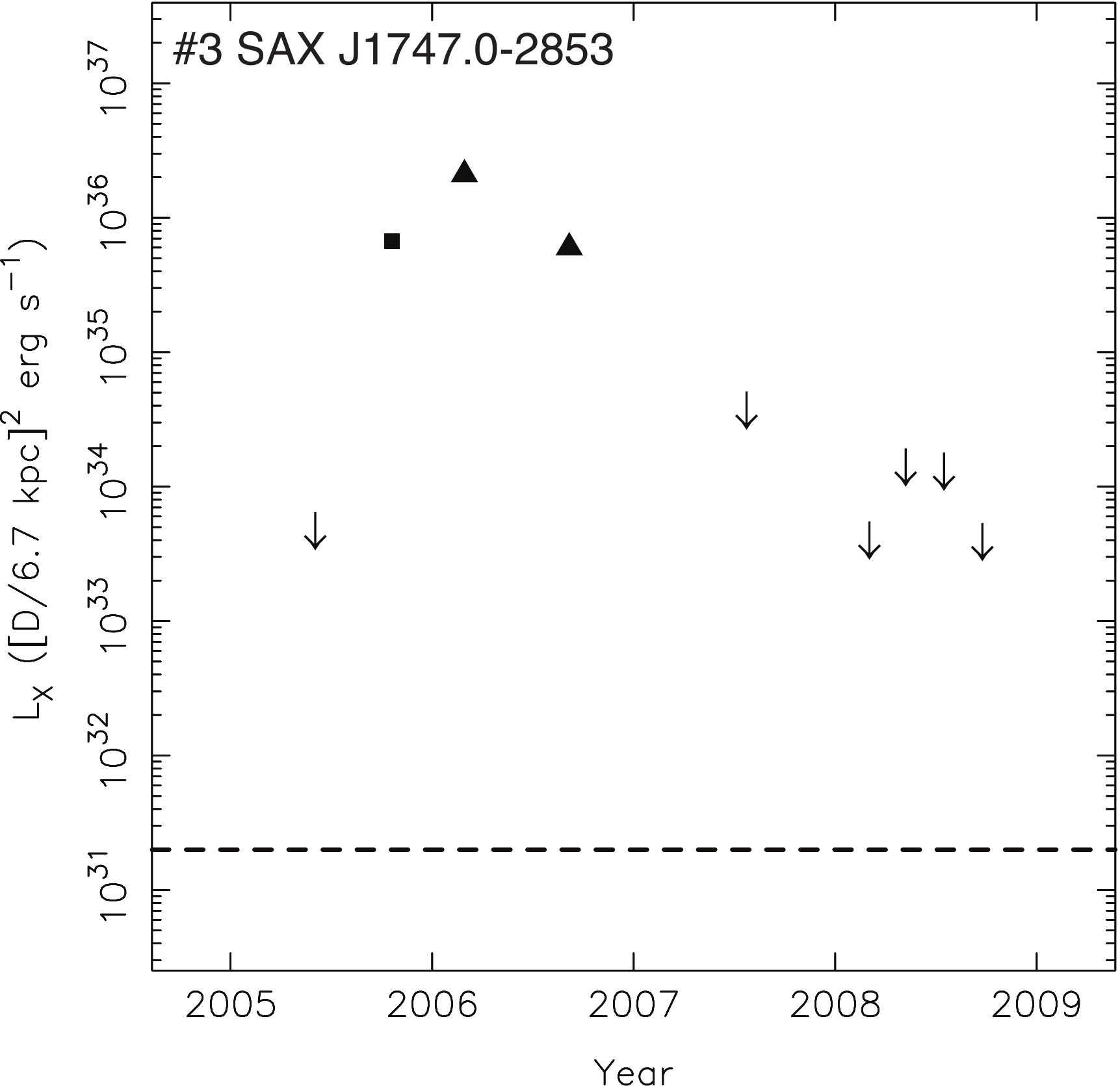}
    \end{center}
    \caption[]{Background-corrected \xmm\ spectra (top) and 2--10 keV luminosity evolution (bottom) of the bursting neutron star LMXB \saxbron. We have included the \xmm/MOS spectra of the two type-I X-ray bursts that were observed from the source on 2006 February 26 (see Section~\ref{subsec:saxlc}). In the lightcurve the square indicates \chan/HRC data and the triangles \xmm\ observations. The upper limit symbols represent a $2\sigma$ confidence level and the horizontal dashed line indicates the quiescent luminosity of the source.  }
 \label{fig:saxbron}
\end{figure}

The source was not detected during subsequent observations (2007 July), from which we can estimate an upper limit of $L_X\lesssim 3.7 \times10^{34}~\lum$. This indicates that the source activity had ceased by that time (see also Fig.~\ref{fig:saxbron}).\footnote{We note that the field in which \saxbron\ is closest to aimpoint, GC-3, was not observed in 2007 July, but during that epoch the source region was covered by the observations of GC-1.} If the source was continuously active  between our monitoring observations of 2005 October and 2006 September, the 2005--2006 outburst had a duration of $\gtrsim11$~months. Non-detections in our data obtained in 2005 June and 2007 July constrain the duration of the active period to be $\lesssim2.1$~yr (see Fig.~\ref{fig:saxbron}).

Although we did not detect \saxbron\ in 2007 July, it was seen active with \inte\ in 2007 October \citep[][]{brandt2007}. \swift/XRT follow-up observations carried out a few days later detected the source at $L_X\sim10^{36}~(D/6.7~\mathrm{kpc})^2~\lum$, confirming the renewed activity \citep[][]{cackett2007}. The source region was not covered by our monitoring observations of 2007 September, but \saxbron\ was not detected in \xmm\ data obtained in 2008 March. This suggests that the new 2007-outburst had a duration of $\lesssim7$~months. \\

\noindent{\bf X-ray spectra:} We fitted the \xmm\ spectra obtained in 2006 February (shown in Fig.~\ref{fig:saxbron}) and September to constrain the spectral shape. The February observation contained two X-ray bursts (see Section~\ref{subsec:saxlc}), which were removed from the data for the purpose of fitting the persistent emission. A simultaneous fit to the two spectra yields $N_{H}=(9.5\pm0.2)\times10^{22}~\nh$ and $\Gamma=2.0\pm0.1$, $2.6\pm0.1$ (see Table~\ref{tab:spec}). 

The inferred 2--10 keV luminosities for the 2006 February and September observations are $L_X\sim2.1\times10^{36}$ and $\sim6.0\times10^{35}~(D/6.7~\mathrm{kpc})^2~\lum$, respectively. Using $N_{H}=9.5\times10^{22}~\nh$ and $\Gamma=2.6$, we can estimate a 2--10 keV luminosity of $L_X\sim6.7\times10^{35}~(D/6.7~\mathrm{kpc})^2~\lum$ for the HRC-I observation of 2005 October and upper limits for other epochs of $L_X\lesssim(4-37)\times 10^{33}~(D/6.7~\mathrm{kpc})^2~\lum$ (see Fig.~\ref{fig:saxbron}). In Section~\ref{subsec:saxlc} we further constrained the quiescent level of \saxbron\ to $L_X\sim2\times10^{31}~(D/6.7~\mathrm{kpc})^2~\lum$. This confirms its classification as a transient X-ray source, as proposed by \citet{werner2004}.\\

\noindent{\bf Activity after our campaign:} In 2009 February, \inte\ picked up an X-ray burst from \saxbron\ without detectable persistent emission, suggesting an outburst luminosity of $L_X\lesssim4\times10^{35}~(D/6.7~\mathrm{kpc})^2~\lum$ \citep[][]{chenevez2009}. Follow-up \swift/XRT observations performed a few days later confirmed that the source was undergoing a faint outburst with  $L_X\sim2\times10^{35}~(D/6.7~\mathrm{kpc})^2~\lum$ \citep[][]{campana2009}. 

In 2011 January--February, renewed activity was seen by \maxi\ \citep[][]{suwa2011}, \swift/XRT \citep[][]{kennea2011} and \inte\ \citep[][]{kuulkers2011}. During this time \inte\ detected the first superburst ever observed from this source \citep[][]{chenevez2011_superburst}. The reported fluxes translate into luminosities of $L_X\sim3\times10^{37}~(D/6.7~\mathrm{kpc})^2~\lum$ \citep[][]{kennea2011,kuulkers2011,chenevez2011_superburst}. 

Summarising, in recent years \saxbron\ has exhibited outbursts in 2006--2007, 2007, 2009 and 2011. The source is therefore very frequently active. \\

 \begin{figure}
 \begin{center}
\includegraphics[width=8.0cm,angle=0]{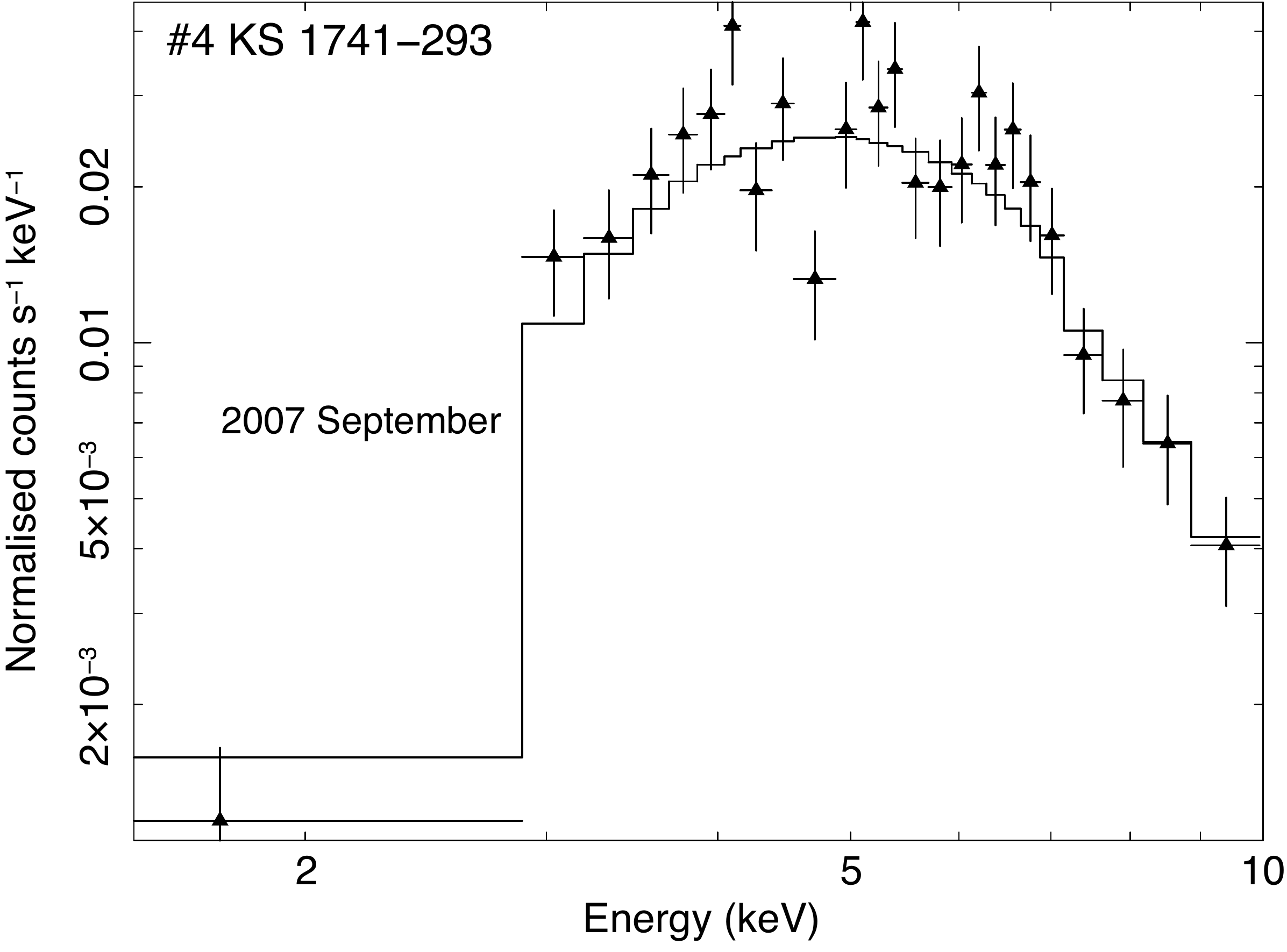}\vspace{0.1cm}
\includegraphics[width=8.0cm]{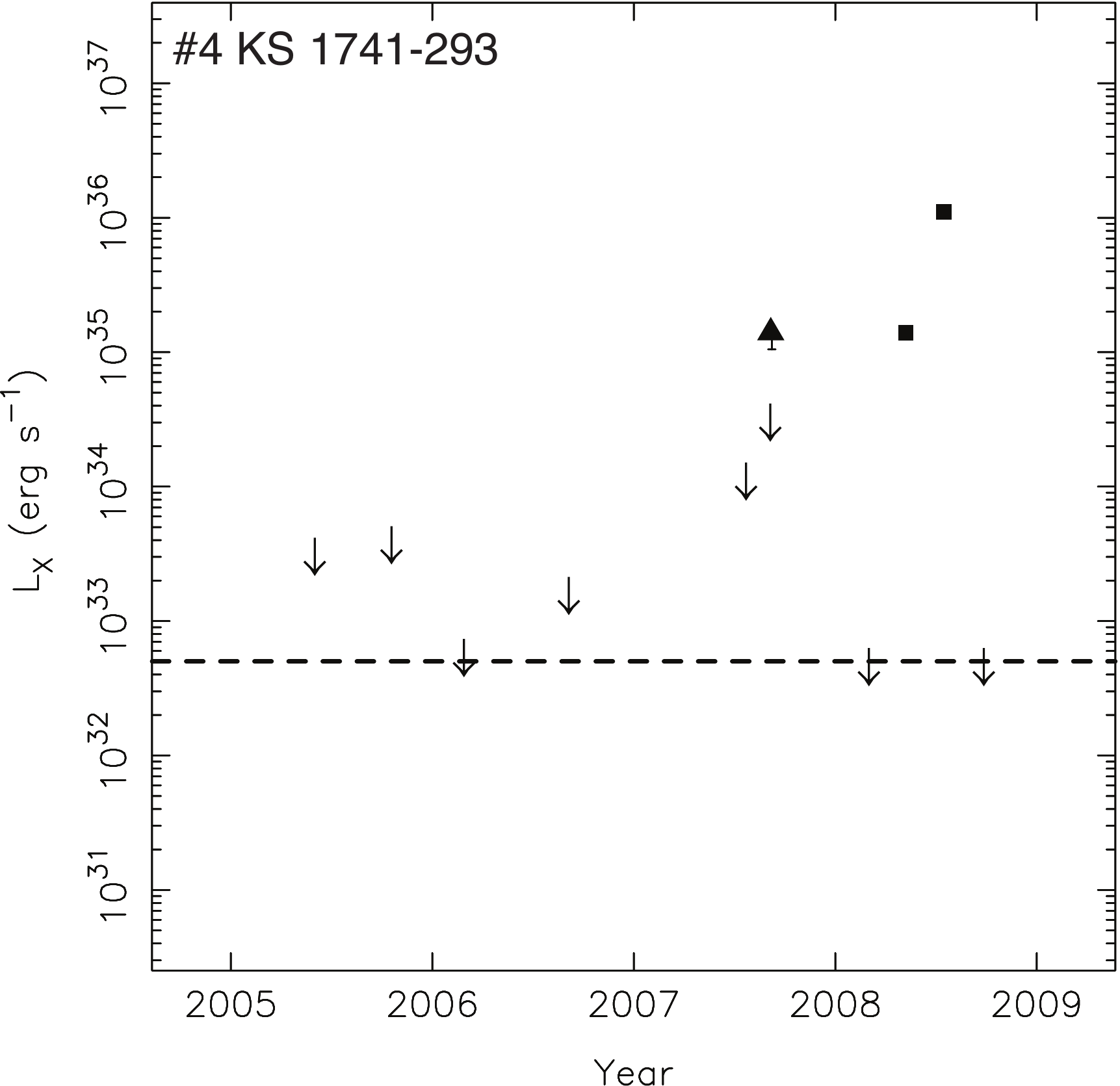}
    \end{center}
    \caption[]{Background-corrected \xmm/PN spectrum (top) and 2--10 keV luminosity evolution (bottom) of the bursting neutron star LMXB \ksbron. In the lightcurve the squares indicate \chan/HRC observations and the triangle \xmm\ data. The upper limit symbols represent a $2\sigma$ confidence level and the horizontal dashed line indicates the quiescent luminosity of the source.  }
 \label{fig:ksbron}
\end{figure}


\subsection{\ksbron}\label{subsec:ksbron}
\noindent{\bf Brief historic overview:} \ksbron\ is a transient neutron star LMXB that was discovered in 1989 August by the TTM onboard the KVANT module of the Mir space station \citep[][]{zand1991}. At that time, the source displayed two type-I X-ray bursts, which revealed its binary nature and testified to the presence of a neutron star. \ksbron\ was again detected with \bepposax\ in 1998 March \citep[][]{sidoli01} and with \asca\ in 1998 September \citep[][]{sakano02}, exhibiting a luminosity of $L_X\sim10^{35-36}~\lum$ on both occasions (assuming $D=8$~kpc). It is likely that these two observations caught the same outburst, which implies a duration of several months. 

\inte\ detected \ksbron\ in outburst in 2003 March and in 2004 March, during which two type-I X-ray bursts were observed \citep[][]{cesare2007}. Furthermore, in 2005 July the source was serendipitously observed with \chan\ and detected at a luminosity of $L_X\sim1\times10^{36}~\lum$ \citep[][]{degenaar08_atel_gc_chan}. \inte\ revealed hard X-ray activity from the source in 2005 August--September \citep[][]{kuulkers07}, and \suzaku\ detected the source in outburst in 2005 September \citep[][]{yuasa2008}. It therefore seems that the source was active for several months in 2005, with an intensity similar to the 1989 and 1998 outburst levels. \\

\noindent{\bf Activity during our campaign:} \ksbron\ is located in the field GC-7 of our campaign and was found active during two different epochs (2007 and 2008). We first detected the source in outburst with \xmm\ on 2007 September 6. It was not detected in subsequent \xmm\ observations carried out on 2008 March 4, with an upper limit on the PN count rate of $\lesssim 3\times10^{-3}~\cnts$. Using the spectral parameters given in Table~\ref{tab:spec}, we can infer an upper limit on the 2--10 keV luminosity of $L_X\lesssim 1 \times10^{33}~\lum$. This demonstrates that the outburst had ended and the source had returned to quiescence.\\

\noindent{\bf X-ray spectrum:} We obtained a source spectrum from the 2007 \xmm\ observation, which is shown in Fig.~\ref{fig:ksbron} (only the PN data is displayed). The source was faint enough not to cause pile-up in the EPIC instruments. The spectral data can be fitted with $N_{\mathrm{H}}=(16.6\pm1.8)\times10^{22}~\nh$ and $\Gamma=1.8\pm0.3$, resulting in a luminosity of $L_X\sim1.4\times10^{35}~\lum$ (Table~\ref{tab:spec}). \\

\noindent{\bf Outburst constraints:} We searched through the \swift/XRT data archive in an attempt to constrain the duration of the 2007 outburst of \ksbron. The source field was covered during a number of pointings performed between 2007 May 23 and August 9. A total of 55 observations were carried out during this episode with an accumulated exposure time of $\sim$57~ks. 

Examination of the \swift\ data reveals that \ksbron\ resided in quiescence for the larger part of this epoch, but that it exhibited enhanced activity during a short four-day interval between 2007 June 11--15, when its intensity rose to $L_X\sim 3 \times 10^{34}~\lum$. The source is not detected in composite images both before (total exposure of $\sim$10~ks) and after ($\sim$43~ks) these dates. These considerations suggest that the 2007 outburst of \ksbron\ was confined within an interval between 2007 August 9 and 2008 March 4, implying a duration of $\lesssim6$~months. 

The short, sub-luminous outburst observed from \ksbron\ is reminiscent of a one-week period of enhanced X-ray activity that was detected from the neutron star LMXB \grsbron\ approximately four months before it started a major outburst \citep[][see also Section~\ref{subsec:grsbron}]{degenaar09_gc}. The mini-outburst of \ksbron\ occurred 1--2 months before the source erupted in its brighter and longer 2007 outburst.

The source was again active during our \chan/HRC observations of 2008 May 10 \citep[][]{degenaar08_atel_gc_chan} and July 16 (see Fig.~\ref{fig:ksbron}). Using the spectral parameters reported in Table~\ref{tab:spec}, the observed HRC count rates can be translated into luminosities of $L_X\sim1\times10^{35}$ and $\sim1\times10^{36}~\lum$, respectively. We obtained several \swift/XRT ToO follow-up observations spread $\sim$2~weeks apart to constrain the properties of this outburst \citep[][]{degenaar2008}. These suggest that the source remained active for approximately four more months following the HRC detection of 2008 May, and returned to quiescence between 2008 August 21 and September 4. We can therefore constrain the duration of the 2008 outburst to be $\sim$3--6~months. The average luminosity inferred from the \swift\ observations is $L_X\sim10^{35}~\lum$, peaking at $L_X\sim2\times10^{36}~\lum$. \\

\noindent{\bf Activity after our campaign:} The results from our monitoring observations demonstrate that \ksbron\ is frequently active and undergoes relatively long outbursts that last several months, which is also reflected by the historic detections of the source. \ksbron\ was again detected in outburst with \inte\ in 2010 March \citep[][]{chenevez2010}. Reneweed activity was observed in 2011 September with \swift\ \citep[2--10 keV luminosity of $L_X\sim2\times10^{35}~\lum$;][]{linares2011} and \inte\ \citep[][]{chenevez2011_transients}. In the past decade \ksbron\ was thus detected in an active state in 2003, 2004, 2005, 2007, 2008, 2010, and 2011. It appears that the outbursts of the source are typically several months in duration and reach 2--10 keV luminosities of $L_X\sim10^{35-36}~\lum$.


\subsection{\grobron}\label{subsec:grobron}
\noindent{\bf Brief historic overview:} The  ``bursting pulsar" \grobron\ is a transient neutron star LMXB that was discovered in 1995 with the BATSE on-board the \textit{Compton Gamma Ray Observatory} \citep[][]{fishman1995}. Its neutron star nature was established by the detection of coherent X-ray pulsations with a frequency of 2.1 Hz, which allowed determining the orbital period of the binary \citep[11.8~d;][]{finger1996}. Major outbursts reaching $L_X\sim10^{37-38}~\lum$ were observed in 1995 and 1996 \citep[for $D=8$~kpc,][]{woods1999}. The source has been observed in quiescence at a 0.5--10 keV luminosity of $L_X\sim(1-3)\times10^{33}~\lum$ \citep[][]{wijnands2002_gro1744,daigne2002}, but enhanced activity above the quiescent level has also been reported \citep[][see also Section~\ref{subsec:weak}]{muno07_atel1013}.\\

\begin{figure}
 \begin{center}
\includegraphics[width=8.0cm,angle=0]{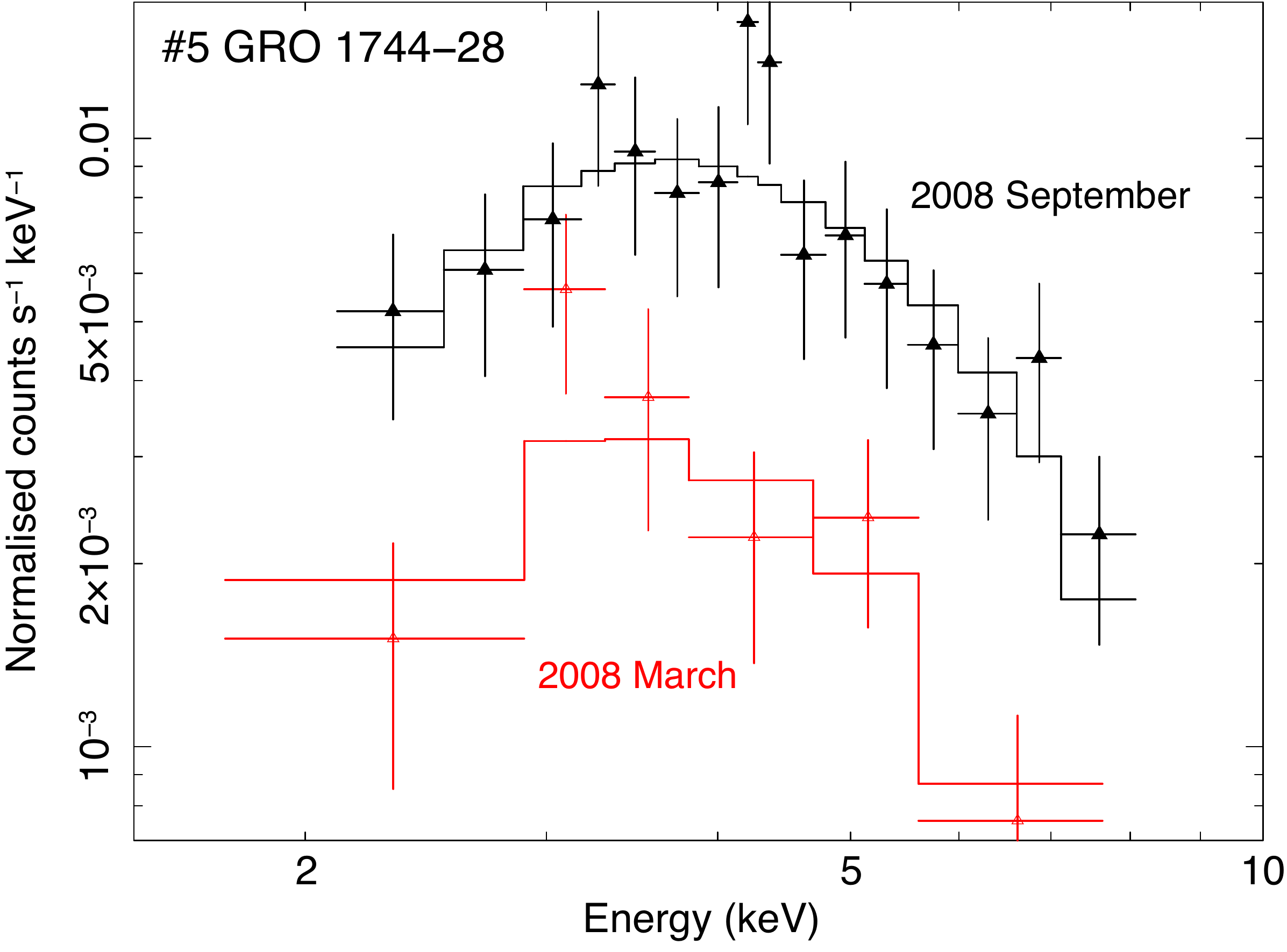}\vspace{0.1cm}
\includegraphics[width=8.0cm]{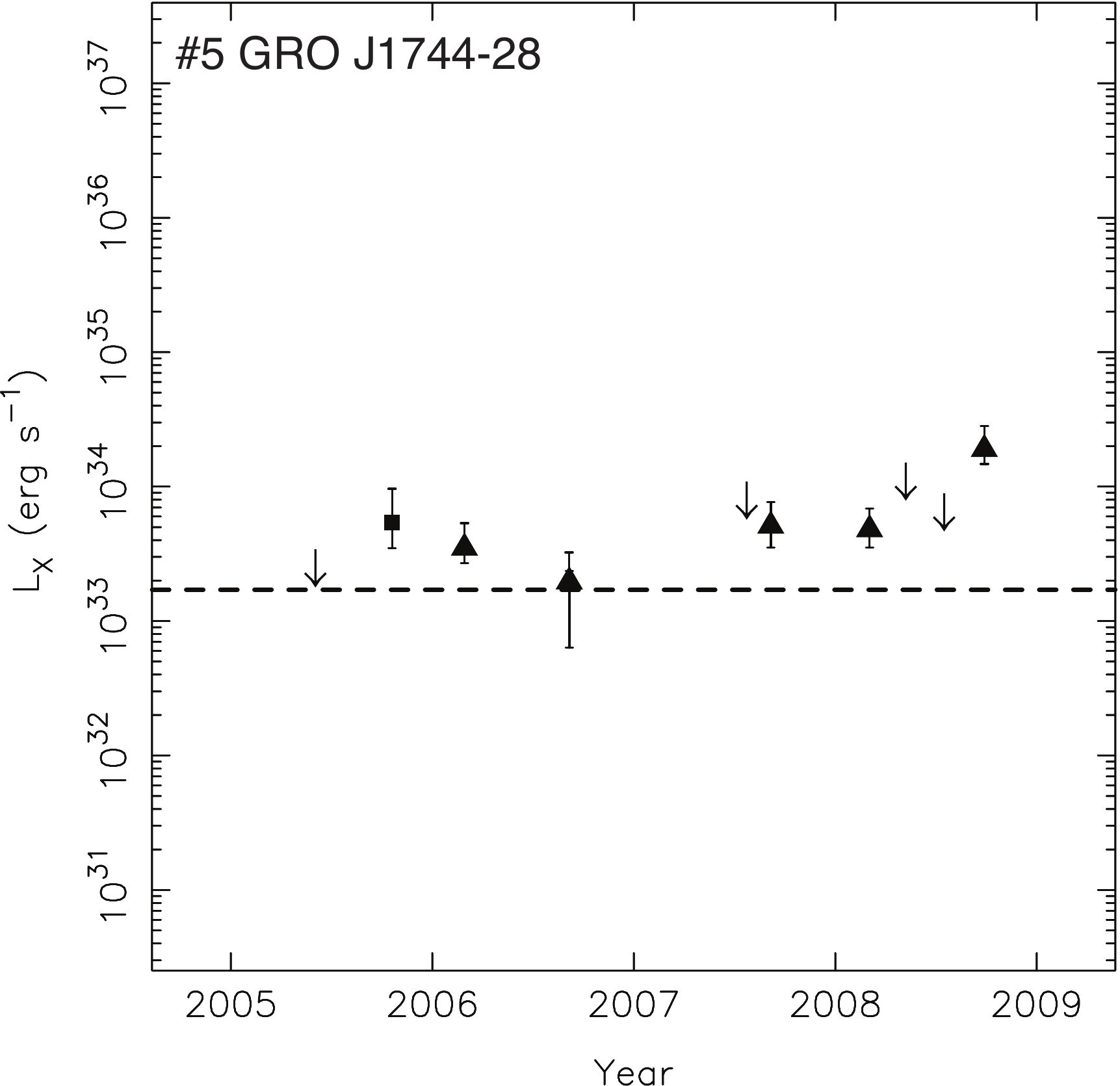}
    \end{center}
    \caption[]{Background-corrected \xmm/PN spectra (top) and 2--10 keV luminosity evolution (bottom) of the pulsating neutron star LMXB \grobron. In the lightcurve the square indicates \chan/HRC data and the triangles \xmm\ observations. The upper limit symbols represent a $2\sigma$ confidence level and the horizontal dashed line indicates the quiescent luminosity of the source.  }
 \label{fig:grobron}
\end{figure}

\noindent{\bf Activity during our campaign:} \grobron\ was detected during all our \xmm\ observations, which went sufficiently deep to detect this source at its quiescent level. The source intensity was, however, near the detection limit of our HRC data (see Fig.~\ref{fig:grobron}). \\

\noindent{\bf X-ray spectra:} For the spectral fits we omitted the \xmm\ data obtained in 2006 September, because they collected only few photons from \grobron. Using an absorbed powerlaw model to fit the other four \xmm\ observations simultaneously, we obtain $N_{H}=(9.4\pm3.0)\times10^{22}~\nh$ and $\Gamma=2.5-3.2$ (see Table~\ref{tab:spec}). The inferred 2--10 keV luminosities of the first observations lie between $L_X\sim(1.5-6.5)\times10^{33}~\lum$, whereas the final observation (2008 September) suggests enhanced activity at $L_X\sim1.9\times10^{34}~\lum$ (see Fig.~\ref{fig:grobron}). Fig.~\ref{fig:grobron} compares the PN spectra obtained in 2008 March and September and illustrates the enhanced activity seen in the September observation. 

With the exception of the last observation, the luminosities inferred for the \xmm\ data are similar to what has been found for \grobron\ in quiescence \citep[][]{wijnands2002_gro1744}. Since (non-pulsating) neutron star LMXBs often display a thermal spectrum in quiescence that is thought to be heat radiated from the surface of the neutron star, we also fitted a blackbody model to the spectral data. We excluded the last observation (2008 September) that hinted at enhanced activity. 

A blackbody fit yields $N_{H}=(5.9\pm2.7)\times10^{22}~\nh$, a temperature $kT_{\mathrm{bb}}=1.0\pm0.3$~keV and an emitting radius of about $R_{\mathrm{bb}}\sim0.1$~km ($\chi_{\nu}=0.86$ for 40 d.o.f.). The inferred temperature is unusually high for a quiescent neutron star, because these are typically found at $kT_{\mathrm{bb}}\sim0.2-0.3$~keV \citep[e.g.,][]{rutledge1999}. Moreover, the detection of this soft thermal component from \grobron\ suffers considerably from the high Galactic extinction in the direction of the source. The mechanism responsible for the quiescent emission of \grobron\ is therefore a subject of debate \citep[][]{wijnands2002_gro1744,daigne2002}.

The spectral index for the powerlaw fit is very similar to that of the hard emission tails that are often detected for neutron star LMXBs above $\sim$2--3~keV \citep[e.g.,][]{rutledge1999}. Its origin is not well understood, but possible explanations include accretion onto the surface or the magnetic field of the neutron star, or a shock generated between the pulsar wind and the material flowing out of the donor star \citep[e.g.,][]{campana1998}. 

We used the results of the powerlaw fits to our \xmm\ data to convert the HRC-I count rates and upper limits to 2--10 keV fluxes and luminosities (Table~\ref{tab:spec} and Fig.~\ref{fig:grobron}).

\subsection{\xmmbron}\label{subsec:xmmbron}
\noindent{\bf Brief historic overview:} \xmmbron\ is an unclassified X-ray transient that was first detected in outburst in 2001 \citep[][]{sakano05} and has been seen active on numerous occasions since \citep[][]{wijnands06,muno07_atel1013,degenaar09_gc,degenaar2010_gc, degenaar2010_atel_grs_xmm}. Its quiescent luminosity is $L_X\sim10^{32}~\lum$ \citep[][]{sakano05}, and it exhibits outbursts with a peak luminosity of $L_X\sim10^{36}~\lum$. However, the source is often detected at intermediate intensities of $L_X\sim10^{33-34}~\lum$ \citep[][see also Section~\ref{subsec:weak}]{degenaar2010_gc}. \\

\noindent{\bf Activity during our campaign:} We detected activity from \xmmbron\ several times during our campaign. As reported by \citet{wijnands06}, \xmmbron\ was detected both during the HRC observations performed on 2005 June 5 \citep[see also][]{wijnands05_atel512} and the ACIS follow-up pointing carried out on July 1. The  luminosities inferred from these observations are $L_X\sim 8\times10^{35}$ and $\sim 3 \times 10^{33}~\lum$, respectively (assuming $D=8$~kpc). This indicates that the intensity decayed by two orders of magnitude within four weeks. 

\begin{figure}
 \begin{center}
\includegraphics[width=8.0cm,angle=0]{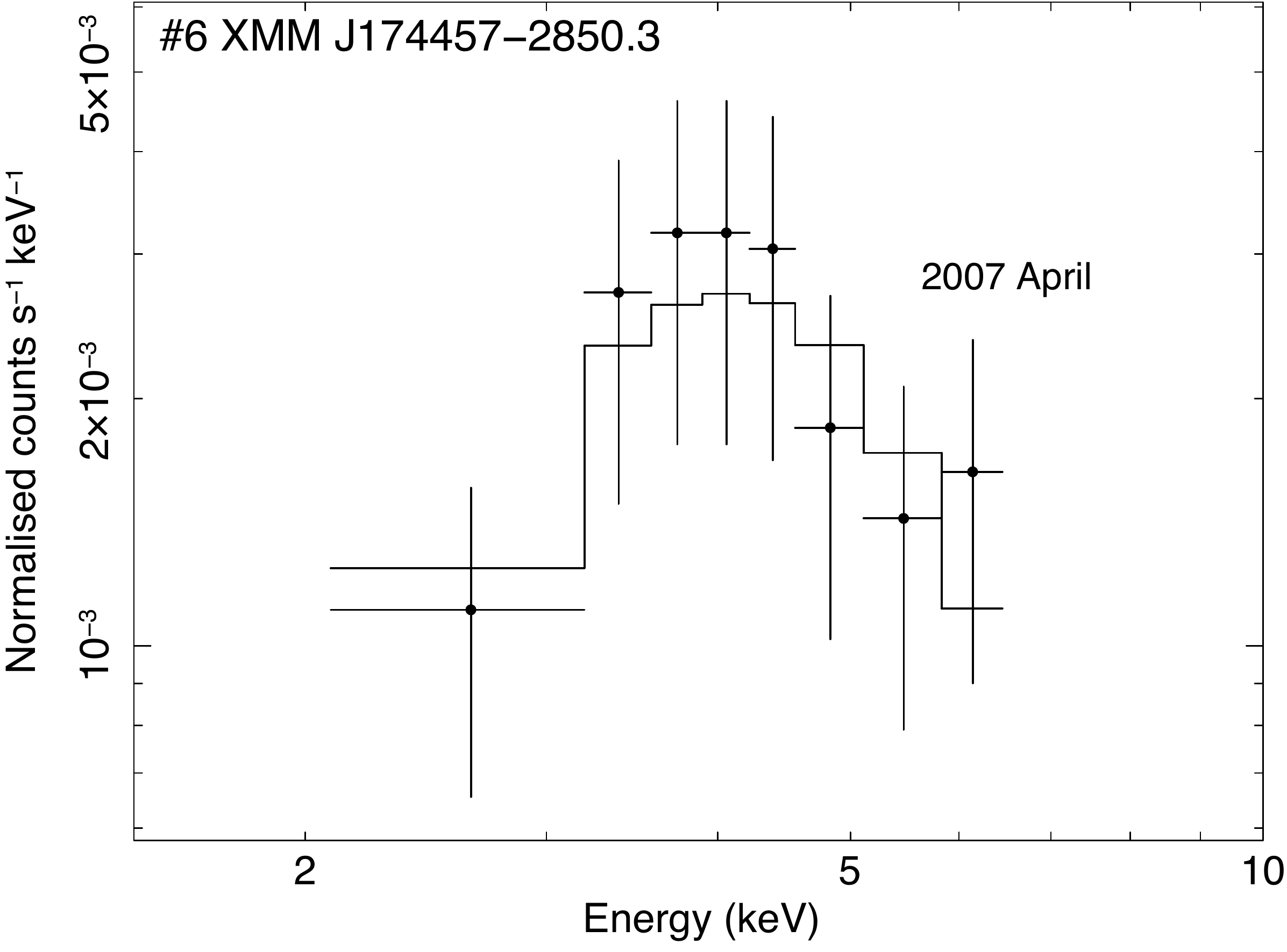}\vspace{0.1cm}
\includegraphics[width=8.0cm]{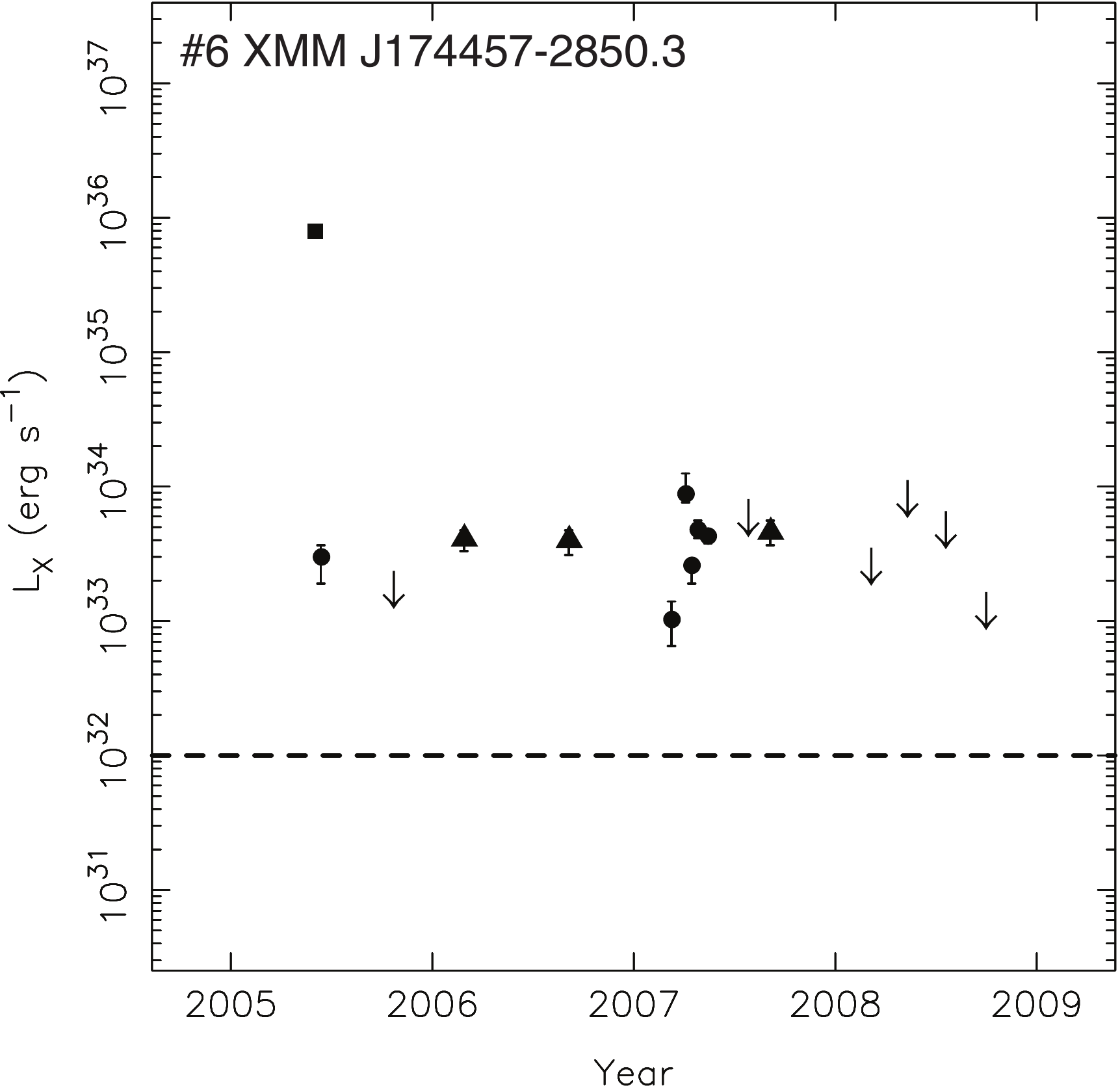}
    \end{center}
    \caption[]{Background-corrected \chan/ACIS spectrum (top) and 2--10 keV luminosity evolution (bottom) of the unclassified X-ray transient \xmmbron. In the lightcurve the square (HRC) and bullets (ACIS) indicate \chan\ data, whereas triangles are used for \xmm\ observations. The upper limit symbols represent a $2\sigma$ confidence level and the horizontal dashed line indicates the quiescent luminosity of the source.  }
 \label{fig:xmmbron}
\end{figure}

Although we detected the source in several other observations during the campaign, we never found it to be as bright as in 2005. In 2006 February and September (\xmm), 2007 March--May (\chan/ACIS) and 2007 September (\xmm) we detected it at $L_X\sim (1-9) \times 10^{33}~\lum$ (see Table~\ref{tab:spec}). As argued in Sections~\ref{subsec:weak} and~\ref{subsec:lowlum_activity}, these low-intensity states likely represent low-level accretion of the source. \xmmbron\ is not detected during the other epochs of our campaign, but the inferred upper limits are comparable to the above mentioned detections (see Fig.~\ref{fig:xmmbron}). \\

\noindent{\bf X-ray spectra:} \xmmbron\ is so faint that most of our observations collect only $\sim$20--30 source photons, prohibiting accurate spectral modelling. We chose to only fit the spectra from the \chan/ACIS observations 6603 and 6605, which collected the largest number of source photons ($\sim$50--60 each). Since the hydrogen column density remains unconstrained when left to vary freely, we fixed this parameter to the value inferred from \swift/XRT data \citep[$N_{\mathrm{H}}=7.5\times10^{22}~\nh$;][]{degenaar2010_gc}. This resulted in $\Gamma=1.3\pm1.0$ and $1.7\pm1.1$ for 2007 March 12 and April 30, respectively. 

Within the errors this spectral shape is consistent with results obtained in other works \citep[][]{sakano02,wijnands06,degenaar2010_gc}. For the remaining observations we converted the detected count rates into 2--10 keV unabsorbed fluxes by adopting the above mentioned hydrogen absorption column density and a photon index of $\Gamma=1.5$. The spectrum extracted from the \chan/ACIS observation performed on 2007 April 30 is plotted in Fig.~\ref{fig:xmmbron}.\\

\noindent{\bf Constraints on the bright 2008 outburst:} \swift/XRT uncovered a relatively bright outburst from \xmmbron\ in 2008 in late June. During the first observation in which the source was within FOV, it was detected at a luminosity of $L_X\sim1\times10^{36}~\lum$. Subsequent observations revealed a decline in source intensity down to $L_X\sim 5 \times 10^{33}~\lum$ within one week \citep[][]{degenaar2010_gc}. The \swift/GC monitoring observations did not cover the source region before the outburst peak, leaving the duration of this relatively bright ($L_X\gtrsim  10^{34}~\lum$) episode unconstrained \citep[][]{degenaar2010_gc}. 

Since \xmmbron\ is not detected in our \chan/HRC data obtained on 2008 May 10, yielding an upper limit on the 2--10 keV luminosity of $L_X\sim 8\times10^{33}~\lum$, we can infer that the luminous phase seen by \swift/XRT must have had a duration of $<49$~days. \\

\noindent{\bf Activity after our campaign:} In 2010 in late July \swift/XRT again detected a relatively bright episode from \xmmbron\ with an observed peak luminosity of $L_X\sim 2 \times 10^{35}~\lum$ and a duration of 5--16 days \citep[][]{degenaar2010_atel_grs_xmm}. In 2011, two weak flares were observed with \swift\ that had a maximum duration of a few days and an intensity of $L_X\sim 1 \times 10^{34}~\lum$. 

In summary, three relatively bright ($L_X\sim10^{35-36}~\lum$) outbursts have been observed from \xmmbron\ in 2005 (\chan), 2008 and 2010 (\swift), whereas it is frequently detected at $L_X\sim10^{33-34}~\lum$. Our monitoring observations support previous findings that the bright episodes of this source last only short times and that the source spends most of its time at luminosities that are a factor $\sim$10--100 above its quiescent level (see also Section~\ref{subsec:weak}). This peculiar behaviour has led to the speculation that this unclassified X-ray source is possibly a wind-accreting X-ray binary \citep[][]{degenaar2010_gc}. 

 \begin{figure}
 \begin{center}
\includegraphics[width=8.0cm,angle=0]{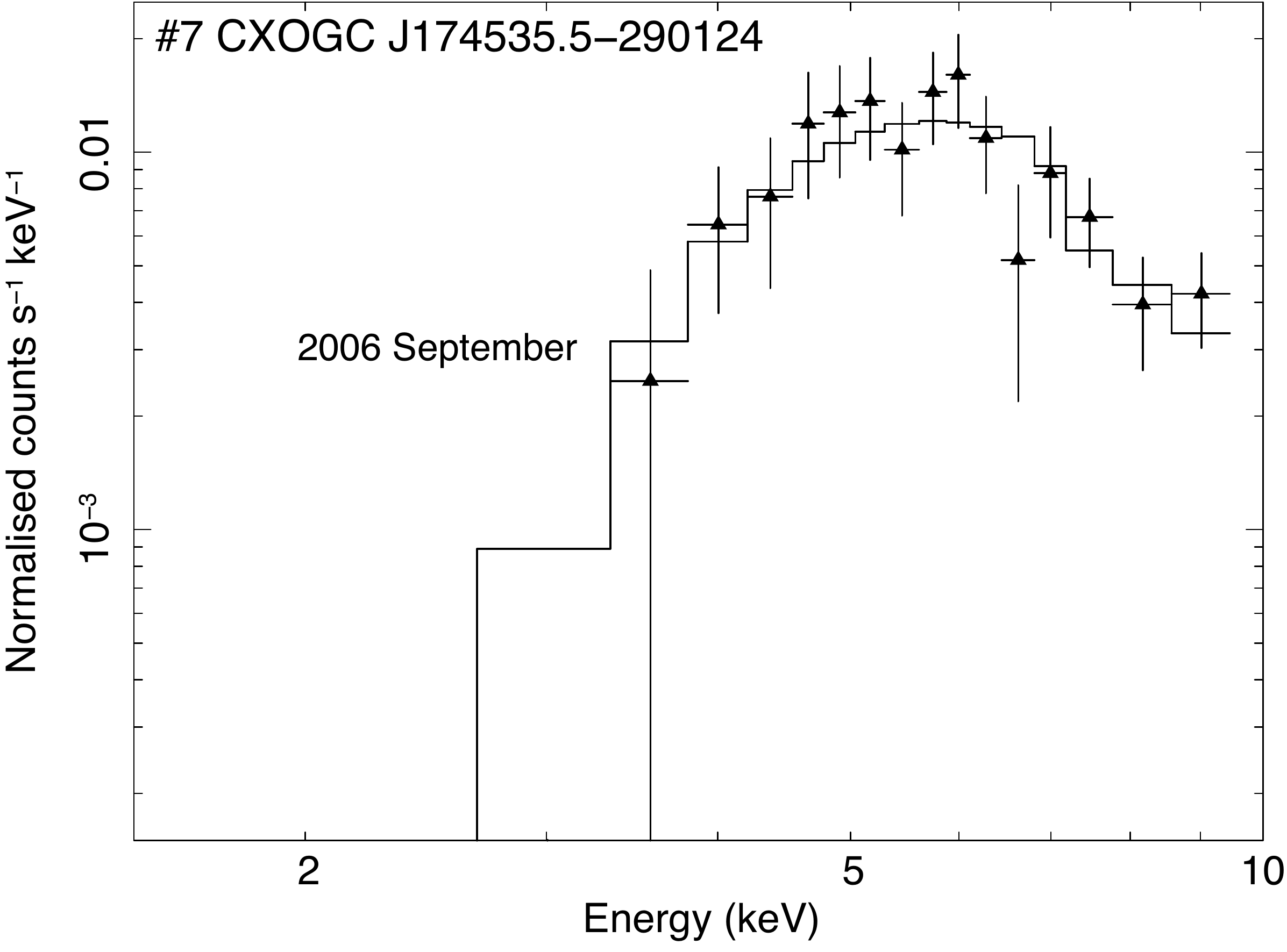}\vspace{0.1cm}
\includegraphics[width=8.0cm]{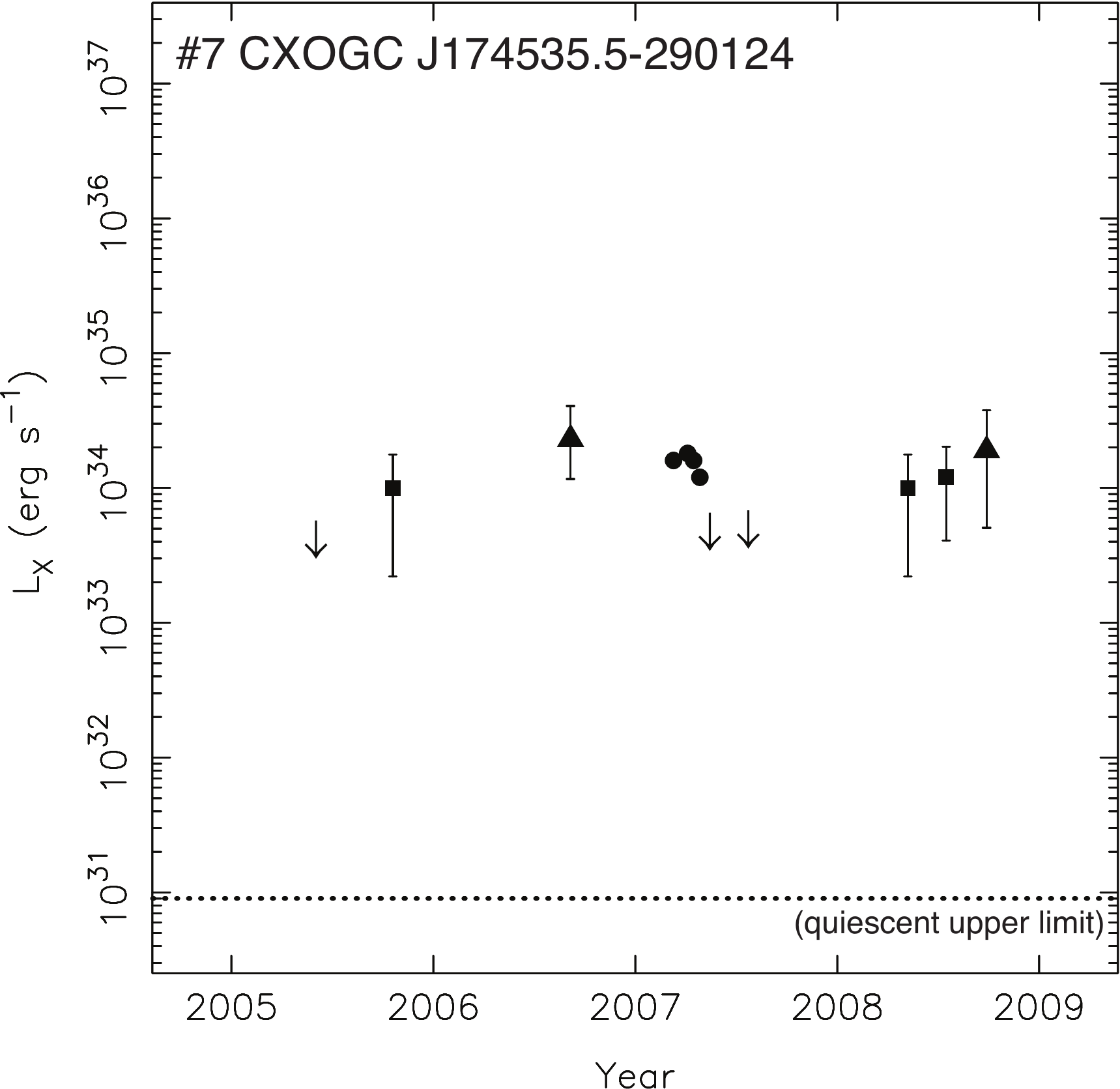}
    \end{center}
    \caption[]{Background-corrected \xmm/PN spectrum (top) and 2--10 keV luminosity evolution (bottom) of the unclassified X-ray transient \brontwee. In the lightcurve the squares (HRC) and bullets (ACIS) indicate \chan\ data, whereas triangles are used for \xmm\ observations. The upper limit symbols represent a $2\sigma$ confidence level and the horizontal dotted line indicates the upper limit on the quiescent luminosity of the source.  }
 \label{fig:brontwee}
\end{figure} 

\subsection{\brontwee}\label{subsec:brontwee}
\noindent{\bf Brief historic overview:} Several of our observations reveal activity of the unclassified transient X-ray source \brontwee. This object was first discovered during \chan\ observations in 2001 \citep[][]{muno03} and has frequently been observed at $L_X\sim10^{33-34}~\lum$ ever since \citep[for $D=8$~kpc;][]{muno05_apj622,wijnands05_atel638,wijnands06_atel892,degenaar08_atel_gc_chan,degenaar09_gc}. The source is not detected in quiescence down to a limiting luminosity of $L_X\lesssim9\times 10^{30}~\lum$ \citep[2--8 keV;][]{muno05_apj622}. 

Studies of this X-ray transient are complicated because of its close proximity to the neutron star LMXB \ascabron, which is a factor $\sim$100 brighter during outburst than \brontwee. The two sources are separated by only $\sim$$14''$, (see Fig.~\ref{fig:chan}). Therefore, sub-arcsecond spatial resolution is required to resolve the two sources whenever \ascabron\ is active \citep[see][]{degenaar2010_gc}. On all occasions that \ascabron\ is active (see Section~\ref{subsec:ascabron}), we were therefore only able to obtain information on \brontwee\ from \chan\ data.\\

\noindent{\bf Activity during our campaign:} This transient was active during our \chan\ observations performed in 2005 October \citep[HRC;][]{wijnands05_atel638}, 2007 March--May (ACIS), and 2008 May and July \citep[both HRC;][]{degenaar08_atel_gc_chan}. \ascabron\ resided in quiescence during our \xmm\ observations performed in 2006 and 2008 September (see Section~\ref{subsec:ascabron}): on both occasions we detected activity from \brontwee\ \citep[see also][]{wijnands06_atel892}. \\

\noindent{\bf X-ray spectra:} For the spectral fitting, we chose to use only the \xmm\ data since that excludes contamination from \ascabron, which resided in quiescence during those observations. The simultaneous fit results in $N_{\mathrm{H}}=(30.4\pm8.6)\times10^{22}~\nh$ and $\Gamma=1.7-2.5$ (see Table~\ref{tab:spec}), yielding $L_X\sim(1.9-2.4)\times10^{34}~\lum$. The spectral properties are therefore very similar at the two different epochs. We used these spectral parameters to convert the count rates and upper limits from the other observations into 2--10 keV unabsorbed fluxes (see Table~\ref{tab:spec}). Fig.~\ref{fig:brontwee} shows the \xmm/PN spectrum obtained on 2006 September 8.\\

\noindent{\bf Outburst constraints:} The long-term lightcurve of \brontwee\ obtained during our campaign is displayed Fig.~\ref{fig:brontwee}. This plot is suggestive of the source exhibiting two different outbursts in 2005--2007 and 2008, although the upper limits for the non-detections are close to the displayed level of activity. This makes it difficult to constrain the outburst and quiescent time scales from our observations. 

\swift/XRT observations are not conclusive either, because the spatial resolution of \swift\ does not allow us to separate \brontwee\ from the frequently active \ascabron. The \swift\ data set indicates that \brontwee\ was continuously active between 2006 July--November and from 2008 September till October (when \ascabron\ resided in quiescence).

In summary, the source was first detected during our campaign on 2005 October 20 and seen active during several observations performed between 2006--2008 with \chan, \xmm\ and \swift. The last detection during our campaign was obtained with \xmm\ on 2008 September 9 (see Table~\ref{tab:spec}). \swift\ resumed its monitoring observations of the source region on 2009 June 4 and did not detect activity from the source. This gives an upper limit on its intensity of $L_X\sim10^{33}~\lum$ and suggests that the source was quiescent at that time \citep[][]{degenaar2010_gc}. If the activity seen between 2005 and 2008 was part of the same outburst, it had a duration of $\gtrsim35$~months ($\gtrsim3$~yr; the time between the first and last detection during our campaign). If the source instead exhibited two different outbursts in 2005--2007 and 2008 (see Fig.~\ref{fig:brontwee}), the duration of these outbursts would be $\gtrsim$18 and $\sim$$ 6-25$~months, respectively.\\


\noindent{\bf Activity after our campaign:} \brontwee\ was not detected during the entire sample of 2009 \swift\ monitoring observations \citep[][]{degenaar2010_gc}, nor in 2010--2011. However, \ascabron\ was active from 2010 June--October, implying that during that epoch no information on the activity of \brontwee\ could be obtained from the \swift\ observations \citep[][]{degenaar2010_atel_asca}.


\subsection{\bronnegen}\label{subsec:bronnegen}
\noindent{\bf Brief historic overview:} \bronnegen\ is a weak unclassified X-ray transient discovered during \chan\ observations of the GC in 1999 \citep[][]{muno03}. This source was detected multiple times between 1999 and 2005, displaying luminosities of a few times $10^{33}~\lum$, whereas an upper limit on its quiescent level of $L_X\lesssim8\times10^{31}~\lum$ was inferred \citep[2--8 keV and assuming $D=8$~kpc;][]{muno05_apj622}. \\

\noindent{\bf Activity during and after our campaign:} We detected activity from \bronnegen\ during our \chan/HRC observations performed in 2005 October \citep[][]{wijnands05_atel638}. Using the spectral parameters reported by \citet[][]{muno04_apj613}, $N_{\mathrm{H}}=18.8\times10^{22}~\nh$ and $\Gamma=2.0$, we can convert the HRC count rate into a luminosity of $L_X\sim1.2\times10^{34}~\lum$. This is a factor $\sim$2 higher than the maximum intensity previously seen with \chan\ \citep[][]{muno04_apj613}. 

\begin{figure}[b]
 \begin{center}
\includegraphics[width=8.0cm]{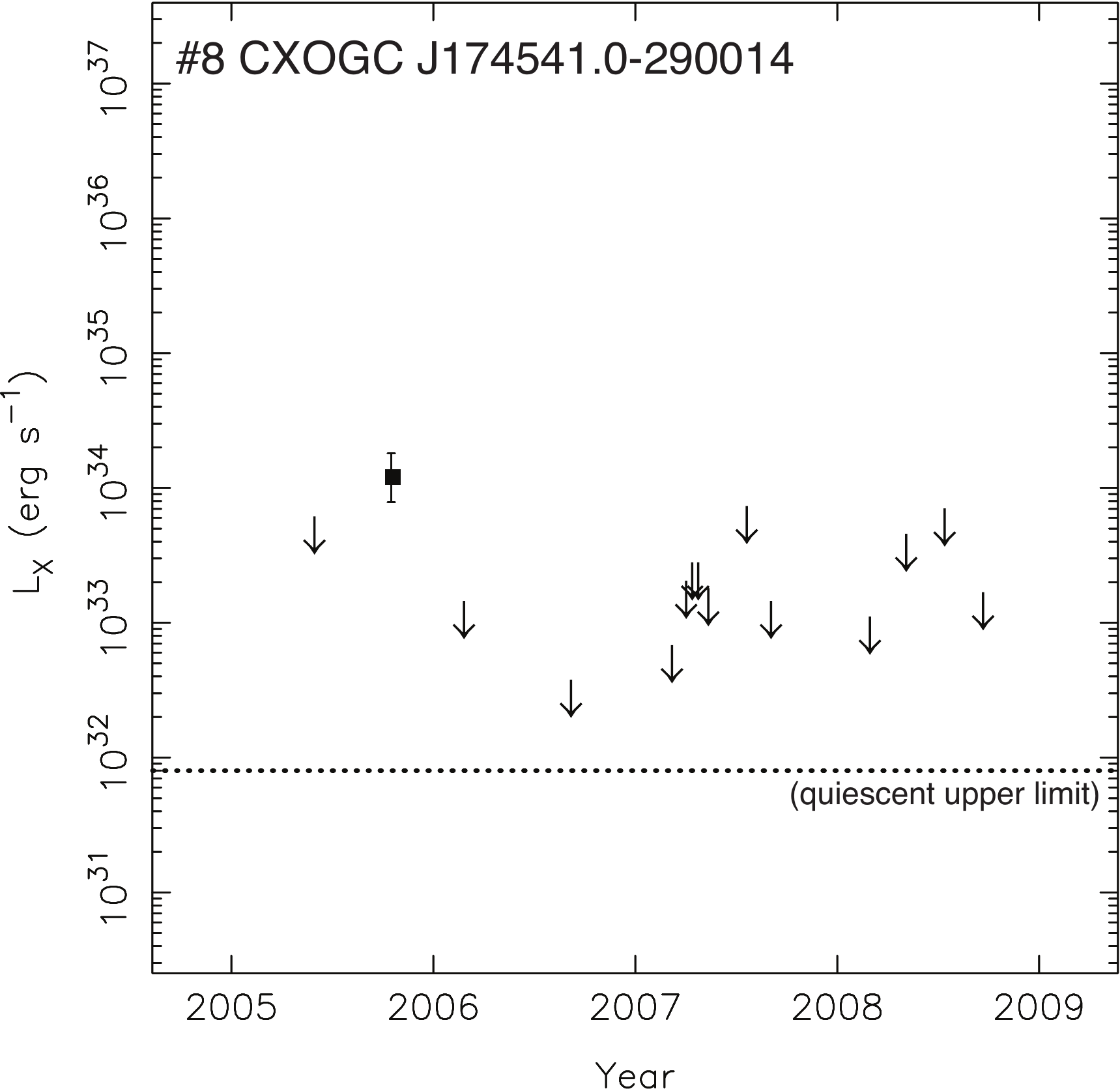}
    \end{center}
    \caption[]{Evolution of the 2--10 keV luminosity of the unclassified X-ray transient \bronnegen. The square in the lightcurve indicates \chan/HRC data. The upper limit symbols represent a $2\sigma$ confidence level and the horizontal dotted line indicates the upper limit on the quiescent luminosity of the source.  }
 \label{fig:bronnegen}
\end{figure}

Although \bronnegen\ was frequently active between 1999 and 2005, we detected the source only once during our campaign. Nevertheless, our obtained upper limits of $L_X\lesssim(0.5-5)\times10^{33}~\lum$ (see Fig.~\ref{fig:bronnegen}) are similar to the level of activity seen previously \citep[][]{muno05_apj622}. No activity has been reported from this source after our campaign ended.

\end{appendix}

\label{lastpage}
\end{document}